\documentclass{article}
\usepackage{amssymb}
\usepackage{amsmath}

\newtheorem{theorem}{Theorem}

\newtheorem{proposition}[theorem]{Proposition}
\newtheorem{remark}[theorem]{Remark}

\newenvironment{proof}[1][Proof]{\noindent\textbf{#1.} }{\ \rule{0.5em}{0.5em}}
\input{tcilatex}
\begin{document}

\begin{center}
{\large \textbf{Lie Algebra of Hamiltonian Vector Fields and the
Poisson-Vlasov Equations\footnote{%
This is an expanded (with the additions of more remarks and, sections 4 and
5.4) version of the article "Geometry of plasma dynamics II: Lie algebra of
Hamiltonian vector fields" to appear in Journal of Geometric Mechanics, 2012.%
}}}

\bigskip

O\u{g}ul Esen and Hasan G\"{u}mral\\[2mm]

Department of Mathematics, Yeditepe University

34755 Ata\c{s}ehir, \.{I}stanbul, Turkey

oesen@yeditepe.edu.tr \ \ \ \ hgumral@yeditepe.edu.tr
\end{center}

\bigskip

\textbf{Abstract: }We introduce natural differential geometric structures
underlying the Poisson-Vlasov equations in momentum variables. First, we
decompose the space of all vector fields over particle phase space into a
semi-direct product algebra of Hamiltonian vector fields and its complement.
The latter is related to dual space of the Lie algebra. We identify
generators of homotheties as dynamically irrelevant vector fields in the
complement. Lie algebra of Hamiltonian vector fields is isomorphic to the
space of all Lagrangian submanifolds with respect to Tulczyjew symplectic
structure. This is obtained as tangent space at the identity of the group of
canonical diffeomorphisms represented as space of sections of a trivial
bundle. We obtain the momentum-Vlasov equations as vertical equivalence, or
representative, of complete cotangent lift of Hamiltonian vector field
generating particle motion. Vertical representatives can be described by
holonomic lift from a Whitney product to a Tulczyjew symplectic space. We
show that vertical representatives of complete cotangent lifts form an
integrable subbundle of this Tulczyjew space. A generalization of complete
cotangent lift is obtained by a Lie algebra homomorphism from the algebra of
symmetric contravariant tensor fields with Schouten concomitant to the Lie
algebra of Hamiltonian vector fields. Momentum maps for particular
subalgebras of symmetric contravariant tensors result in plasma-to-fluid map
in momentum variables of Vlasov equations.\ We exhibit dynamical relations
between Lie algebras of Hamiltonian vector fields and of contact vector
fields, in particular; infinitesimal quantomorphisms on their quantization
bundle. A diagram connecting these kinetic and fluid theories is presented.
Gauge symmetries of particle motion are extended to tensorial objects
including complete lift of particle motion. Poisson equation is then
obtained as zero value of momentum map for the Hamiltonian action of gauge
symmetries for kinematical description.

\bigskip

\pagebreak

\bigskip

\section{Introduction}

The dynamics of collisionless plasma is governed by the Poisson-Vlasov
equations%
\begin{equation}
\nabla _{q}^{2}\phi _{f}(\mathbf{q})=-e\int f(\mathbf{q},\mathbf{p})d^{3}%
\mathbf{p}\text{ }  \label{poi}
\end{equation}%
\begin{equation}
\frac{\partial f}{\partial t}+\frac{\mathbf{p}}{m}\cdot \nabla _{q}f-e\nabla
_{q}\phi _{f}\cdot \nabla _{p}f=0  \label{denvlasov}
\end{equation}%
where $f(\mathbf{q},\mathbf{p})$ is the density of plasma particles and $%
\phi _{f}(\mathbf{q})$ is the electric potential depending on the density
through the Poisson equation (\ref{poi}). The underlying geometric structure
of the Poisson-Vlasov system of differential equations was made available by
their Hamiltonian formulation \cite{mor80}, \cite{mw81}, \cite{mg80}, \cite%
{mor81}, \cite{gib81}, \cite{mor82}. The Vlasov equation was shown to be the
Lie-Poisson equation on the dual of Lie algebra of group of canonical
diffeomorphisms of particle phase space $T^{\ast }\mathcal{Q}$, identified
with the space of densities \cite{mw82}, \cite{mar82}, \cite{mwrss83}. This
identification was obtained as the dual of the isomorphism between the Lie
bracket algebra of Hamiltonian vector fields and the Poisson bracket algebra
of functions on $T^{\ast }\mathcal{Q}$ defined up to addition of a constant.
Based on this and with reference to the work of Van Hove in \cite{vh51}, it
was already (foot-)noted in \cite{mw82} that the correct configuration space
for plasma dynamics is the group of transformations of $T^{\ast }\mathcal{Q}%
\times \mathbb{R}$ preserving the contact one-form, also known as the group
of quantomorphisms.

In \cite{gpd1}, adapting the group of canonical diffeomorphisms as
configuration space, we obtained Poisson and Vlasov equations in Eulerian
momentum variables which, by symmetry reduction, define the plasma density
function $f$ and give Eqs.(\ref{poi}) and (\ref{denvlasov}) as well as their
Hamiltonian structure. In this work we shall elaborate geometric structures
underlying plasma dynamics in momentum variables and, we shall indicate, at
infinitesimal level, connection with quantomorphisms in the framework of
kinetic theories.

\subsection{Preliminaries and motivation}

The Lie-Poisson construction starts with the kinematical description of
particle motion on the phase space $T^{\ast }\mathcal{Q}$ where $\mathcal{Q}%
\subset \mathbb{R}^{3}$ is the configuration space of particles \cite{mar82}%
. Take a curve $\varphi _{t}$ in the group $G=Diff_{can}(T^{\ast }\mathcal{Q}%
)$ of all canonical diffeomorphisms of $T^{\ast }\mathcal{Q}$ preserving the
canonical symplectic two-form $\Omega _{T^{\ast }\mathcal{Q}}$ (see \cite%
{ban78}, \cite{RaSc}, \cite{RSA}, \cite{ban97}, \cite{ms98} for aspects of
diffeomorphism groups). Given a point $\mathbf{Z}\in T^{\ast }\mathcal{Q}$
regarded as a Lagrangian label, let $\mathbf{z}=(\mathbf{q},\mathbf{p}%
)=\varphi _{t}(\mathbf{Z})=\varphi (\mathbf{Z},t)$ denote the Eulerian
coordinates of plasma particles. The phase space velocity 
\begin{equation}
\mathbf{\dot{z}}=\frac{d}{dt}\varphi _{t}(\mathbf{Z})=X(\mathbf{z}%
,t)=X_{t}(\varphi _{t}(\mathbf{Z}))
\end{equation}%
generates the flow $\varphi _{t}$. Since $\varphi _{t}$ is canonical, $X$ is
locally Hamiltonian. We assume that it is globally Hamiltonian and write $h(%
\mathbf{z},t)$ for the corresponding Hamiltonian function so that $%
X=X_{h}=\Omega _{T^{\ast }\mathcal{Q}}^{\sharp }(-dh)$ on $T^{\ast }\mathcal{%
Q}$.

\begin{remark}
For diffeomorphisms groups, exponential map assigns to each vector the
time-one map of its flow. This is neither surjective nor injective around
identity \cite{ban97}. However, restricted to compact support, any smooth
Hamiltonian vector field on $T^{\ast }\mathcal{Q}$ generates a flow in $G$.
The group $G$ of canonical (or Hamiltonian) diffeomorphisms is a subgroup in
the identity component of the group of symplectomorphisms, that is,
diffeomorphisms preserving a symplectic two-form. It is normal and
path-connected \cite{ms98}. See \cite{ban78} for a proof that $G$ is a
simple group. In fact, groups of symplectomorphisms, diffeomorphisms, volume
preserving diffeomorphisms and contactomorphisms are simple indecomposable
Lie groups in Cartan`s list \cite{ms98}.
\end{remark}

The gauge group of particle motion is the additive group $\mathcal{F}(%
\mathcal{Q})$ of functions on $\mathcal{Q}$ acting on $T^{\ast }\mathcal{Q}$
by fiber translations. The Lie algebra 
\begin{equation*}
\mathfrak{g}=(\mathfrak{X}_{ham}(T^{\ast }\mathcal{Q});-[,])
\end{equation*}%
of $G$ consists of (smooth) Hamiltonian vector fields on $T^{\ast }\mathcal{Q%
}$ and $[\;,\;]$ denotes the standard Jacobi-Lie bracket, with conventions
as in \cite{amr88}. The dual vector space $\mathfrak{g}^{\ast }$ of the Lie
algebra $\mathfrak{X}_{ham}(T^{\ast }\mathcal{Q})$ is the non-closed
one-form densities on $T^{\ast }\mathcal{Q}$. Equivalently, the space of
one-form densities whose symplectic duals have non-vanishing divergences%
\begin{equation}
\mathfrak{g}^{\ast }=\{\Pi _{id}\otimes d\mu \in \Lambda ^{1}(T^{\ast }%
\mathcal{Q})\otimes Den(T^{\ast }\mathcal{Q})\mid div_{\Omega _{T^{\ast }%
\mathcal{Q}}}\Pi _{id}^{\sharp }\neq \text{constant}\}  \label{gst}
\end{equation}%
so that the pairing of $\mathfrak{g}$ and $\mathfrak{g}^{\ast }$ is weakly
nondegenerate with respect to the $L^{2}$-norm \cite{gpd1}, \cite{og11}. The
further requirement in Eq.(\ref{gst}) that they be different from constants
is of dynamical origin and will be explained later in section 2.2.

Reduction of canonical bracket on $T^{\ast }G$ by right invariant extension,
that is the invariance under particle relabelling symmetry, of functions on $%
T^{\ast }G$ gives the $+$Lie-Poisson bracket 
\begin{equation}
\left\{ K\left( \Pi _{id}\right) ,H\left( \Pi _{id}\right) \right\}
_{LP}=\int_{T^{\ast }\mathcal{Q}}\Pi _{id}\left( \mathbf{z}\right) \cdot %
\left[ \frac{\delta K}{\delta \Pi _{id}\left( \mathbf{z}\right) },\frac{%
\delta H}{\delta \Pi _{id}\left( \mathbf{z}\right) }\right] d\mu \left( 
\mathbf{z}\right)  \label{LP}
\end{equation}%
on $\mathfrak{g}^{\ast },$ where $\delta K/\delta \Pi _{id}\left( \mathbf{z}%
\right) $ and $\delta H/\delta \Pi _{id}\left( \mathbf{z}\right) $ are
regarded to be elements of $\mathfrak{g}$ \cite{mr94}, $d\mu \left( \mathbf{z%
}\right) =d^{3}\mathbf{q}d^{3}\mathbf{p}$ is the Liouville volume element on 
$T^{\ast }\mathcal{Q}$ and the bracket inside the integral is the Lie
algebra bracket. In particular, for the right invariant Hamiltonian
functional 
\begin{equation}
H_{LP}\left( \Pi _{id}\right) =\int_{T^{\ast }\mathcal{Q}}\left\langle \Pi
_{id}\left( \mathbf{z}\right) ,X_{h_{f}}\left( \mathbf{z}\right)
\right\rangle d\mu \left( \mathbf{z}\right) \text{, \ \ }  \label{hampi}
\end{equation}%
involving the particle Hamiltonian 
\begin{equation}
h_{f}(\mathbf{z})=\frac{p^{2}}{2m}+\frac{e}{2}\phi _{f}(\mathbf{q}),
\label{hamf}
\end{equation}%
which depends on the density, and the non-closed one-form $\Pi _{id}(\mathbf{%
z})=\mathbf{\Pi }_{q}\cdot d\mathbf{q}+\mathbf{\Pi }_{p}\cdot d\mathbf{p}$,
the Lie-Poisson equations on $\mathfrak{g}^{\ast }$ are%
\begin{align}
\mathbf{\dot{\Pi}}_{q}& =-X_{h}(\mathbf{\Pi }_{q})+e\left( \mathbf{\Pi }%
_{p}\cdot \nabla _{q}\right) \left( \nabla _{q}\phi _{f}\right)  \label{mv1}
\\
\mathbf{\dot{\Pi}}_{p}& =-X_{h}(\mathbf{\Pi }_{p})-{\frac{1}{m}}\mathbf{\Pi }%
_{q}  \label{mv}
\end{align}%
which are the momentum-Vlasov equations \cite{gpd1}. The unconventional
factor $1/2$ in the potential term of the function $h_{f}$ is a
manifestation of the nonlinearity arising from the constraint imposed by the
Poisson equation \cite{mor80}, \cite{mor81}, \cite{kd84}. By definition, the
momentum variables $(\mathbf{\Pi }_{q},\mathbf{\Pi }_{p})$ represent
equivalence classes up to additions of the terms $\nabla _{q}k(\mathbf{z})$
and $\nabla _{p}k(\mathbf{z}),$ respectively, for arbitrary function $k(%
\mathbf{z})$. Thus, the reduced dynamics on $\mathfrak{g}^{\ast }$ has a
further symmetry given by the action of the additive group $\mathcal{F}%
(T^{\ast }\mathcal{Q})$ of functions on $T^{\ast }\mathcal{Q}$. The momentum
map $\mathfrak{g}^{\ast }\rightarrow \mathcal{F}^{\ast }(T^{\ast }\mathcal{Q}%
)=Den(T^{\ast }\mathcal{Q})$ given by the differential substituition%
\begin{equation}
f(\mathbf{z})=div_{\Omega _{T^{\ast }\mathcal{Q}}}\Pi _{id}^{\sharp }=\nabla
_{p}\cdot \mathbf{\Pi }_{q}(\mathbf{z})-\nabla _{q}\cdot \mathbf{\Pi }_{p}(%
\mathbf{z})  \label{density}
\end{equation}%
defines the plasma density function $f$ \cite{gpd1}, \cite{og11}. The
reduction of the Lie-Poisson structure gives the Vlasov equation (\ref%
{denvlasov}) in density variable as well as the non-canonical Hamiltonian
structure defined by the Lie-Poisson bracket 
\begin{equation*}
\left\{ K\left( f\right) ,H\left( f\right) \right\} _{LP}=\int_{T^{\ast }%
\mathcal{Q}}f\left( \mathbf{z}\right) \cdot \left[ \frac{\delta K}{\delta
f\left( \mathbf{z}\right) },\frac{\delta H}{\delta f\left( \mathbf{z}\right) 
}\right] d\mu \left( \mathbf{z}\right)
\end{equation*}%
on $Den(T^{\ast }\mathcal{Q})$ and the Hamiltonian functional%
\begin{equation*}
H_{LP}\left( f\right) =\int_{T^{\ast }\mathcal{Q}}h_{f}(\mathbf{z})f\left( 
\mathbf{z}\right) d\mu \left( \mathbf{z}\right) .
\end{equation*}

\begin{remark}
The symmetries of the momentum-Vlasov equations were used, in \cite{gpd1},
to cast them into a canonical Hamiltonian formalism with a quadratic
Hamiltonian functional. This leads Eqs. (\ref{mv1}) and (\ref{mv}) to admit
a variational formulations. Namely, the Lagrangian functional%
\begin{equation*}
L_{0}\left[ \mathbf{\Pi }_{p}\right] =\int_{T^{\ast }\mathcal{Q}}\left( 
\frac{m}{2}|X_{h}(\mathbf{\Pi }_{p})+\frac{d\mathbf{\Pi }_{p}}{dt}|^{2}-%
\frac{e}{2}\frac{\partial ^{2}\phi _{\Pi }}{\partial q^{i}\partial q^{j}}\Pi
^{i}\Pi ^{j}\right) \left( \mathbf{z}\right) \text{\textbf{\ }}d\mu \left( 
\mathbf{z}\right)
\end{equation*}%
involving the velocity $d\mathbf{\Pi }_{p}/dt$ shifted by the term $-X_{h}(%
\mathbf{\Pi }_{p})$, gives the Euler-Lagrange equations%
\begin{equation*}
\ddot{\Pi}^{i}(\mathbf{z})+2X_{h}(\dot{\Pi}^{i}(\mathbf{z}))+X_{h}^{2}(\Pi
^{i}(\mathbf{z}))+\frac{e}{m}\delta ^{ij}\frac{\partial ^{2}\phi _{\Pi }(%
\mathbf{q})}{\partial q^{k}\partial q^{j}}\Pi ^{k}(\mathbf{z})=0
\end{equation*}%
which can also be obtained from Eqs.(\ref{mv1}) and (\ref{mv}) by
eliminating the variables $\Pi _{i}$.
\end{remark}

In \cite{gpd1}, we presented Lie-Poisson structures in momentum and density
variables and establish the relations between the two. The formulation of
dynamics in density variable is obtained by further reduction of
momentum-Vlasov equations by the symmetry defining gauge equivalence classes
of momentum variables. The gauge algebra is, as a vector space, shown to be
the same as $\mathfrak{g}$ but with an action different from the coadjoint
action. The Eulerian velocity and momentum variables are elements of $%
\mathfrak{g}\mathbf{=}$ $\mathfrak{X}_{ham}(T^{\ast }\mathcal{Q})$ and $%
\mathfrak{g}^{\ast }$, respectively. These variables are complementary in
the vector space $TT^{\ast }\mathcal{Q}$. Obviously, this and other
geometric properties disappear upon identification of $\mathfrak{g}$ and $%
\mathfrak{g}^{\ast }$ with function spaces $\mathcal{F}(T^{\ast }\mathcal{Q}%
) $ and $Den(T^{\ast }\mathcal{Q})$, respectively. Introduction of a
formulation in the variables $\Pi _{id}$ provides a computational advantage
and this also prevents us from confusion in the geometry which may arise
upon identification with function spaces. For example, the function $h_{f}$
and the density $f$ appear symmetrically in the Hamiltonian functional of
the Lie-Poisson structure whereas the corresponding variables $X_{h_{f}}$
and $\Pi _{id}$ in $\mathfrak{g}$ and $\mathfrak{g}^{\ast }$ are
complementary in the sense that $\Omega _{T^{\ast }\mathcal{Q}}^{\flat }(%
\mathfrak{g})$ and $\mathfrak{g}^{\ast }$ decompose the space of one-forms
on $T^{\ast }\mathcal{Q}$ into spaces of exact and non-closed one-forms,
respectively (c.f. section $2.1$). The momentum-Vlasov equations in
components of $\Pi _{id}$ expresses the evolution of a volume cell, that is
the density $f$, in the phase space $T^{\ast }\mathcal{Q}$ in terms of its
boundaries, that is, surfaces of the momenta $\Pi _{id}$. This
interpretation was first given by Ye and Morrison in \cite{ym92} for the
Clebsch variables $(\alpha ,\beta )$ defined by $\left\{ \alpha ,\beta
\right\} _{T^{\ast }\mathcal{Q}}=f$. In the present context, they form a
non-closed one-form $\alpha d\beta $ and can be identified with $\Pi _{id}$.
The momentum formulation clarifies the geometric relation between the
motions of plasma particles and the Lie-Poisson description of dynamics. It
does become necessary to investigate the plasma dynamics described by the
more basic momentum-Vlasov equations.

This observation is the starting point of the present work and motivates the
elaboration of geometric setting underlying the momentum-Vlasov equations (%
\ref{mv1}) and (\ref{mv}). Our aim is to analyse in detail the structures of
Lie algebra of Hamiltonian vector fields and its dual in order to prepare a
suitable background for application of Tulczyjew construction to orbits of
canonical diffeomorphisms. As will be seen in the sequel, present
formulation of dynamics on higher order tangent and cotangent bundles over $%
T^{\ast }\mathcal{Q}$ constitutes a useful model for investigation of
orbital dynamics.

\subsection{Content of the work}

In this work, we shall elaborate the central part of the following diagram 
\begin{equation*}
\begin{array}{ccc}
& TT^{\ast }\mathcal{Q}\text{ \ \ \ }%
\begin{array}{c}
\underleftrightarrow{\Omega _{T^{\ast }\mathcal{Q}}^{\sharp },\Omega
_{T^{\ast }\mathcal{Q}}^{\flat }} \\ 
\text{ \ \ \ \ \ }%
\end{array}%
\text{\ \ }T^{\ast }T^{\ast }\mathcal{Q} &  \\ 
& \text{ \ \ \ } &  \\ 
\begin{array}{c}
X_{\varphi }\text{\ }{\huge \nearrow }\text{ } \\ 
\mathstrut%
\end{array}
& X_{h}\Big\uparrow\Big\downarrow\tau _{T^{\ast }\mathcal{Q}}\text{ \ \ \ \
\ \ \ \ \ }\pi _{T^{\ast }\mathcal{Q}}\Big\downarrow\Big\uparrow\Pi _{id} & 
\begin{array}{c}
{\huge \nwarrow }\Pi _{\varphi } \\ 
\mathstrut%
\end{array}
\\ 
T^{\ast }\mathcal{Q}\text{ \ \ \ \ \ \ }%
\begin{array}{c}
\text{ \ \ } \\ 
\text{\ }\overrightarrow{\text{ \ \ }\varphi \text{ \ \ }}%
\end{array}
& T^{\ast }\mathcal{Q}\text{ \ \ \ \ \ \ \ \ }\equiv \text{ \ \ \ \ \ \ \ \
\ \ }T^{\ast }\mathcal{Q} & 
\begin{array}{c}
\text{ \ \ } \\ 
\overleftarrow{\text{ \ \ }\varphi \text{ \ \ }}%
\end{array}%
\text{\ \ \ \ \ \ \ \ }T^{\ast }\mathcal{Q}\text{ }%
\end{array}%
\text{ }
\end{equation*}%
which summarizes the mapping properties with reference to particle phase
space $T^{\ast }\mathcal{Q}$ of elements of $T_{\varphi }G$, $T_{\varphi
}^{\ast }G$, $\mathfrak{g}$ and $\mathfrak{g}^{\ast }$; namely, $X_{\varphi
},\Pi _{\varphi },X_{h}$ and $\Pi _{id}$, respectively. This may serve as
the main diagram to relate the constructions of present work for Eulerian
variables to the Lagrangian variables. We shall establish the precise
relation between the particle motion, its symmetries and the Poisson-Vlasov
equations. We shall obtain Poisson equation as a consequence of gauge
symmetries of Hamiltonian description of motions of plasma particles.

In the next section, we shall first decompose the algebra of vector fields
on $T^{\ast }\mathcal{Q}$ into a semi-direct product algebra of Hamiltonian
vector fields and its complement in $\mathfrak{X(}T^{\ast }\mathcal{Q)}$.
The latter, as a vector space, is isomorphic to dual space of the Lie
algebra. Further properties of this decomposition will be studied. In
particular, we shall identify homotheties as non-dynamical part of the dual
of Lie algebra. We shall then present, according to the diagram, 
\begin{equation*}
\begin{array}{c}
T^{\ast }T^{\ast }\mathcal{Q}\text{ \ \ \ \ \ \ }%
\begin{array}{c}
\text{ \ \ }\underleftarrow{\text{ }\Omega _{T^{\ast }\mathcal{Q}}^{\flat }%
\text{\ }} \\ 
\mathstrut%
\end{array}%
\text{ \ \ \ \ \ \ \ \ \ \ }TT^{\ast }\mathcal{Q}\text{\ \ \ \ \ \ \ \ }%
\begin{array}{c}
\underrightarrow{\text{ \ \ }\alpha _{\mathcal{Q}}\text{ \ \ }}\text{ \ \ \
\ } \\ 
\mathstrut%
\end{array}%
\text{ \ \ }T^{\ast }T\mathcal{Q} \\ 
\text{ \ \ }\pi _{T^{\ast }\mathcal{Q}}\searrow \nwarrow -dh\text{ \ \ \ \ \
\ \ \ }\swarrow \tau _{T^{\ast }\mathcal{Q}}\text{ \ \ \ \ }T\pi _{\mathcal{Q%
}}\searrow \text{ \ \ \ \ \ \ \ }dl\nearrow \swarrow \pi _{T\mathcal{Q}} \\ 
\text{ \ \ \ \ \ }T^{\ast }\mathcal{Q}\text{ \ \ \ \ \ \ \ \ \ \ \ \ \ \ \ \
\ \ \ \ \ \ \ \ \ \ \ \ \ }T\mathcal{Q}\text{ \ }%
\end{array}%
\end{equation*}%
two special symplectic structures on $TT^{\ast }\mathcal{Q}$ with the
underlying symplectic manifold being endowed with the Tulczyjew two-form.
Considering a description of configuration space as the space of sections of
a trivial bundle, we shall show that the algebra of Hamiltonian vector
fields is isomorphic to the space of Lagrangian submanifolds of Tulczyjew
symplectic manifold.

In section three, starting with the Hamiltonian vector field generating the
particle motion, we shall obtain the momentum-Vlasov equations as the
vertical equivalence of its complete cotangent lift. This will be shown to
be the same as the vertical lift of coadjoint action on momentum variables.
As a result, we shall realize the commutative diagram%
\begin{equation}
\begin{array}{ccc}
\begin{array}{c}
\text{canonical Hamiltonian} \\ 
\text{motion of particles} \\ 
\text{on }T^{\ast }\mathcal{Q}%
\end{array}
& 
\begin{array}{c}
\underrightarrow{\text{ \ \ \ \ complete \ \ \ }} \\ 
\text{cotangent lift}%
\end{array}
& 
\begin{array}{c}
\text{Hamiltonian} \\ 
\text{motion on }T^{\ast }T^{\ast }\mathcal{Q}%
\end{array}
\\ 
\text{ \ \ \ }%
\begin{array}{c}
\mathstrut \\ 
\begin{array}{c}
\text{vertical lift of} \\ 
\text{coadjoint action}%
\end{array}%
\searrow%
\end{array}
&  & 
\begin{array}{c}
\mathstrut \\ 
\swarrow 
\begin{array}{c}
\text{vertical (jet)} \\ 
\text{equivalence}%
\end{array}%
\end{array}%
\text{ \ \ \ } \\ 
& 
\begin{array}{c}
\text{momentum-} \\ 
\text{Vlasov equations } \\ 
\text{on }VT^{\ast }T^{\ast }\mathcal{Q}%
\end{array}
& 
\end{array}
\label{geodyn}
\end{equation}%
connecting motion of individual plasma particles to Eulerian dynamics in
momentum variables. Complete cotangent lifts are Hamiltonian vector fields
with a degenerate Hamiltonian function for the canonical symplectic
structure. We shall point out a Lagrangian formulation for them with a Morse
family on certain Whitney product. We shall give a geometric description of
vertical representative of cotangent lift in terms of holonomic lift
operator from this Whitney product into the Tulczyjew symplectic space $%
TT^{\ast }T^{\ast }\mathcal{Q}$.

In section four, we will define a Lie algebra homomorphism from the algebra
of symmetric contravariant tensor fields with Schouten concomitant to the
algebra $\mathfrak{X}_{ham}\left( T^{\ast }\mathcal{Q}\right) =\mathfrak{g}$
of Hamiltonian vector fields. This will generalize the complete cotangent
lift of vector fields to symmetric contravariant tensors. We will then
obtain the moments of momentum-Vlasov dynamical variables. For particular
subalgebras of symmetric contravariant tensors these moments will give
plasma-to-fluid map in momentum variables of $\mathfrak{g}^{\ast }$.

In section five, we establish a correspondence between the Lie algebra of
Hamiltonian vector fields on $T^{\ast }\mathcal{Q}$ and the Lie algebra of
infinitesimal strict contact transformations, also called quantomorphisms,
of quantization bundle of $T^{\ast }\mathcal{Q}$. Relying on our recent work 
\cite{og11}, we first present kinetic equations, both in momentum and
density variables, of particles moving according to contact transformations
of standard three-dimensional contact manifold. We then restrict the group
of contactomorphisms to strict contact transformations and obtain a system
of kinetic equations equivalent to the momentum -Vlasov equations in one
dimension. This section will be concluded with a diagram summarizing the
relations between various kinetic and fluid theories.

In section six, we expand on our earlier result in \cite{gpd1} where we
described the Poisson equation as a kinematical constraint on the dynamics
of Eulerian variables. More precisely, we shall show that the Poisson
equation characterizes the set of zero values of the momentum map associated
with the action of additive group of functions $\mathcal{F}(\mathcal{Q})$ on
the position space $\mathcal{Q}$ of particles. This is the gauge group of
particle motion on the canonical phase space $T^{\ast }\mathcal{Q}$.
Momentum map realization of the Poisson equation implies that the true
configuration space for the Poisson-Vlasov dynamics must be the semi-direct
product space $\mathcal{F}(\mathcal{Q})\circledS Diff_{can}(T^{\ast }%
\mathcal{Q})$ with the action of $\mathcal{F}(\mathcal{Q})$ given by fiber
translation on $T^{\ast }\mathcal{Q}$ and, by composition on right with the
canonical transformations.

Section seven will be devoted to a summary and discussion of the results as
well as some future work to be addressed elsewhere. See also the
introductions of each section where we summarize contents in more technical
terms.

\subsection{Notations}

General definitions will be given with reference to an arbitrary smooth
manifold $\mathcal{M}$. $\theta _{\mathcal{M}}$, $\Omega _{\mathcal{M}}$
will be used for canonical one-form and symplectic two-form defined on $%
\mathcal{M}$. This convention of showing the space of definition as a
subscript will be extended to other objects when necessary. $\Gamma (pr)$
will usually denote the space of sections of a bundle $pr:E\longrightarrow B$%
. For spaces of sections of tangent and cotangent bundles of a manifold $%
\mathcal{M}$ we will use $\mathfrak{X}(\mathcal{M})$ and $\Lambda ^{1}(%
\mathcal{M})$, respectively. The bracket $<,>_{\mathcal{M}}$ will be used
for natural pairing between differential forms and vector fields over $%
\mathcal{M}$. $\mathcal{F}(\mathcal{M})$ and $Den(\mathcal{M})$ will denote
spaces of functions (zero-forms) and volume forms on $\mathcal{M}$. $i_{X}$
and $\mathcal{L}_{X}$ will be used for the interior product (contraction) \
and the Lie derivative with respect to the vector field $X$. Throughout the
work $G$, $\mathfrak{g}$ and $\mathfrak{g}^{\ast }$ will be used frequently
for $Diff_{can}(T^{\ast }\mathcal{Q})$, its Lie algebra $\mathfrak{X}%
_{ham}(T^{\ast }\mathcal{Q)}$ and the dual of the latter, respectively. If $%
\mathbf{q}\in \mathcal{Q}$ then, we will use 
\begin{eqnarray}
(\mathbf{q},\mathbf{\dot{q}}) &\in &T_{q}\mathcal{Q}\text{, \ \ \ \ \ \ \ \
\ \ \ \ }\mathbf{z}=(\mathbf{q},\mathbf{p})\in T_{q}^{\ast }\mathcal{Q},%
\text{\ }  \notag \\
(\mathbf{z},\boldsymbol{\pi }) &\in &T_{z}^{\ast }T_{q}^{\ast }\mathcal{Q},%
\text{ \ \ \ }(\mathbf{q},\mathbf{\dot{q}},\boldsymbol{\lambda }_{q},%
\boldsymbol{\lambda }_{\dot{q}})\in T_{z}^{\ast }T_{q}\mathcal{Q}, \\
(\mathbf{z},\mathbf{\dot{z}}) &\in &T_{z}T_{q}^{\ast }\mathcal{Q}\text{, \ \
\ \ \ \ \ }(\mathbf{z},\mathbf{X}_{h})\in T_{z}T_{q}^{\ast }\mathcal{Q}\text{%
.}  \notag
\end{eqnarray}%
The canonical one-form on $T^{\ast }\mathcal{Q}$ will be $\theta _{T^{\ast }%
\mathcal{Q}}(\mathbf{z})=\mathbf{p}\cdot d\mathbf{q}$ and the symplectic
two-form is $\Omega _{T^{\ast }\mathcal{Q}}(\mathbf{z})=d\theta _{T^{\ast }%
\mathcal{Q}}(\mathbf{z})=d\mathbf{p\wedge }d\mathbf{q}$.

\section{Lie Algebra of Hamiltonian Vector Fields}

We shall show that the algebra of all vector fields on particle phase space
can be decomposed into a semi-direct product of Hamiltonian vector fields
and space of symplectic duals of all non-closed one-forms on particle phase
space. Among the latter, those vector fields with constant divergence are
related to homotheties on particle phase space and they correspond to
constant plasma densities. We then introduce special symplectic structures
and present the Tulczyjew symplectic structure on $TT^{\ast }\mathcal{Q}$
relevant to a description of particle dynamics as Lagrangian submanifolds.
Finally, we shall expand on the remark in \cite{gpd1} that the configuration
space $G$ of plasma dynamics can be represented as space of Lagrangian
submanifolds in the space of sections of a trivial bundle, and show that the
Lie algebra of Hamiltonian vector fields can be obtained, as tangent space
over the identity, to be spaces of all Lagrangian submanifolds of Tulczyjew
symplectic space for particle dynamics.

\subsection{Algebra of vector fields in $TT^{\ast }\mathcal{Q}$}

Let $\mathfrak{X(}T^{\ast }\mathcal{Q)}$ and $\Lambda ^{1}(T^{\ast }\mathcal{%
Q)}$ denote the spaces of smooth sections of $TT^{\ast }\mathcal{Q}%
\rightarrow T^{\ast }\mathcal{Q}$ and $T^{\ast }T^{\ast }\mathcal{Q}%
\rightarrow T^{\ast }\mathcal{Q}$, respectively. The nondegeneracy of
canonical symplectic form $\Omega _{T^{\ast }\mathcal{Q}}$ on $T^{\ast }%
\mathcal{Q}$ leads to the musical isomorphism $\Omega _{T^{\ast }\mathcal{Q}%
}^{\flat }:\mathfrak{X(}T^{\ast }\mathcal{Q)}\rightarrow \Lambda
^{1}(T^{\ast }\mathcal{Q)}$ defined, for arbitrary vector fields $X,Y\in 
\mathfrak{X(}T^{\ast }\mathcal{Q)}$, by $\Omega _{T^{\ast }\mathcal{Q}%
}^{\flat }\left( X\right) (Y)=\Omega _{T^{\ast }\mathcal{Q}}\left(
X,Y\right) $ or, alternatively, by $\Omega _{T^{\ast }\mathcal{Q}}^{\flat
}\left( X\right) =i_{X}\Omega _{T^{\ast }\mathcal{Q}}$. The isomorphism $%
\Omega _{T^{\ast }\mathcal{Q}}^{\sharp }:\Lambda ^{1}(T^{\ast }\mathcal{Q)}%
\rightarrow \mathfrak{X(}T^{\ast }\mathcal{Q)}$ is obtained by fiberwise
inversion of $\Omega _{T^{\ast }\mathcal{Q}}^{\flat }$. In the local
coordinates introduced above we have 
\begin{eqnarray}
\Omega _{T^{\ast }\mathcal{Q}}^{\flat }(\mathbf{q},\mathbf{p};\mathbf{\dot{q}%
},\mathbf{\dot{p}}) &=&(\mathbf{q},\mathbf{p};\mathbf{\dot{p}},-\mathbf{\dot{%
q}}),  \notag \\
\Omega _{T^{\ast }\mathcal{Q}}^{\sharp }(\mathbf{q},\mathbf{p};\boldsymbol{%
\pi }_{q}\boldsymbol{,\pi }_{p}) &=&(\mathbf{q},\mathbf{p};-\boldsymbol{\pi }%
_{p},\boldsymbol{\pi }_{q}).
\end{eqnarray}

If the image of a vector field $X$ by the mapping $\Omega _{T^{\ast }%
\mathcal{Q}}^{\flat }$ is closed then $X$ is a locally Hamiltonian vector
field, and hence, the space $\mathfrak{X}_{ham}(T^{\ast }\mathcal{Q)}$ of
Hamiltonian vector fields on $T^{\ast }\mathcal{Q}$ is isomorphic to the
space $kerd$ of all closed one-forms on $T^{\ast }\mathcal{Q}$. If $\Omega
_{T^{\ast }\mathcal{Q}}^{\flat }\left( X\right) $ is exact, we write $\Omega
_{T^{\ast }\mathcal{Q}}^{\flat }\left( X_{h}\right) =-dh$ and $X_{h}$ is
said to be globally Hamiltonian \cite{amr88},\cite{mr94}.

\begin{proposition}
Let $\mathfrak{g}^{\flat }=\Omega _{T^{\ast }\mathcal{Q}}^{\flat }\left( 
\mathfrak{g}\right) $ and $\left( \mathfrak{g}^{\ast }\right) ^{\sharp
}\subset \mathfrak{X(}T^{\ast }\mathcal{Q)}$ denote the vector space of
closed one-forms on $T^{\ast }\mathcal{Q}$ and the image of $\mathfrak{g}%
^{\ast }$ under the mapping $\Omega _{T^{\ast }\mathcal{Q}}^{\sharp }$,
respectively. With the isomorphism 
\begin{equation}
\Omega _{T^{\ast }\mathcal{Q}}^{\flat }:\mathfrak{g}=\mathfrak{X}%
_{ham}(T^{\ast }\mathcal{Q)}\leftrightarrow \mathfrak{g}^{\flat }=kerd\cap
\Lambda ^{1}(T^{\ast }\mathcal{Q)}
\end{equation}%
we have the decompositions%
\begin{eqnarray}
\Lambda ^{1}(T^{\ast }\mathcal{Q)} &=&kerd\oplus \mathfrak{g}^{\ast }=%
\mathfrak{g}^{\flat }\oplus \mathfrak{g}^{\ast }  \label{isobir} \\
\mathfrak{X(}T^{\ast }\mathcal{Q)} &=&\mathfrak{g}\oplus \left( \mathfrak{g}%
^{\ast }\right) ^{\sharp }  \label{isoiki}
\end{eqnarray}%
of the spaces of one-forms and vector fields on $T^{\ast }\mathcal{Q}$.
Moreover, $\mathfrak{g}$ and $\left( \mathfrak{g}^{\ast }\right) ^{\sharp }$
are Lie subalgebras, and $\left( \mathfrak{g}^{\ast }\right) ^{\sharp }$ is
an ideal of $\mathfrak{X(}T^{\ast }\mathcal{Q)}$ 
\begin{equation*}
\left[ \mathfrak{g}\mathbf{,}\mathfrak{g}\right] \subset \mathfrak{g}\text{
, \ \ \ \ }\left[ \left( \mathfrak{g}^{\ast }\right) ^{\sharp }\mathbf{,}%
\left( \mathfrak{g}^{\ast }\right) ^{\sharp }\right] \subset \left( 
\mathfrak{g}^{\ast }\right) ^{\sharp }\text{, \ \ \ }\left[ \mathfrak{g}%
\mathbf{,}\left( \mathfrak{g}^{\ast }\right) ^{\sharp }\right] \subset
\left( \mathfrak{g}^{\ast }\right) ^{\sharp }.
\end{equation*}

\begin{proof}
Eq.(\ref{isobir}) follows from the definitions above. Since the dual $%
\mathfrak{g}^{\ast }$ of the Lie algebra $\mathfrak{g}$ of Hamiltonian
vector fields is defined to be non-closed one-form densities on $T^{\ast }%
\mathcal{Q}$, the remaining elements of $\Lambda ^{1}(T^{\ast }\mathcal{Q)}$%
, namely, closed one-forms constitute the underlying vector space of $%
\mathfrak{g}^{\flat }$. For decomposition in Eq.(\ref{isoiki}) we have that
Hamiltonian vector fields are divergence-free with respect to the symplectic
or Liouville volume $d\mu =\Omega _{T^{\ast }\mathcal{Q}}^{3}$. Let $\left( 
\mathfrak{g}^{\ast }\right) ^{\sharp }$ denote the image of dual $\mathfrak{g%
}^{\ast }$ of Lie algebra $\mathfrak{g}$ of Hamiltonian vector fields under
the isomorphism $\Omega _{T^{\ast }\mathcal{Q}}^{\sharp }$. For a
non-degenerate $L^{2}-$pairing of $\mathfrak{g}$ and $\mathfrak{g}^{\ast }$, 
$\left( \mathfrak{g}^{\ast }\right) ^{\sharp }$ contains vector fields with
nonvanishing divergences. In other words, $\alpha ^{\sharp }\equiv \Omega
_{T^{\ast }\mathcal{Q}}^{\sharp }(\alpha )\in \left( \mathfrak{g}^{\ast
}\right) ^{\sharp }$ is not Hamiltonian in any sense. $\left[ \mathfrak{g}%
\mathbf{,}\mathfrak{g}\right] \subset \mathfrak{g}$ follows from definition
of (locally) Hamiltonian vector fields. For vector fields with non-constant
divergences second property in the last conclusion can be obtained by direct
computation. For the last property, if $X$ is locally Hamiltonian, then for
the non-Hamiltonian vector field $\alpha ^{\sharp }$ we compute 
\begin{equation}
i_{\left[ X,\alpha ^{\sharp }\right] }\Omega _{T^{\ast }\mathcal{Q}}=%
\mathcal{L}_{X}i_{\alpha ^{\sharp }}\Omega _{T^{\ast }\mathcal{Q}}=d\Omega
_{T^{\ast }\mathcal{Q}}\left( \alpha ^{\sharp },X\right) +i_{X}di_{\alpha
^{\sharp }}\Omega _{T^{\ast }\mathcal{Q}}
\end{equation}%
which need not be closed for arbitrary choices of $X$ and $\alpha ^{\sharp }$
and hence, not Hamiltonian. Last conclusion implies that the algebraic
structure on sections of $TT^{\ast }\mathcal{Q}$ is a semi-direct product
algebra%
\begin{equation*}
\mathfrak{X(}T^{\ast }\mathcal{Q)}=\mathfrak{g}\mathbf{\circledS }\left( 
\mathfrak{g}^{\ast }\right) ^{\sharp }
\end{equation*}%
of vector fields with the Hamiltonian vector fields in $\mathfrak{g}$ acting
on the second factor $\left( \mathfrak{g}^{\ast }\right) ^{\sharp }$ by Lie
derivative. This, of course, is a consequence of the coadjoint action of $%
\mathfrak{g}$ on its dual $\mathfrak{g}^{\ast }$ which, in turn, produces
Lie-Poisson dynamics.
\end{proof}
\end{proposition}

\subsection{Homotheties}

Defining the divergence of an element of $\left( \mathfrak{g}^{\ast }\right)
^{\sharp }$ to be a density, we obtain the identification of $\left( 
\mathfrak{g}^{\ast }\right) ^{\sharp }$ with the space $Den(T^{\ast }%
\mathcal{Q})$ of densities on $T^{\ast }\mathcal{Q}$. In particular, vector
fields $\Pi _{c}^{\sharp }$ with constant divergences (with respect to
Liouville volume and for $n-$dimensional plasma) 
\begin{equation}
\mathcal{L}_{\Pi _{c}^{\sharp }}\Omega ^{n}=(div_{\Omega }\Pi _{c}^{\sharp
})\Omega ^{n}=c\Omega ^{n}\text{, \ \ \ \ }c=\text{constant}  \label{div}
\end{equation}%
correspond to constant plasma densities. In this case, we have either
Lagrangian description of kinematics or, no dynamics in an Eulerian
description. Since $\Omega $ is nondegenerate, Eq.(\ref{div}) implies%
\begin{equation*}
\mathcal{L}_{\Pi _{c}^{\sharp }}\Omega =\frac{c}{n}\Omega \text{, \ \ \ \ }c=%
\text{constant.}
\end{equation*}%
That means, vector fields with constant divergences are infinitesimal
homotheties of the symplectic form $\Omega $ \cite{izu85}. Although, $\Pi
_{c}^{\sharp }$ is not even locally Hamiltonian, it follows from the
identity 
\begin{equation}
\mathcal{L}_{\left[ X,Y\right] }=\mathcal{L}_{X}\mathcal{L}_{Y}-\mathcal{L}%
_{Y}\mathcal{L}_{X}  \label{lxy}
\end{equation}%
that the Lie bracket of two vectors with constant divergence is locally
Hamiltonian. If we denote the set in $\left( \mathfrak{g}^{\ast }\right)
^{\sharp }$ of vector fields with constant divergences by $\left( \mathfrak{g%
}_{c}^{\ast }\right) ^{\sharp }$ then, straightforward computations prove

\begin{proposition}
\label{lm}$\left[ \mathfrak{g}\mathbf{,}\left( \mathfrak{g}_{c}^{\ast
}\right) ^{\sharp }\right] \subset \mathfrak{g}\mathbf{,}$ $\left[ \left( 
\mathfrak{g}^{\ast }\right) ^{\sharp }\mathbf{,}\left( \mathfrak{g}%
_{c}^{\ast }\right) ^{\sharp }\right] \subset \left( \mathfrak{g}^{\ast
}\right) ^{\sharp },$ $\left[ \left( \mathfrak{g}_{c}^{\ast }\right)
^{\sharp }\mathbf{,}\left( \mathfrak{g}_{c}^{\ast }\right) ^{\sharp }\right]
\subset \mathfrak{g}\mathbf{.}$

\begin{proof}
For the first assertion we have%
\begin{eqnarray*}
i_{\left[ X_{h},\Pi _{c}^{\sharp }\right] }\Omega &=&i_{X_{h}}di_{\Pi
_{c}^{\sharp }}\Omega +di_{X_{h}}i_{\Pi _{c}^{\sharp }}\Omega =i_{X_{h}}%
\frac{c}{n}\Omega +d\Omega \left( \Pi _{c}^{\sharp },X_{h}\right) \\
&=&d\left( -\frac{c}{n}h+\Omega \left( \Pi _{c}^{\sharp },X_{h}\right)
\right)
\end{eqnarray*}%
where we used the identity 
\begin{equation}
i_{[X,Y]}=\mathcal{L}_{X}i_{Y}-i_{Y}\mathcal{L}_{X}.  \label{ident}
\end{equation}%
If we replace $X_{h}$ with a locally Hamiltonian vector field then a similar
computation implies that the bracket is again locally Hamiltonian. For the
second, we compute, from the definition of locally Hamiltonian vector fields 
\begin{eqnarray*}
di_{\left[ \Pi _{id}^{\sharp },\Pi _{c}^{\sharp }\right] }\Omega &=&\mathcal{%
L}_{\Pi _{id}^{\sharp }}di_{\Pi _{c}^{\sharp }}\Omega -di_{\Pi _{c}^{\sharp
}}\frac{1}{n}div_{\Omega }\Pi _{id}^{\sharp }\Omega \\
&=&\frac{c}{n^{2}}div_{\Omega }\Pi _{id}^{\sharp }\Omega -d\left( \frac{1}{n}%
div_{\Omega }\Pi _{id}^{\sharp }\right) \wedge i_{\Pi _{c}^{\sharp }}\Omega -%
\frac{1}{n}\left( div_{\Omega }\Pi _{id}^{\sharp }\right) di_{\Pi
_{c}^{\sharp }}\Omega \\
&=&-d\left( \frac{1}{n}div_{\Omega }\Pi _{id}^{\sharp }\right) \wedge i_{\Pi
_{c}^{\sharp }}\Omega .
\end{eqnarray*}%
This can be zero only if $\Pi _{c}^{\sharp }$ is globally Hamiltonian with
divergence of the arbitrary element $\Pi _{id}^{\sharp }$ of $\left( 
\mathfrak{g}^{\ast }\right) ^{\sharp },$ which is not possible.
\end{proof}
\end{proposition}

The action of homotheties on plasma density function may be computed using
the identity in Eq.(\ref{lxy})%
\begin{eqnarray}
\mathcal{L}_{\left[ \Pi _{id}^{\sharp },\Pi _{c}^{\sharp }\right] }\left(
d\mu \right) &=&\mathcal{L}_{\Pi _{id}^{\sharp }}\left( cd\mu \right) -%
\mathcal{L}_{\Pi _{c}^{\sharp }}\left( fd\mu \right)  \notag \\
&=&cfd\mu -df\wedge i_{\Pi _{c}^{\sharp }}\left( d\mu \right) -cfd\mu  \notag
\\
&=&-df\wedge i_{\Pi _{c}^{\sharp }}\left( d\mu \right) =-i_{\Pi _{c}^{\sharp
}}\left( df\right) d\mu =-\Pi _{c}^{\sharp }\left( f\right) d\mu
\end{eqnarray}%
and thus, is described by 
\begin{equation}
\Pi _{id}\rightarrow fd\mu \text{, \ \ \ \ \ \ \ \ \ }\left[ \Pi
_{c}^{\sharp },\Pi _{id}^{\sharp }\right] \rightarrow \Pi _{c}^{\sharp
}\left( f\right) d\mu .
\end{equation}%
This can be neglected by a redefinition of density. We thus restrict the
definition of $\mathfrak{g}^{\ast }$ to one-forms $\Pi _{id}$ for which $%
div_{\Omega }\Pi _{id}^{\sharp }=f\neq $constant.

\begin{remark}
Proposition \ref{lm} opens up the possibility to apply the following
Lebedev-Manin construction \cite{lm80} to plasma dynamics. Assume that we
have $\mathfrak{a=g}\mathbf{\oplus }\left( \mathfrak{g}^{\ast }\right)
^{\sharp }$ as a vector space with $\mathfrak{g}\mathbf{,}\left( \mathfrak{g}%
^{\ast }\right) ^{\sharp }$\textbf{\ }and $\left( \mathfrak{g}_{c}^{\ast
}\right) ^{\sharp }$ satisfying%
\begin{equation*}
\left[ \mathfrak{g}\mathbf{,}\mathfrak{g}\right] \subset \mathfrak{g}\text{
, \ \ \ \ }\left[ \left( \mathfrak{g}^{\ast }\right) ^{\sharp }\mathbf{,}%
\left( \mathfrak{g}^{\ast }\right) ^{\sharp }\right] \subset \left( 
\mathfrak{g}^{\ast }\right) ^{\sharp }\text{, \ \ \ }\left[ \mathfrak{g}%
\mathbf{,}\left( \mathfrak{g}^{\ast }\right) ^{\sharp }\right] \subset
\left( \mathfrak{g}^{\ast }\right) ^{\sharp }.
\end{equation*}%
Let $\left\langle ,\right\rangle $ be an invariant non-degenerate scalar
product on $\mathfrak{a}$ with 
\begin{equation*}
\left\langle \mathfrak{g,g}\right\rangle =\left\langle \left( \mathfrak{g}%
^{\ast }\right) ^{\sharp },\left( \mathfrak{g}^{\ast }\right) ^{\sharp
}\right\rangle =0.
\end{equation*}
For $F:\left( \mathfrak{g}^{\ast }\right) ^{\sharp }\longrightarrow 
\mathbb{R}
$, define $\delta F(\Pi _{id}^{\sharp })/\delta \Pi _{id}^{\sharp }\in 
\mathfrak{g}$ by%
\begin{equation*}
\left\langle \Pi ^{\sharp },\frac{\delta F(\Pi _{id}^{\sharp })}{\delta \Pi
_{id}^{\sharp }}\right\rangle =\frac{d}{d\epsilon }F(\Pi _{id}^{\sharp
}+\epsilon \Pi ^{\sharp })|_{\epsilon =0},\text{ \ }\forall \Pi
_{id}^{\sharp },\Pi ^{\sharp }\in \left( \mathfrak{g}^{\ast }\right)
^{\sharp }.
\end{equation*}%
Let $F$ be an invariant function on $\mathfrak{a}$, that is, $\left[
X,\delta F(X)/\delta X\right] =0$ for all $X\in \mathfrak{a}$. For $\Pi
_{c}^{\sharp }\in \left( \mathfrak{g}_{c}^{\ast }\right) ^{\sharp }$, set $%
F_{\Pi _{c}^{\sharp }}(\Pi _{id}^{\sharp })=F(\Pi _{id}^{\sharp }+\Pi
_{c}^{\sharp })$ for all $\Pi _{id}^{\sharp }\in \left( \mathfrak{g}^{\ast
}\right) ^{\sharp }$. Then, for two invariant functions $F,G$ we have $%
\{F_{\Pi _{c}^{\sharp }},G_{\Pi _{c}^{\sharp }}\}_{LP}=0$ on $\left( 
\mathfrak{g}^{\ast }\right) ^{\sharp }$ (with Lie-Poisson bracket adapted
from $\mathfrak{g}^{\ast }$). The Lie-Poisson equations 
\begin{equation*}
\dot{\Pi}_{id}^{\sharp }=\left[ \Pi _{id}^{\sharp },\delta F_{\Pi
_{c}^{\sharp }}(\Pi _{id}^{\sharp })/\delta \Pi _{id}^{\sharp }\right]
\end{equation*}
can be written in equivalent Lax form%
\begin{equation*}
\frac{d}{dt}(\Pi _{id}^{\sharp }+\Pi _{c}^{\sharp })=\left[ \Pi
_{id}^{\sharp }+\Pi _{c}^{\sharp },\frac{\delta F_{\Pi _{c}^{\sharp }}(\Pi
_{id}^{\sharp })}{\delta \Pi _{id}^{\sharp }}\right] .
\end{equation*}
\end{remark}

\begin{remark}
It has been argued that the physical initial conditions must satisfy $f(%
\mathbf{z},0)>0$ \cite{mor81}.The restrictions on the definition of momentum
variables may further be expanded to the physical requirement that the
density function be positive. We remark that this condition is intimately
related to the non-degeneracy of symplectic structure on coadjoint orbit of
canonical diffeomorphisms. The condition $f(\mathbf{z},0)>0$ requires the
description of density by elements $\Pi _{id}\in \mathfrak{g}^{\ast }$ with $%
div_{\Omega _{T^{\ast }\mathcal{Q}}}\Pi _{id}^{\sharp }>0$. Equivalently, in
the language of differential forms, we have $d(\Pi _{id}\wedge \Omega
_{T^{\ast }\mathcal{Q}}^{2})>0$. Consider a six dimensional domain $\mathit{D%
}$ in $T^{\ast }\mathcal{Q}$ with boundary $\partial \mathit{D}$. Then, the
positive divergence implies 
\begin{equation}
\int_{\partial \mathit{D}}\Pi _{id}(\mathbf{z})\wedge \Omega _{T^{\ast }%
\mathcal{Q}}^{2}(\mathbf{z})>0  \label{sscs}
\end{equation}%
so that we have a volume element or, an orientation, for the five
dimensional boundary of the region $\mathit{D}$. This can now be related to
the nondegeneracy of the coadjoint orbit symplectic structure on $\mathfrak{g%
}^{\ast }$. An element of the tangent space to the coadjoint orbit through $%
\Pi _{id}$ will be of the form $\mathcal{L}_{X_{k}}(\Pi _{id})$. By
definition, the orbit symplectic structure is 
\begin{eqnarray*}
\Omega _{\Pi _{id}}\left( \mathcal{L}_{X_{k}}(\Pi _{id}),\mathcal{L}%
_{X_{g}}(\Pi _{id})\right) &=&\int_{\mathit{D}}\Pi _{id}(\mathbf{z})\cdot
\lbrack X_{k}(\mathbf{z}),X_{g}(\mathbf{z})]\text{ }\Omega _{T^{\ast }%
\mathcal{Q}}^{3}(\mathbf{z}) \\
&=&\int_{\partial \mathit{D}}\{g(\mathbf{z}),k(\mathbf{z})\}\text{ }\Pi
_{id}(\mathbf{z})\wedge \text{ }\Omega _{T^{\ast }\mathcal{Q}}^{2}(\mathbf{z}%
)
\end{eqnarray*}%
which, by Eq.(\ref{sscs}) does not vanish for arbitrary functions $g$ and $k$%
.
\end{remark}

\subsection{\label{alg}Lie algebra of one-forms over $T^{\ast }\mathcal{Q}$}

To study the algebraic structure on $\Lambda ^{1}(T^{\ast }\mathcal{Q)}=%
\mathfrak{g}^{\flat }\oplus \mathfrak{g}^{\ast }$ we define the bracket of
one-forms 
\begin{equation*}
\{\alpha ,\beta \}_{\Omega _{T^{\ast }\mathcal{Q}}^{-1}}=\mathfrak{L}%
_{\alpha ^{\sharp }}\beta -\mathfrak{L}_{\beta ^{\sharp }}\alpha -d\Omega
_{T^{\ast }\mathcal{Q}}^{-1}\left( \alpha ,\beta \right)
\end{equation*}%
where $\Omega _{T^{\ast }\mathcal{Q}}^{-1}$ denotes the Poisson bi-vector
obtained by inverting the matrix of symplectic two-form $\Omega _{T^{\ast }%
\mathcal{Q}}$ and $\alpha ^{\sharp }=\Omega _{T^{\ast }\mathcal{Q}}^{\sharp
}\left( \alpha \right) $ is a vector field on $T^{\ast }\mathcal{Q}$
corresponding to the one-form $\alpha $ in $\Lambda ^{1}\left( T^{\ast }%
\mathcal{Q}\right) $. If $\alpha $ and $\beta $ are closed forms in $%
\mathfrak{g}^{\flat }$ corresponding to Hamiltonian vector fields then we
have 
\begin{equation*}
\{\alpha ,\beta \}_{\Omega _{T^{\ast }\mathcal{Q}}^{-1}}=d\left( i_{\alpha
^{\sharp }}\beta -i_{\beta ^{\sharp }}\alpha -\Omega _{T^{\ast }\mathcal{Q}%
}^{-1}\left( \alpha ,\beta \right) \right)
\end{equation*}%
which is exact and hence in $\mathfrak{g}^{\flat }.$ If $\alpha $ is closed
and $d\Pi _{id}\neq 0$, then 
\begin{equation*}
\{\alpha ,\Pi _{id}\}_{\Omega _{T^{\ast }\mathcal{Q}}^{-1}}=i_{\alpha
^{\sharp }}d\Pi _{id}+d\left( i_{\alpha ^{\sharp }}\Pi _{id}-i_{^{\Pi
_{id}^{\sharp }}}\alpha -\Omega _{T^{\ast }\mathcal{Q}}^{-1}\left( \alpha
,\Pi _{id}\right) \right)
\end{equation*}%
where the condition that the first term be closed requires the invariance
relations $di_{\alpha ^{\sharp }}d\Pi _{id}=\mathfrak{L}_{\alpha ^{\sharp
}}d\Pi _{id}=d\mathfrak{L}_{\alpha ^{\sharp }}\Pi _{id}=0$ for arbitrary $%
\alpha \in \mathfrak{g}^{\flat }$ and $\Pi _{id}\in \mathfrak{g}^{\ast }.$
The same argument applies for two arbitrary elements of $\mathfrak{g}^{\ast
}.$ Thus we have the following proposition, which summurizes the
calculations above.

\begin{proposition}
$\left\{ \mathfrak{g}^{\flat }\mathbf{,}\mathfrak{g}^{\flat }\right\}
_{\Omega _{T^{\ast }\mathcal{Q}}^{-1}}\subset \mathfrak{g}^{\flat }\mathbf{,}
$ $\left\{ \mathfrak{g}^{\ast }\mathbf{,}\mathfrak{g}^{\ast }\right\}
_{\Omega _{T^{\ast }\mathcal{Q}}^{-1}}\subset \mathfrak{g}^{\ast },$ $%
\left\{ \mathfrak{g}\mathbf{^{\flat },}\mathfrak{g}^{\ast }\right\} _{\Omega
_{T^{\ast }\mathcal{Q}}^{-1}}\subset \mathfrak{g}^{\ast }.$
\end{proposition}

According to this result, it is obvious that $\mathfrak{g}^{\flat }$ is a
subalgebra of $\Lambda ^{1}(T^{\ast }\mathcal{Q)}$ with respect to the
bracket $\left\{ \text{ }\mathbf{,}\text{ }\right\} _{\Omega _{T^{\ast }%
\mathcal{Q}}^{-1}}.$ In particular, for locally Hamiltonian vector fields in 
$\mathfrak{g}_{lh}$, $\mathfrak{g}_{lh}^{\flat }$ consists of closed but
non-exact one froms which are elements of the first de Rham cohomology space
of the particle phase space $T^{\ast }\mathcal{Q}$. These cohomological
one-forms satisfy%
\begin{equation*}
\{\mathfrak{g}_{lh}^{\flat },\mathfrak{g}^{\flat }\}_{\Omega _{\mathcal{Q}%
}^{-1}}\text{ }\subset \mathfrak{g}^{\flat }\text{, \ \ \ \ \ \ }\{\mathfrak{%
g}_{lh}^{\flat },\mathfrak{g}^{\ast }\}_{\Omega _{\mathcal{Q}}^{-1}}\subset 
\mathfrak{g}^{\ast }.
\end{equation*}

\begin{remark}
The equation $\func{div}\Pi _{id}^{\sharp }=f$ offers an alternative
notation for the elements $\Pi _{f},$ $\Pi _{g}$ in $\mathfrak{g}^{\ast }$
satisfying $\func{div}\Pi _{f}^{\sharp }=f$ and $\func{div}\Pi _{g}^{\sharp
}=g,$ respectively. We can pull-back the canonical Poisson structure on $%
T^{\ast }\mathcal{Q}$ by the map $\mathfrak{g}^{\ast }\rightarrow F\left(
T^{\ast }\mathcal{Q}\right) :\Pi _{f}\rightarrow f$ hence define a Lie
algebra structure 
\begin{equation}
\left[ \Pi _{f},\Pi _{g}\right] =\Pi _{\left\{ f,g\right\} }
\label{Liealgg*}
\end{equation}%
on $\mathfrak{g}^{\ast }$. This is the reduced Poisson structure on $%
\mathfrak{g}^{\ast }$ given in proposition 7. Eq.(\ref{Liealgg*}) gives also
that the map $\Pi _{f}\rightarrow f$ is a Poisson map.
\end{remark}

\subsection{Tulczyjew symplectic structure on $TT^{\ast }\mathcal{Q}$}

The space $TT^{\ast }\mathcal{Q}$ admits a symplectic structure first
described by Tulczyjew \cite{tul77}, \cite{tul77r}, \cite{tulc77}, \cite%
{tul81}, \cite{tul99}. A special symplectic structure is a quintuple 
\begin{equation*}
(\mathcal{P},\pi _{\mathcal{M}}^{\mathcal{P}},\mathcal{M},\vartheta _{%
\mathcal{P}},\chi )
\end{equation*}
where $\pi _{\mathcal{M}}^{\mathcal{P}}:\mathcal{P}\rightarrow \mathcal{M}$
is a fibre bundle, $\vartheta _{\mathcal{P}}$ is a one-form on $\mathcal{P}$%
, and $\chi :\mathcal{P}\rightarrow T^{\ast }\mathcal{M}$ is a fiber
preserving diffeomorphism such that $\chi ^{\ast }\theta _{T^{\ast }\mathcal{%
M}}=\vartheta _{\mathcal{P}}$ for $\theta _{T^{\ast }\mathcal{M}}$ being the
canonical one-form on $T^{\ast }\mathcal{M}$. $\chi $ can be characterized
uniquely by the condition $\left\langle \chi (p),X_{\mathcal{M}%
}(x)\right\rangle =\left\langle \vartheta _{\mathcal{P}}(p),X_{\mathcal{P}%
}(p)\right\rangle $ for each $p\in \mathcal{P}$, $\pi _{\mathcal{M}}^{%
\mathcal{P}}(p)=x$ and for vector fields $X_{\mathcal{M}}:\mathcal{M}%
\rightarrow T\mathcal{M}$, $X_{\mathcal{P}}:\mathcal{P}\rightarrow T\mathcal{%
P}$ satisfying $\left( \pi _{\mathcal{M}}^{\mathcal{P}}\right) _{\ast }X_{%
\mathcal{P}}=X_{\mathcal{M}}$. $\left( \mathcal{P},d\vartheta _{\mathcal{P}%
}\right) $ is the underlying symplectic manifold of the special symplectic
structure.

\begin{proposition}
The space $TT^{\ast }\mathcal{Q}$ is the underlying symplectic manifold for
two different special symplectic structures%
\begin{equation}
(TT^{\ast }\mathcal{Q},\tau _{T^{\ast }\mathcal{Q}},T^{\ast }\mathcal{Q}%
,\vartheta _{1},\Omega _{T^{\ast }\mathcal{Q}}^{\flat }),\ \ (TT^{\ast }%
\mathcal{Q},T\pi _{\mathcal{Q}},T\mathcal{Q},\vartheta _{2},\alpha _{%
\mathcal{Q}})  \label{sss12}
\end{equation}%
where the one-forms $\vartheta _{1}$ and $\vartheta _{2}$ are, in the
adapted coordinates,%
\begin{eqnarray}
\vartheta _{1}(\mathbf{z},\mathbf{\dot{z}}) &=&((\Omega _{T^{\ast }\mathcal{Q%
}}^{\flat })^{\ast }\theta _{T^{\ast }T^{\ast }\mathcal{Q}})(\mathbf{z},%
\mathbf{\dot{z}})=\mathbf{\dot{p}}\cdot d\mathbf{q}-\mathbf{\dot{q}}\cdot d%
\mathbf{p}  \label{tet1} \\
\vartheta _{2}(\mathbf{z},\mathbf{\dot{z}}) &=&\alpha _{\mathcal{Q}}^{\ast
}(\theta _{T^{\ast }T\mathcal{Q}})(\mathbf{z},\mathbf{\dot{z}})=\mathbf{\dot{%
p}}\cdot d\mathbf{q}+\mathbf{p}\cdot d\mathbf{\dot{q}}  \label{tet2}
\end{eqnarray}%
and the Tulczyjew two-form of the underlying symplectic manifold is 
\begin{equation}
\Omega _{TT^{\ast }\mathcal{Q}}(\mathbf{z},\mathbf{\dot{z}})=d\vartheta _{1}(%
\mathbf{z},\mathbf{\dot{z}})=d\vartheta _{2}(\mathbf{z},\mathbf{\dot{z}})=d%
\mathbf{\dot{p}\wedge }d\mathbf{q}+d\mathbf{p\wedge }d\mathbf{\dot{q}.}
\label{tulcsym}
\end{equation}
\end{proposition}

These are constructed by means of two different fibrations of $TT^{\ast }%
\mathcal{Q}$ over $T^{\ast }\mathcal{Q}$ and $T\mathcal{Q}$ which can be
represented by the diagram%
\begin{equation}
\begin{array}{c}
T^{\ast }T^{\ast }\mathcal{Q}\text{ \ \ \ \ \ \ }%
\begin{array}{c}
\text{ \ \ }\underleftarrow{\text{ }\Omega _{T^{\ast }\mathcal{Q}}^{\flat }%
\text{\ }} \\ 
\mathstrut%
\end{array}%
\text{ \ \ \ \ \ \ \ \ \ \ }TT^{\ast }\mathcal{Q}\text{\ \ \ \ \ \ \ \ }%
\begin{array}{c}
\underrightarrow{\text{ \ \ }\alpha _{\mathcal{Q}}\text{ \ \ }}\text{ \ \ \
\ } \\ 
\mathstrut%
\end{array}%
\text{ \ \ }T^{\ast }T\mathcal{Q} \\ 
\text{ \ \ }\pi _{T^{\ast }\mathcal{Q}}\searrow \nwarrow -dh\text{ \ \ \ \ \
\ \ \ }\swarrow \tau _{T^{\ast }\mathcal{Q}}\text{ \ \ \ \ }T\pi _{\mathcal{Q%
}}\searrow \text{ \ \ \ \ \ \ \ }dl\nearrow \swarrow \pi _{T\mathcal{Q}} \\ 
\text{ \ \ \ \ \ }T^{\ast }\mathcal{Q}\text{ \ \ \ \ \ \ \ \ \ \ \ \ \ \ \ \
\ \ \ \ \ \ \ \ \ \ \ \ \ }T\mathcal{Q}\text{ \ } \\ 
\text{ \ \ \ \ } \\ 
\text{\ \ }\pi _{\mathcal{Q}}\searrow \text{ \ \ \ \ \ \ \ \ \ \ }\swarrow
\tau _{\mathcal{Q}}\text{\ } \\ 
\mathcal{Q}%
\end{array}%
\end{equation}%
known as the Tulczyjew triple. Here, $\tau _{\mathcal{Q}}$, $\pi _{\mathcal{Q%
}}$, $\tau _{T^{\ast }\mathcal{Q}}$ and $\pi _{T^{\ast }\mathcal{Q}}$ are
natural projections, $\Omega _{T^{\ast }\mathcal{Q}}^{\flat }$ is the
induced map from the symplectic two-form $\Omega _{T^{\ast }\mathcal{Q}}$ on 
$T^{\ast }\mathcal{Q}$, $\alpha _{\mathcal{Q}}$ is a diffeomorphism
constructed as a dual of canonical involution $\kappa _{\mathcal{Q}}$ of $TT%
\mathcal{Q}$. $\alpha _{\mathcal{Q}}$ is a canonical description of the
equivalence of functors $TT^{\ast }$ and $T^{\ast }T$ while $\kappa _{%
\mathcal{Q}}$ describes the canonical flip of the first derivatives with
respect to two different parametrizations for second order tangent bundle.
In coordinates, we have $\alpha _{\mathcal{Q}}\left( \mathbf{q},\mathbf{p};%
\mathbf{\dot{q}},\mathbf{\dot{p}}\right) =\left( \mathbf{q},\mathbf{\dot{q}};%
\mathbf{\dot{p}},\mathbf{p}\right) $. The triangular diagrams on left and
right define special symplectic structures on $TT^{\ast }\mathcal{Q}$ by
pull-back of canonical one-forms $\theta _{T^{\ast }T^{\ast }\mathcal{Q}}$
and $\theta _{T^{\ast }T\mathcal{Q}}$ on the cotangent bundles $T^{\ast
}T^{\ast }\mathcal{Q}$ and $T^{\ast }T\mathcal{Q}$, respectively. $\Omega
_{T^{\ast }\mathcal{Q}}^{\flat }$ and $\alpha _{\mathcal{Q}}$ are
symplectomorphisms from $TT^{\ast }\mathcal{Q}$ to the canonical symplectic
manifolds $T^{\ast }T\mathcal{Q}$ and $T^{\ast }T^{\ast }\mathcal{Q}$.

Hamiltonian and Lagrangian formulations can then be realized as Lagrangian
submanifolds of $TT^{\ast }\mathcal{Q}$. A submanifold $\mathcal{S}$ of a
symplectic manifold $\left( \mathcal{M},\Omega \right) $ is a Lagrangian
submanifold, if its dimension is half the dimension of $\mathcal{M}$ and the
restriction of $\Omega $ on $\mathcal{S}$ vanishes, that is $\left. \Omega
\right\vert _{S}=0$ \cite{aw71}, \cite{aw73}.

Consider a special symplectic structure $(\mathcal{P},\pi _{\mathcal{M}}^{%
\mathcal{P}},\mathcal{M},\vartheta _{\mathcal{P}},\chi )$ and let $g:%
\mathcal{M\rightarrow \mathbb{R}}$ be a real valued function. Then, the set 
\begin{equation*}
\mathcal{S}_{\mathcal{P}}=\left\{ p\in \mathcal{P}:\left\langle dg(x),T\pi _{%
\mathcal{M}}^{\mathcal{P}}\circ X_{\mathcal{P}}(p)\right\rangle
=\left\langle \vartheta _{\mathcal{P}}(p),X_{\mathcal{P}}(p)\right\rangle
,\forall X_{\mathcal{P}}\in \mathfrak{X}(\mathcal{P})\right\}
\end{equation*}%
is a Lagrangian submanifold of the underlying symplectic manifold $(\mathcal{%
P},d\vartheta _{\mathcal{P}})$ and the function $g$ is called the generating
function \cite{st72}. It follows from the definition of $\mathcal{S}_{%
\mathcal{P}}$ that, the one-form $\vartheta _{\mathcal{P}}$ is characterized
by the relation $(\pi _{\mathcal{M}}^{\mathcal{P}})^{\ast }dg=\vartheta _{%
\mathcal{P}}$. Since $\chi $ is a symplectic diffeomorphism, it maps $%
\mathcal{S}_{\mathcal{P}}$ to the space $Im\left( dg\right) $ which is a
Lagrangian submanifold of $T^{\ast }\mathcal{M}$. In general, the image of a
closed one-form on $\mathcal{M}$ is a Lagrangian submanifold of $T^{\ast }%
\mathcal{M}$ and its pull-back to $\mathcal{P}$ by $\chi $ is a Lagrangian
submanifold of $\mathcal{P}$.

Let $l:T\mathcal{Q}\rightarrow 
\mathbb{R}
$. The image of mapping $dl:T\mathcal{Q}\rightarrow T^{\ast }T\mathcal{Q}$
is described by the equations $\boldsymbol{\lambda }_{\dot{q}}=\nabla _{q}l(%
\mathbf{q,\dot{q}})$ and\ $\boldsymbol{\lambda }_{q}=\nabla _{\dot{q}}l(%
\mathbf{q,\dot{q}})$. Pull back of this to $TT^{\ast }\mathcal{Q}$ gives the
dynamical equations $(T\pi _{\mathcal{Q}})^{\ast }dl=\vartheta _{2}$ which,
in coordinates, read $\nabla _{\dot{q}}l(\mathbf{q,\dot{q}})=\mathbf{p}$, $%
\nabla _{q}l(\mathbf{q,\dot{q}})=\mathbf{\dot{p}}$ . For a Hamiltonian
function $h:T^{\ast }\mathcal{Q}\rightarrow \mathcal{%
\mathbb{R}
}$, the image $Im(-dh)$ is a Lagrangian submanifold of $T^{\ast }T^{\ast }%
\mathcal{Q}$. The Hamilton's equations on $TT^{\ast }\mathcal{Q}$ are
obtained from the relation $\vartheta _{1}=\tau _{T^{\ast }\mathcal{Q}%
}^{\ast }(-dh)=-d(h\circ \tau _{T^{\ast }\mathcal{Q}})$, which, in
coordinates, are expressed as%
\begin{equation}
-dh(\mathbf{z})=\mathbf{\dot{p}}\cdot d\mathbf{q}-\mathbf{\dot{q}}\cdot d%
\mathbf{p},\;\;\;\;\mathbf{\dot{q}}=\nabla _{p}{h}(\mathbf{z}),\;\;\;\mathbf{%
\dot{p}}=-\nabla _{q}{h}(\mathbf{z}){.}  \label{hameqs}
\end{equation}%
Since the derivative of $\vartheta _{1}=\tau _{T^{\ast }\mathcal{Q}}^{\ast
}(-dh)$ vanishes, the Hamiltonian dynamics becomes a Lagrangian submanifold
of $(TT^{\ast }\mathcal{Q},d\vartheta _{1})$ generated by the function $-h$.
If $X$ is locally Hamiltonian then the one form $i_{X}\Omega _{T^{\ast }%
\mathcal{Q}}$ is still closed by definition and $Im\left( X\right) $ defines
a Lagrangian submanifold of $TT^{\ast }\mathcal{Q}$, as well. Thus, we have
the identification of the vector space $\mathfrak{X}_{ham}(T^{\ast }\mathcal{%
Q)}$ with the space of all Lagrangian submanifolds of the Tulczyjew
symplectic space $(TT^{\ast }\mathcal{Q},\Omega _{TT^{\ast }\mathcal{Q}})$.
In the next subsection, we shall obtain this space as tangent space over
identity of configuration space of plasma.

\subsection{Spaces of Lagrangian submanifolds}

As the Hamiltonian dynamics of a single particle described by a
diffeomorphism $\varphi \in Diff_{can}(T^{\ast }\mathcal{Q})$ corresponds to
a Lagrangian submanifold of the Tulczyjew symplectic manifold $TT^{\ast }%
\mathcal{Q}$, it is possible to describe all such motions, that is, each
configuration of plasma by a Lagrangian submanifold in $TT^{\ast }\mathcal{Q}
$ \cite{gpd1}. We shall show that the space $Lag(TT^{\ast }\mathcal{Q}$,$%
\Omega _{TT^{\ast }\mathcal{Q}})$ of all Lagrangian submanifolds can be
obtained as the tangent space over identity of a suitable representation of
the group $Diff_{can}(T^{\ast }\mathcal{Q})$ of all canonical
transformations. We rely on the fact that the configuration space $%
Diff_{can}(T^{\ast }\mathcal{Q})$, as a manifold of maps \cite{Mi78}, \cite%
{TS01}, can also be given a description in terms of sections $\Gamma
(pr_{0}) $ of the trivial bundle $pr_{0}:T^{\ast }\mathcal{Q}_{0}\mathcal{%
\times }T^{\ast }\mathcal{Q\rightarrow }T^{\ast }\mathcal{Q}_{0}$ where $%
T^{\ast }\mathcal{Q}_{0}$ is the particle phase space with Lagrangian
coordinates $\mathbf{Z}$ and $T^{\ast }\mathcal{Q}$ carries Eulerian
coordinates $\mathbf{z}$. The total space $T^{\ast }\mathcal{Q}_{0}\mathcal{%
\times }T^{\ast }\mathcal{Q}$ is then symplectic with the two-form \cite%
{st72}, \cite{tul74}, \cite{wei77}, \cite{bt80}%
\begin{equation*}
\Omega _{-}(\mathbf{Z,z})=\Omega _{T^{\ast }\mathcal{Q}_{0}}(\mathbf{Z}%
)-\Omega _{T^{\ast }\mathcal{Q}}(\mathbf{z})=d\mathbf{P}\wedge d\mathbf{Q}-d%
\mathbf{p}\wedge d\mathbf{q}.
\end{equation*}

\begin{proposition}
$Diff_{can}(T^{\ast }\mathcal{Q})$ can be identified with the space $%
Lag\Gamma (pr_{0},\Omega _{-}\mathcal{)}$ of all Lagrangian sections of the
trivial bundle $(pr_{0},\Omega _{-}\mathcal{)}$. In this case, the Lie
algebra $\mathfrak{X}_{ham}(T^{\ast }\mathcal{Q})$ of Hamiltonian vector
fields corresponds to the space $Lag(TT^{\ast }\mathcal{Q}$,$\Omega
_{TT^{\ast }\mathcal{Q}})$ of all Lagrangian submanifolds of the Tulczyjew
symplectic space.

\begin{proof}
A diffeomorphism $\varphi :T^{\ast }\mathcal{Q}_{0}\mathcal{\rightarrow }%
T^{\ast }\mathcal{Q}$ is canonical if $\Omega _{T^{\ast }\mathcal{Q}%
_{0}}-\varphi ^{\ast }\Omega _{T^{\ast }\mathcal{Q}}=0$. It follows that $%
\Omega _{-}$ vanishes when restricted to the graphs 
\begin{equation*}
Gr\varphi =\{(\mathbf{Z},\varphi (\mathbf{Z)}):\mathbf{Z}\in T^{\ast }%
\mathcal{Q}_{0}\}\subset \Gamma (pr_{0})
\end{equation*}%
of canonical diffeomorphisms \cite{wei77}, \cite{bt80}. For a base point $%
\mathbf{Z}\in T^{\ast }\mathcal{Q}_{0}$, the total space is twelve
dimensional and $Gr\varphi $ is a six dimensional subspace. When $\varphi $
is canonical, $\Omega _{-}$ vanishes on $Gr\varphi $ and this is a
Lagrangian submanifold in $(T^{\ast }\mathcal{Q}_{0}\mathcal{\times }T^{\ast
}\mathcal{Q},\Omega _{-}\mathcal{)}$. If we denote the space of all sections
of the trivial bundle on which the restriction of $\Omega _{-}$ vanishes,
namely, the space of all Lagrangian sections by $Lag\Gamma (pr_{0},\Omega
_{-}\mathcal{)}$, then we have the bijective correspondence 
\begin{equation*}
Diff_{can}(T^{\ast }\mathcal{Q})\longleftrightarrow Lag\Gamma (pr_{0},\Omega
_{-}\mathcal{)}:\varphi \longleftrightarrow Gr\varphi \text{.}
\end{equation*}%
To find the tangent space over the identity mapping, we proceed as follows.
Corresponding to a curve $\varphi _{t}\in Diff_{can}(T^{\ast }\mathcal{Q})$
with $\varphi _{0}(\mathbf{Z})=\mathbf{z}$, we have the curve $t\mapsto
Gr\varphi _{t}$ in $Lag\Gamma (pr_{0},\Omega _{-}\mathcal{)}$ with $%
Gr\varphi _{0}=\{(\mathbf{Z},\mathbf{z}):\mathbf{Z}\in T^{\ast }\mathcal{Q}%
_{0}\}$. The tangent space $T_{Gr\varphi _{t}}Lag\Gamma (pr_{0},\Omega _{-}%
\mathcal{)}$ consists of vectors 
\begin{equation}
X_{Gr\varphi _{t}}(\mathbf{Z})={\frac{d}{dt}}Gr\varphi _{t}(\mathbf{Z})=(%
\mathbf{Z},\varphi _{t}(\mathbf{Z)};0,\frac{d\varphi _{t}(\mathbf{Z)}}{dt})=(%
\mathbf{Z},\varphi _{t}(\mathbf{Z)};0,\mathbf{X}_{\varphi _{t}}(\mathbf{Z}))
\end{equation}%
tangent to $Gr\varphi _{t}$. For each $\mathbf{Z\in }T^{\ast }\mathcal{Q}%
_{0} $, this is a vector tangent to the fiber $T^{\ast }\mathcal{Q}$ over $%
\mathbf{Z}$. That means, $X_{Gr\varphi _{t}}(\mathbf{Z})$ is in the vertical
tangent space 
\begin{equation}
V_{Gr\varphi _{t}(\mathbf{Z})}(T^{\ast }\mathcal{Q}_{0}\mathcal{\times }%
T^{\ast }\mathcal{Q)}.  \label{vertan}
\end{equation}%
Over the identity, $t=0$, we have%
\begin{equation*}
X_{Gr\varphi _{0}}(\mathbf{Z})={\frac{d}{dt}}Gr\varphi _{t}(\mathbf{Z}%
)|_{t=0}=(\mathbf{Z},\mathbf{z};0,\frac{d\varphi _{t}(\mathbf{Z)}}{dt}%
|_{t=0})=(\mathbf{Z},\mathbf{z};0,\mathbf{X}_{h}(\mathbf{z}))
\end{equation*}%
where $X_{h}(\mathbf{z})=$ $\mathbf{X}_{h}(\mathbf{z})\cdot \nabla _{z}$ is
the Hamiltonian vector field generating $\varphi _{t}\in Diff_{can}(T^{\ast }%
\mathcal{Q})$. Thus, over the identity mapping we have 
\begin{equation*}
V_{(\mathbf{Z,z)}}(T^{\ast }\mathcal{Q}_{0}\mathcal{\times }T^{\ast }%
\mathcal{Q)}\longleftrightarrow T_{\mathbf{z}}T^{\ast }\mathcal{Q}%
:X_{Gr\varphi _{0}}(\mathbf{Z})\longleftrightarrow X_{h}(\mathbf{z}).
\end{equation*}%
In fact, for each $\varphi $\ the vertical tangent space in (\ref{vertan})
is isomorphic to a copy of $TT^{\ast }\mathcal{Q}$. To this end, we recall
the definition of pull-back bundle. Given $E\longrightarrow N$ and a
continuous map $\Phi :M\longrightarrow N$, the pull-back of $E$ by $\Phi $
is the bundle $\Phi ^{\ast }E\longrightarrow M$ whose fiber $(\Phi ^{\ast
}E)_{x}$ over $x\in M$ is the fiber $E_{\Phi (x)}$ of $E\longrightarrow N$
over $\Phi (x)$. In our case, $E_{\Phi (x)}=V_{Gr\varphi _{t}(\mathbf{Z}%
)}(T^{\ast }\mathcal{Q}_{0}\mathcal{\times }T^{\ast }\mathcal{Q)}$ and hence%
\begin{equation*}
V_{Gr\varphi _{t}(\mathbf{Z})}(T^{\ast }\mathcal{Q}_{0}\mathcal{\times }%
T^{\ast }\mathcal{Q)}\mathcal{=}Gr\varphi _{t}{}^{\ast }(V_{(\mathbf{Z,z)}%
}(T^{\ast }\mathcal{Q}_{0}\mathcal{\times }T^{\ast }\mathcal{Q)}).
\end{equation*}%
Since we are dealing with a trivial bundle, the vertical space is just the
tangent space to the second factor, so we have 
\begin{eqnarray*}
V_{Gr\varphi _{t}(\mathbf{Z})}(T^{\ast }\mathcal{Q}_{0}\mathcal{\times }%
T^{\ast }\mathcal{Q)} &\mathcal{=}&Gr\varphi _{t}{}^{\ast }(TT^{\ast }%
\mathcal{Q)=}(id,\varphi _{t})^{\ast }(TT^{\ast }\mathcal{Q)} \\
&=&\varphi _{t}{}^{\ast }(TT^{\ast }\mathcal{Q)}
\end{eqnarray*}%
as the tangent space over $\varphi \in Diff_{can}(T^{\ast }\mathcal{Q})$.
Thus, the tangent space at $Gr\varphi _{t}$ to the space of Lagrangian
sections $Lag\Gamma (pr_{0},\Omega _{-}\mathcal{)}$ is the space of sections
consisting of \ pull-back by $\varphi $ of (Hamiltonian) vector fields on $%
T^{\ast }\mathcal{Q}$. As $\varphi $ is canonical, these sections are images
of Hamiltonian vector fields on $T^{\ast }\mathcal{Q}$ and hence are
Lagrangian submanifolds in $\varphi {}^{\ast }(TT^{\ast }\mathcal{Q)}$. For $%
\varphi $ being the identity element of $G$, we obtain the space of
Lagrangian submanifolds $Lag(TT^{\ast }\mathcal{Q)}$ as the Lie algebra of
the group $Diff_{can}(T^{\ast }\mathcal{Q})$.

\begin{remark}
Having the correspondence $Diff_{can}(T^{\ast }\mathcal{Q}%
)\longleftrightarrow Lag\Gamma (pr_{0},\Omega _{-}\mathcal{)}$ for
configuration space of plasma dynamics, one needs operations between
Lagrangian submanifolds of $Lag\Gamma (pr_{0},\Omega _{-}\mathcal{)}$
similar to right and left multiplications of the group $Diff_{can}(T^{\ast }%
\mathcal{Q})$ producing particle relabelling symmetry and kinematical
motion, respectively. This can be achieved in the framework of symplectic
relations. The Lagrangian submanifolds of symplectic spaces of the form $(%
\mathcal{M}_{2}\times \mathcal{M}_{1},\Omega _{-}=\Omega _{2}-\Omega _{1})$
were defined as symplectic relations and their composition rules were proved
in \cite{st72}.
\end{remark}
\end{proof}
\end{proposition}

\bigskip

\bigskip

\pagebreak

\section{From Particle Dynamics to Vlasov Equation}

We shall describe a purely geometric framework in which one can find a
precise relation between the individual particle motion and the Vlasov
equation. In this framework, the one-form $\vartheta _{2}(\mathbf{z},\mathbf{%
\dot{z}})$ of the special symplectic structure in Eq.(\ref{tet2}) and the
Tulczyjew symplectic two-form can be obtained as complete tangent lifts of
canonical forms on particle phase space. Given the infinitesimal generator
of particle motion, its complete cotangent lift describes a Lagrangian
submanifold of certain special symplectic structure. We shall present a
Morse family generating Legendre transformation, in the sense of Tulczyjew,
for the lifted motion. Using a holonomic lift operator, we shall carry the
lifted motion to vertical subspace of a Tulczyjew symplectic space. This
subspace is integrable when restricted to generators lifted from the Lie
algebra of Hamiltonian vector fields. Finally, introducing the vertical
lifts of one-forms we shall obtain the relation between the Hamiltonian
vector fields generating the particle motion and the momentum-Vlasov
equations. We refer to \cite{sas58}, \cite{dom62}, \cite{ton62}, \cite{kn}, 
\cite{ly63}, \cite{yd63}, \cite{yl64}, \cite{st64}, \cite{tan65}, \cite{yk66}%
, \cite{yp66}, \cite{ma67}, \cite{gan81}, \cite{cr}, \cite{pog89}, \cite%
{kob91}, \cite{kms93}, \cite{mr94}, \cite{LeRo} for definitions and various
aspects of lifts of geometric objects some of which we summarize in the
following subsection.

\subsection{Complete lifts \label{completelifts}}

Let $X$ be a vector field on $\mathcal{M}$, $\varphi _{t}:\mathcal{M}%
\rightarrow \mathcal{M}$ be its flow and $\tau _{\mathcal{M}}:T\mathcal{%
M\rightarrow M}$ be the tangent bundle. The tangent lift\textbf{\ }$\varphi
_{t}^{c}:T\mathcal{M}\rightarrow T\mathcal{M}$ of $\varphi _{t}$ is defined
as to satisfy $\tau _{\mathcal{M}}\circ \varphi _{t}^{c}=\varphi _{t}\circ
\tau _{\mathcal{M}}$ and constitutes a one-parameter group of
diffeomorphisms on $T\mathcal{M}$. Differentiating the defining relation we
obtain $T\tau _{\mathcal{M}}\circ X^{c}=X\circ \tau _{\mathcal{M}}$ where $%
T\tau _{\mathcal{M}}$ is the tangent mapping of $\tau _{\mathcal{M}}$. This
means that $X$ and $X^{c}$ are $\tau _{\mathcal{M}}$ related. $X^{c}$ is
called the complete tangent lift of the vector field $X$. In local
coordinates $\left( x^{a},v^{a}\right) $ of $T\mathcal{M}$, the complete
tangent lift of $X(x)=X^{a}(x)\partial /\partial x^{a}$ is given by%
\begin{equation}
X^{c}\left( x,v\right) =X^{a}\left( x\right) \dfrac{\partial }{\partial x^{a}%
}+v^{b}\dfrac{\partial X^{a}(x)}{\partial x^{b}}\dfrac{\partial }{\partial
v^{a}}.
\end{equation}

Similarly, the cotangent lift of the flow $\varphi _{t}$ is a one-parameter
group of diffeomorphisms $\varphi _{t}^{c\ast }$ on $T^{\ast }\mathcal{M}$
satisfying $\pi _{\mathcal{M}}\circ \varphi _{t}^{c\ast }=\varphi _{t}\circ
\pi _{\mathcal{M}}$, where $\pi _{\mathcal{M}}$ is the natural projection of 
$T^{\ast }\mathcal{M}$ to $\mathcal{M}$. The generator $X^{c\ast }$ of $%
\varphi _{t}^{c\ast }$ is the complete cotangent lift of $X$ and is obtained
from the infinitesimal version of the defining relation as 
\begin{equation}
T\pi _{\mathcal{M}}\circ X^{c\ast }=X\circ \pi _{\mathcal{M}},
\label{incotanlift}
\end{equation}%
which means that, $X$ and $X^{c\ast }$ are $\pi _{\mathcal{M}}$ related.
Since Eq.(\ref{incotanlift}) is equivalent to $\pi _{\mathcal{M}}^{\ast
}X=X^{c\ast }$, we have

\begin{proposition}
The cotangent lift $^{c\ast }:\mathfrak{X}\left( \mathcal{M}\right)
\rightarrow \mathfrak{X}\left( T^{\ast }\mathcal{M}\right) $ is a Lie
algebra isomorphism into $\left[ X,Y\right] ^{c\ast }=\left[ X^{c\ast
},Y^{c\ast }\right] ,$ for all $X,Y\in \mathfrak{X}\left( \mathcal{M}\right) 
$.
\end{proposition}

In fact, this is the homomorphism that leads to the so called
plasma-to-fluid map (c.f section 4.4).

The complete cotangent lift $X^{c\ast }$ of a vector field $X$ on $\mathcal{M%
}$ is a Hamiltonian vector field on the canonically symplectic manifold $%
T^{\ast }\mathcal{M}$. Indeed, the lifted flow $\varphi _{t}^{c\ast }$
preserves the canonical one-form $\theta _{T^{\ast }\mathcal{M}}$ on the
cotangent bundle $\left( \varphi _{t}^{c\ast }\right) ^{\ast }\theta
_{T^{\ast }\mathcal{M}}=\theta _{T^{\ast }\mathcal{M}}$ \cite{mr94},\cite%
{st64},\cite{LeRo}. Differentiating at $t=0$ gives $\mathcal{L}_{X^{c\ast
}}\theta _{T^{\ast }\mathcal{M}}=0$. Using the identity $\mathcal{L}%
_{X}=di_{X}+i_{X}d$ we obtain the Hamilton's equations%
\begin{equation}
i_{X^{c\ast }}\left( \Omega _{T^{\ast }\mathcal{M}}\right) =-d\left(
i_{X^{c\ast }}\theta _{T^{\ast }\mathcal{M}}\right)  \label{hameq}
\end{equation}%
for $X^{c\ast }$ with the Hamiltonian function $i_{X^{c\ast }}\theta
_{T^{\ast }\mathcal{M}}$. Taking $\theta _{T^{\ast }\mathcal{M}}\left(
x,y\right) =y_{a}dx^{a}$ and $X=X^{a}(x)\partial /\partial x^{a}$ we obtain 
\begin{equation}
X^{c\ast }\left( x,y\right) =X^{a}(x)\dfrac{\partial }{\partial x^{a}}-y_{b}%
\dfrac{\partial X^{b}(x)}{\partial x^{a}}\dfrac{\partial }{\partial y_{a}}
\label{colift}
\end{equation}%
and the Hamiltonian function $y_{a}X^{a}(x)$ is degenerate in the fiber
variables $y_{a}$. We observe that complete cotangent lifts can be given a
variational formulation as well. Define the Whitney product%
\begin{equation}
T^{\ast }\mathcal{M}\times _{\mathcal{M}}T\mathcal{M}=\{\left( \alpha
,X\right) \in T^{\ast }\mathcal{M}\times T\mathcal{M}\text{ \ \ }:\pi _{%
\mathcal{M}}\left( \alpha \right) =\tau _{\mathcal{M}}\left( X\right) \} 
\notag
\end{equation}%
which may be viewed as a submanifold of $TT^{\ast }\mathcal{M}$ given by $%
\dot{y}=0$. Then, the following result is straightforward.

\begin{proposition}
\label{liftlag}Associated to the cotangent lift in Eq.(\ref{colift}), the
first order differential equations%
\begin{equation*}
\dot{x}^{a}=X^{a}(x),\text{ \ \ \ \ \ }\dot{y}_{a}=-y_{b}\dfrac{\partial
X^{b}(x)}{\partial x^{a}}
\end{equation*}%
are the Euler-Lagrange equations for the (degenerate) Lagrangian density 
\begin{equation*}
L(x,\dot{x},y)=y_{a}(\dot{x}^{a}-X^{a}(x))=y_{a}\dot{x}^{a}-H(x,y)
\end{equation*}%
defined on the Whitney product $T^{\ast }\mathcal{M\times }_{\mathcal{M}}T%
\mathcal{M}$.

\begin{remark}
In \cite{nanni}, the Whitney product was shown to be isomorphic to the
restriction of $TT^{\ast }\mathcal{M}$ to zero section of $T^{\ast }\mathcal{%
M}$. Then, a generalized tangent bundle and a generalized complex structure
on $\mathcal{M}$ were introduced as the Whitney product being a bundle over $%
\mathcal{M}$ and, as a comlex structure on Whitney product.
\end{remark}
\end{proposition}

Complete tangent lift $f^{c}\in \mathcal{F}\left( T\mathcal{M}\right) $ of a
function $f\in \mathcal{F}\left( \mathcal{M}\right) $ is simply the
directional derivative $f^{c}\left( x,v\right) =df\left( x\right) \cdot v$
and is given in coordinates as $f^{c}=v^{a}\partial f/\partial x^{a}.$ Let $%
\omega _{\mathcal{M}}\in \Lambda ^{k}\left( \mathcal{M}\right) $ be a
differential k-form on $\mathcal{M}$. Its complete tangent lift $\omega _{%
\mathcal{M}}^{c}\in \Lambda ^{k}\left( T\mathcal{M}\right) $ is a
differential k-form on $T\mathcal{M}$ and is defined by means of the lifts
of vector fields and functions, namely,%
\begin{equation}
\omega _{\mathcal{M}}^{c}(X_{1}^{c},...,X_{k}^{c})=(\omega _{\mathcal{M}%
}(X_{1},...,X_{k}))^{c}.
\end{equation}%
For a one-form $\theta _{\mathcal{M}}=\theta _{a}dx^{a}\in \Lambda
^{1}\left( \mathcal{M}\right) $, we compute 
\begin{equation}
\theta _{\mathcal{M}}^{c}=\frac{\partial \theta _{a}}{\partial x^{b}}%
v^{b}dx^{a}+\theta _{a}dv^{a}
\end{equation}%
and for a two-form $\Omega _{\mathcal{M}}=(1/2)\Omega _{ab}dx^{a}\wedge
dx^{b}\in \Lambda ^{2}\left( \mathcal{M}\right) $, we find%
\begin{equation}
\Omega _{\mathcal{M}}^{c}(x,v)=\frac{1}{2}(v^{a}\frac{\partial \Omega _{ab}}{%
\partial x^{d}}dx^{d}\wedge dx^{b}+\Omega _{ab}dv^{a}\wedge dx^{b}+\Omega
_{ab}dx^{a}\wedge dv^{b}).
\end{equation}%
For a constant matrix $\Omega _{ab}$, this reduces to 
\begin{equation}
\Omega _{\mathcal{M}}^{c}(x,v)=\Omega _{ab}dx^{a}\wedge dv^{b}=dx^{a}\wedge
d(\Omega _{ab}v^{b}).
\end{equation}%
If $\Omega _{\mathcal{M}}$ defines a constant symplectic structure on $%
\mathcal{M}$, Hamiltonian vector fields are of the form $\Omega
_{ab}v^{b}=\partial k(x)/\partial x^{a}$ for functions $k(x)$ on $\mathcal{M}
$. Restricting the lifted two-form to Hamiltonian vector fields we find%
\begin{equation}
\Omega _{\mathcal{M}}^{c}(x,v)=dx^{a}\wedge d\frac{\partial k(x)}{\partial
x^{a}}=dx^{a}\wedge \frac{\partial dk(x)}{\partial x^{a}}=d^{2}k(x)\equiv 0
\end{equation}%
which means that, with respect to the lift $\Omega _{\mathcal{M}}^{c}$ of
(constant) symplectic structure $\Omega _{\mathcal{M}}$, Hamiltonian vector
fields of $\Omega _{\mathcal{M}}$ define Lagrangian submanifolds of $(T%
\mathcal{M},$ $\Omega _{\mathcal{M}}^{c})$. In fact, the Tulczyjew
symplectic two-form is of this sort.

\subsection{Lift of particle dynamics}

We consider the particle dynamics on $\mathcal{M}=T^{\ast }\mathcal{Q}$
described as the flow of Hamiltonian vector field 
\begin{equation}
X_{h}(\mathbf{z})={\frac{1}{m}}\mathbf{p}\cdot \nabla _{q}-e\nabla _{q}\phi
_{f}\left( \mathbf{q}\right) \cdot \nabla _{p}
\end{equation}%
with respect to the symplectic two-form $\Omega _{T^{\ast }\mathcal{Q}}(%
\mathbf{z})=d\mathbf{p}\wedge d\mathbf{q}$ which is exact $\Omega _{T^{\ast }%
\mathcal{Q}}=d\theta _{T^{\ast }\mathcal{Q}}$ with $\theta _{T^{\ast }%
\mathcal{Q}}(\mathbf{z})=\mathbf{p}\cdot d\mathbf{q}$ and, for the
Hamiltonian function ${h(\mathbf{z})}=p^{2}/2m+e\phi _{f}(\mathbf{q})$ which
is the energy of a charged particle. First, we note

\begin{proposition}
Complete tangent lifts of $\theta_{T^{\ast }\mathcal{Q}}$ and $\Omega
_{T^{\ast }\mathcal{Q}}$ are 
\begin{eqnarray}
\theta _{T^{\ast }\mathcal{Q}}^{c}(\mathbf{z},\mathbf{\dot{z}}) &=&\vartheta
_{2}(\mathbf{z},\mathbf{\dot{z}})=\alpha _{\mathcal{Q}}^{\ast }(\theta
_{T^{\ast }T\mathcal{Q}})(\mathbf{z},\mathbf{\dot{z}})  \notag \\
\Omega _{T^{\ast }\mathcal{Q}}^{c}(\mathbf{z},\mathbf{\dot{z}}) &=&\Omega
_{TT^{\ast }\mathcal{Q}}(\mathbf{z},\mathbf{\dot{z}})
\end{eqnarray}%
which are the one-form for the fibration over $T\mathcal{Q}$ and the
Tulczyjew two-form on $TT^{\ast }\mathcal{Q}$, given in Eqs.(\ref{tet2}) and
(\ref{tulcsym}), respectively.
\end{proposition}

The complete cotangent lift of $X_{h}(\mathbf{z})$ is the vector 
\begin{equation}
X_{h}^{c\ast }\left( \mathbf{z},\mathbf{\Pi }_{id}\right) =X_{h}\left( 
\mathbf{z}\right) +e(\mathbf{\Pi }_{p}\cdot \nabla _{q})\nabla _{q}\phi
_{f}\left( \mathbf{q}\right) \cdot \nabla _{\Pi _{q}}-{\frac{1}{m}}\mathbf{%
\Pi }_{q}\cdot \nabla _{\Pi _{p}}  \label{cotlift}
\end{equation}%
on $T_{z}^{\ast }T_{q}^{\ast }\mathcal{Q}$ which is canonically Hamiltonian 
\begin{equation}
i_{X_{h}^{c\ast }}\Omega _{T^{\ast }T^{\ast }\mathcal{Q}}=-dH_{T^{\ast
}T^{\ast }\mathcal{Q}}  \label{coliham}
\end{equation}%
with the canonical two-form%
\begin{equation}
\Omega _{T^{\ast }T^{\ast }\mathcal{Q}}(\mathbf{z},\mathbf{\Pi }_{id})=d(%
\mathbf{\Pi }_{q}\cdot d\mathbf{q}+\mathbf{\Pi }_{p}\cdot d\mathbf{p}%
)=d\theta _{T^{\ast }T^{\ast }\mathcal{Q}}(\mathbf{z},\mathbf{\Pi }_{id})
\end{equation}%
and for the Hamiltonian function%
\begin{equation}
H_{T^{\ast }T^{\ast }\mathcal{Q}}(\mathbf{z},\mathbf{\Pi }_{id})={\frac{1}{m}%
}\mathbf{p}\cdot \mathbf{\Pi }_{q}-e\nabla _{q}\phi (\mathbf{q})\cdot 
\mathbf{\Pi }_{p}=\left\langle X_{h}\left( \mathbf{z}\right) ,\Pi
_{id}\left( \mathbf{z}\right) \right\rangle _{T^{\ast }\mathcal{Q}}.
\end{equation}%
Hence, the constructions for $X_{h}(\mathbf{z})$ can be carried over the
cotangent lift $X_{h}^{c\ast }\left( \mathbf{z},\mathbf{\Pi }_{id}\right) $
by replacing $\mathcal{Q}$ with $T^{\ast }\mathcal{Q}$ but with a degenerate
Hamiltonian function. An invariant way of writing this Hamiltonian is%
\begin{equation}
H_{T^{\ast }T^{\ast }\mathcal{Q}}=i_{X_{h}^{c\ast }}\theta _{T^{\ast
}T^{\ast }\mathcal{Q}}=i_{X_{h}}\theta _{T^{\ast }T^{\ast }\mathcal{Q}}
\end{equation}%
where we used the fact that $\theta _{T^{\ast }T^{\ast }\mathcal{Q}}$ has no
components along vertical directions. It follows that the canonical one-form 
$\theta _{T^{\ast }T^{\ast }\mathcal{Q}}$ is an absolute invariant of the
cotangent lift%
\begin{equation}
\mathcal{L}_{X_{h}^{c\ast }}\theta _{T^{\ast }T^{\ast }\mathcal{Q}%
}=dH_{T^{\ast }T^{\ast }\mathcal{Q}}+i_{X_{h}^{c\ast }}\Omega _{T^{\ast
}T^{\ast }\mathcal{Q}}=0  \label{inv2}
\end{equation}%
which can be regarded to be equivalent to the Hamilton's equations (\ref%
{coliham}).

The image of the section $dH_{T^{\ast }T^{\ast }\mathcal{Q}}:T^{\ast
}T^{\ast }\mathcal{Q}\rightarrow T^{\ast }T^{\ast }T^{\ast }\mathcal{Q}$ is
a Lagrangian submanifold of $\left( T^{\ast }T^{\ast }T^{\ast }\mathcal{Q}%
,\Omega _{T^{\ast }T^{\ast }T^{\ast }\mathcal{Q}}\right) $ and the image of $%
\Omega _{T^{\ast }T^{\ast }\mathcal{Q}}^{\sharp }\left( -dH_{T^{\ast
}T^{\ast }\mathcal{Q}}\right) =X_{h}^{c\ast }$ is a Lagrangian submanifold
of $TT^{\ast }T^{\ast }\mathcal{Q}$ with the Tulczyjew's two-form $\Omega
_{TT^{\ast }T^{\ast }\mathcal{Q}}$. Hence, we have

\begin{proposition}
The image of complete cotangent lift $X_{h}^{c\ast }$ of Hamiltonian vector
field $X_{h}$ is a Lagrangian submanifold of the special symplectic structure%
\begin{equation}
(TT^{\ast }T^{\ast }\mathcal{Q},\tau _{T^{\ast }T^{\ast }\mathcal{Q}%
},T^{\ast }T^{\ast }\mathcal{Q},\Theta _{1}=\left( \Omega _{T^{\ast }T^{\ast
}\mathcal{Q}}^{\flat }\right) ^{\ast }\theta _{T^{\ast }T^{\ast }T^{\ast }%
\mathcal{Q}},\Omega _{T^{\ast }T^{\ast }\mathcal{Q}}^{\flat })  \label{sss1}
\end{equation}%
generated by the Hamiltonian function $-H_{T^{\ast }T^{\ast }\mathcal{Q}}(%
\mathbf{z},\mathbf{\Pi }_{id})$.
\end{proposition}

Since $H_{T^{\ast }T^{\ast }\mathcal{Q}}$ is an invariant of the lift $%
X_{h}^{c\ast }$ we have $i_{X_{h}^{c\ast }}(\Theta _{1})=0$. It follows that 
$\Theta _{1}$ and the Tulczyjew's two-form are invariants of any cotangent
lift. In other words, the cotangent lifts of diffeomorphisms of $T^{\ast }%
\mathcal{Q}$ are symmetries of the special symplectic structure (\ref{sss1}).

We shall now apply Proposition (\ref{liftlag}) to obtain a variational
formulation of the first order equations associated to the cotangent lift in
Eq.(\ref{cotlift}). Since the equations under consideration have the
additional property of being Lagrangian submanifolds described in above
Proposition, we shall, instead, follow an approach based on this property.
Tulczyjew proposed a geometric construction for a generalized Legendre
transformation which also works for degenerate Lagrangians \cite{tul99}. We
now adapt this construction to find an inverse Legendre transformation for
the Hamiltonian system in Eq.(\ref{cotlift}) for which the Hamiltonian
function $H_{T^{\ast }T^{\ast }\mathcal{Q}}$ is degenerate in momenta $%
\mathbf{\Pi }_{id}$. The construction consists of finding an alternative
representation of the Lagrangian submanifold $Im\left( X_{h}^{c\ast }\right) 
$ of the special symplectic structure (\ref{sss1}) with respect to the
special symplectic structure%
\begin{equation}
(TT^{\ast }T^{\ast }\mathcal{Q},T\pi _{T^{\ast }\mathcal{Q}},TT^{\ast }%
\mathcal{Q},\Theta _{2}=\alpha _{T^{\ast }\mathcal{Q}}^{\ast }(\theta
_{T^{\ast }TT^{\ast }\mathcal{Q}}),\alpha _{T^{\ast }\mathcal{Q}})
\label{tuls}
\end{equation}%
underlying the Lagrangian formulation of dynamics. Following \cite{tul99},
we consider a fibration $N\rightarrow TT^{\ast }\mathcal{Q}$ and let $\left(
z^{a},\dot{z}^{b},\pi _{\alpha }\right) ;$ $a,b=1,...,6,$ $\alpha =1,...,m$
be adapted coordinates on $N$. A function $E:N\rightarrow \mathbb{R}$ can be
considered to be a family of functions on the base $TT^{\ast }\mathcal{Q}$
parametrized by the fiber coordinates $\pi _{\alpha }$. This family is
called a Morse family if the rank of the $m\times (m+12)$-matrix%
\begin{equation}
\begin{array}{ccc}
(\frac{\partial ^{2}E}{\partial \pi _{\alpha }\partial \pi _{\beta }} & 
\frac{\partial ^{2}E}{\partial \pi _{\alpha }\partial z^{b}} & \frac{%
\partial ^{2}E}{\partial \pi _{\alpha }\partial \dot{z}^{b}})%
\end{array}
\label{emorse}
\end{equation}%
is maximal \cite{tul74}, \cite{tul77}, \cite{tul77r}, \cite{wei77}, \cite%
{cha95}, \cite{ben}, \cite{LiMa}.

\begin{proposition}
The image of complete cotangent lift $X_{h}^{c\ast }$ of Hamiltonian vector
field $X_{h}$ is a Lagrangian submanifold of the special symplectic
structure in the expression (\ref{tuls}) generated by the Morse family%
\begin{equation}
E\left( \mathbf{z},\mathbf{\Pi }_{id},\mathbf{\dot{z}}\right) =\left( 
\mathbf{\dot{z}}-\mathbf{X}_{h}\left( \mathbf{z}\right) \right) \cdot 
\mathbf{\Pi }_{id}  \label{morse}
\end{equation}%
defined on the Whitney product $T^{\ast }T^{\ast }\mathcal{Q}\times
_{T^{\ast }\mathcal{Q}}TT^{\ast }\mathcal{Q}$.

\begin{proof}
Let $m=6$ and choose the total space $N$ to be the Whitney product 
\begin{eqnarray}
T^{\ast }T^{\ast }\mathcal{Q}\times _{T^{\ast }\mathcal{Q}}TT^{\ast }%
\mathcal{Q} &=&\{\left( \mathbf{\Pi }_{id},X\right) \in T^{\ast }T^{\ast }%
\mathcal{Q}\times TT^{\ast }\mathcal{Q}  \label{whitney} \\
\text{ \ \ } &:&\pi _{T^{\ast }\mathcal{Q}}\left( \mathbf{\Pi }_{id}\right)
=\tau _{T^{\ast }\mathcal{Q}}\left( X\right) \}  \notag
\end{eqnarray}%
for which the local coordinates are $\left( \mathbf{z},\mathbf{\Pi }_{id},%
\mathbf{\dot{z}}\right) $. The Whitney product is a submanifold of $TT^{\ast
}T^{\ast }\mathcal{Q}$ which can locally be described by the equations $%
\mathbf{\dot{\Pi}}_{id}=0$. Consider the function $E\left( \mathbf{z},%
\mathbf{\Pi }_{id},\mathbf{\dot{z}}\right) $ on $TT^{\ast }T^{\ast }\mathcal{%
Q}$ defined by Eq.(\ref{morse}). The matrix in Eq.(\ref{emorse}) becomes $(0$
$-\partial X_{h}^{a}/\partial z^{b}$ $\delta _{b}^{a})$. This has rank $6$,
and so, $E$ is a Morse family. The condition%
\begin{equation}
\alpha _{T^{\ast }\mathcal{Q}}^{\ast }(\theta _{T^{\ast }TT^{\ast }\mathcal{Q%
}})(\mathbf{z},\mathbf{\Pi }_{id},\mathbf{\dot{z}},\mathbf{\dot{\Pi}}%
_{id})=dE\left( \mathbf{z},\mathbf{\Pi }_{id},\mathbf{\dot{z}}\right)
\end{equation}%
for $E$ to generate a Lagrangian submanifold of the special symplectic
structure (\ref{tuls}) gives the components of the complete cotangent lift $%
X_{h}^{c\ast }$.
\end{proof}
\end{proposition}

In the next subsection, we shall carry the dynamics on Whitney product to
the Tulczyjew symplectic space $TT^{\ast }T^{\ast }\mathcal{Q}$ by means of
holonomic lift.

\subsection{Momentum-Vlasov equations}

Define the holonomic lift operator 
\begin{equation}
\Gamma :T^{\ast }T^{\ast }\mathcal{Q}\times _{T^{\ast }\mathcal{Q}}TT^{\ast }%
\mathcal{Q}\rightarrow TT^{\ast }T^{\ast }\mathcal{Q}
\end{equation}
from the Whitney product in Eq.(\ref{whitney}) to the Tulczyjew symplectic
space by $\Gamma \left( \theta _{T^{\ast }\mathcal{Q}},X_{T^{\ast }\mathcal{Q%
}}\right) =T\theta _{T^{\ast }\mathcal{Q}}\left( X_{T^{\ast }\mathcal{Q}%
}\right) $. In coordinates, if $X_{T^{\ast }\mathcal{Q}}(\mathbf{z}%
)=X^{a}\left( \mathbf{z}\right) \partial /\partial z^{a}$ and $\theta
_{T^{\ast }\mathcal{Q}}\left( \mathbf{z}\right) =\pi _{a}(\mathbf{z})dz^{a}$
then, 
\begin{equation}
T\theta _{T^{\ast }\mathcal{Q}}\left( X_{T^{\ast }\mathcal{Q}}\right) (%
\mathbf{z},\mathbf{\pi })=X_{T^{\ast }\mathcal{Q}}^{hol}(\mathbf{z},\mathbf{%
\pi })=X^{a}(\mathbf{z})\left( \frac{\partial }{\partial z^{a}}+\frac{%
\partial \pi _{b}(\mathbf{z})}{\partial z^{a}}\frac{\partial }{\partial \pi
_{b}}\right)
\end{equation}%
which is a generalized vector field of order one \cite{KoSc}, \cite{sa}.
More generally, define the holonomic part $HX_{T^{\ast }T^{\ast }\mathcal{Q}%
} $ of a projectable vector field 
\begin{equation}
X_{T^{\ast }T^{\ast }\mathcal{Q}}(\mathbf{z},\mathbf{\pi })=X^{a}(\mathbf{z})%
\frac{\partial }{\partial z^{a}}+X_{b}\left( \mathbf{z},\mathbf{\pi }\right) 
\frac{\partial }{\partial \pi _{b}}\in T_{(\mathbf{z},\mathbf{\pi })}T_{%
\mathbf{z}}^{\ast }T^{\ast }\mathcal{Q}
\end{equation}%
on $T^{\ast }T^{\ast }\mathcal{Q}$ as the holonomic lift of its projection,
that is, 
\begin{equation}
HX_{T^{\ast }T^{\ast }\mathcal{Q}}=\Gamma \circ T\pi _{T^{\ast }\mathcal{Q}%
}\circ X_{T^{\ast }T^{\ast }\mathcal{Q}}=\left( \left( \pi _{T^{\ast }%
\mathcal{Q}}\right) _{\ast }X_{T^{\ast }T^{\ast }\mathcal{Q}}\right) ^{hol}.
\end{equation}%
The vertical representative is the complement in $TT^{\ast }T^{\ast }%
\mathcal{Q}$ of the holonomic part 
\begin{equation}
VX_{T^{\ast }T^{\ast }\mathcal{Q}}=X_{T^{\ast }T^{\ast }\mathcal{Q}%
}-HX_{T^{\ast }T^{\ast }\mathcal{Q}}=\left( X_{b}\left( \mathbf{z},\mathbf{%
\pi }\right) -X^{a}\left( \mathbf{z}\right) \frac{\partial \pi _{b}\left( 
\mathbf{z}\right) }{\partial z^{a}}\right) \frac{\partial }{\partial \pi _{b}%
}
\end{equation}%
and is a vertical valued generalized vector field of order one as well \cite%
{KoSc}, \cite{sa}, \cite{Tul80}.

The (first) prolongation of a (projectable) generalized vector field of
order one 
\begin{equation}
X_{T^{\ast }T^{\ast }\mathcal{Q}}^{g}\left( \mathbf{z},\mathbf{\pi ,\pi }%
_{z}\right) =X^{a}(\mathbf{z})\frac{\partial }{\partial z^{a}}+X_{b}\left( 
\mathbf{z},\mathbf{\pi ,\pi }_{z}\right) \frac{\partial }{\partial \pi _{b}}
\label{genvec}
\end{equation}%
is defined by 
\begin{equation}
pr^{1}X_{T^{\ast }T^{\ast }\mathcal{Q}}^{g}=X_{T^{\ast }T^{\ast }\mathcal{Q}%
}^{g}+\Phi _{ab}\frac{\partial }{\partial \left( \partial \pi _{b}/\partial
z^{a}\right) }
\end{equation}%
where the coefficient functions are 
\begin{equation}
\Phi _{ab}=\left( \frac{d}{dz^{a}}\left( X_{b}-X^{d}\frac{\partial \pi _{b}}{%
\partial z^{d}}\right) +X^{d}\frac{\partial \pi _{b}}{\partial z^{a}\partial
z^{d}}\right) .
\end{equation}%
Here, the set $\left( \mathbf{z},\mathbf{\pi ,\pi }_{z}\right) $ is the
induced local coordinate system for the jet bundle of the fibration $T^{\ast
}T^{\ast }\mathcal{Q}\rightarrow T^{\ast }\mathcal{Q}$, and $d/dz^{a}$ is\
the total derivative operator with respect to $z^{a}$. Lie bracket of two
first order generalized vector fields $X_{T^{\ast }T^{\ast }\mathcal{Q}}^{g}$
and $Y_{T^{\ast }T^{\ast }\mathcal{Q}}^{g}$ is the unique first order
generalized vector field%
\begin{eqnarray}
\left[ X_{T^{\ast }T^{\ast }\mathcal{Q}}^{g},Y_{T^{\ast }T^{\ast }\mathcal{Q}%
}^{g}\right] _{pro} &=&\left( pr^{1}X_{T^{\ast }T^{\ast }\mathcal{Q}%
}^{g}\left( Y^{a}\right) -pr^{1}Y_{T^{\ast }T^{\ast }\mathcal{Q}}^{g}\left(
X^{a}\right) \right) \frac{\partial }{\partial z^{a}}  \notag \\
&&+\left( pr^{1}X_{T^{\ast }T^{\ast }\mathcal{Q}}^{g}\left( Y_{b}\right)
-pr^{1}Y_{T^{\ast }T^{\ast }\mathcal{Q}}^{g}\left( X_{b}\right) \right) 
\frac{\partial }{\partial \pi _{b}}.  \label{probracet}
\end{eqnarray}%
If $X_{T^{\ast }T^{\ast }\mathcal{Q}}^{g}$ and $Y_{T^{\ast }T^{\ast }%
\mathcal{Q}}^{g}$ are ordinary vector fields on $T^{\ast }T^{\ast }\mathcal{Q%
}$, then $\left[ \text{ },\text{ }\right] _{pro}$ reduces to the Jacobi-Lie
bracket of vector fields \cite{olv86}.

If $X_{T^{\ast }T^{\ast }\mathcal{Q}}$ and $Y_{T^{\ast }T^{\ast }\mathcal{Q}%
} $ are two projectable vector fields on $T^{\ast }T^{\ast }\mathcal{Q}$, a
straightforward calculation gives%
\begin{equation}
\left[ HX_{T^{\ast }T^{\ast }\mathcal{Q}},HY_{T^{\ast }T^{\ast }\mathcal{Q}}%
\right] _{pro}=H\left[ X_{T^{\ast }T^{\ast }\mathcal{Q}},Y_{T^{\ast }T^{\ast
}\mathcal{Q}}\right]  \label{twohorlie}
\end{equation}%
where $\left[ \text{ },\text{ }\right] _{pro}$ is the bracket in Eq.(\ref%
{probracet}). That means, holonomic lift defines an isomorphism between
subspace $HT^{\ast }T^{\ast }\mathcal{Q}=Im\left( \Gamma \right) $ of the
direct sum decomposition $TT^{\ast }T^{\ast }\mathcal{Q}=VT^{\ast }T^{\ast }%
\mathcal{Q}\oplus HT^{\ast }T^{\ast }\mathcal{Q}$ and the space of
projectable vector fields on $T^{\ast }T^{\ast }\mathcal{Q}$ \cite{pog89}, 
\cite{kob91}, \cite{KoSc}. However, for vertical representatives there
appears, in addition, a vector valued two-form.

\begin{proposition}
For vertical representatives, the bracket $\left[ \text{ },\text{ }\right]
_{pro}$ gives 
\begin{equation}
\left[ VX_{T^{\ast }T^{\ast }\mathcal{Q}},VY_{T^{\ast }T^{\ast }\mathcal{Q}}%
\right] _{pro}=V\left[ X_{T^{\ast }T^{\ast }\mathcal{Q}},Y_{T^{\ast }T^{\ast
}\mathcal{Q}}\right] _{pro}+\mathfrak{B}\left( X_{T^{\ast }T^{\ast }\mathcal{%
Q}},Y_{T^{\ast }T^{\ast }\mathcal{Q}}\right) ,
\end{equation}%
where $\mathfrak{B}$ is a vertical-vector valued two-form%
\begin{equation}
\mathfrak{B}\left( X_{T^{\ast }T^{\ast }\mathcal{Q}},Y_{T^{\ast }T^{\ast }%
\mathcal{Q}}\right) =\left[ HY_{T^{\ast }T^{\ast }\mathcal{Q}},VX_{T^{\ast
}T^{\ast }\mathcal{Q}}\right] _{pro}-\left[ HX_{T^{\ast }T^{\ast }\mathcal{Q}%
},VY_{T^{\ast }T^{\ast }\mathcal{Q}}\right] _{pro}.  \label{Bis}
\end{equation}
\end{proposition}

If, on the other hand, $X_{T^{\ast }T^{\ast }\mathcal{Q}}$ and $Y_{T^{\ast
}T^{\ast }\mathcal{Q}}$ are restricted to be complete cotangent lifts of
vector fields on $T^{\ast }\mathcal{Q}$, then the two-form $\mathfrak{B}$ in
Eq.(\ref{Bis}) vanishes \cite{og11}, and we obtain

\begin{proposition}
\label{proviso}Let $X_{T^{\ast }\mathcal{Q}},Y_{T^{\ast }\mathcal{Q}}\in 
\mathfrak{X}\left( T^{\ast }\mathcal{Q}\right) $ and denote by $X_{T^{\ast }%
\mathcal{Q}}^{c\ast },Y_{T^{\ast }\mathcal{Q}}^{c\ast }$ their complete
cotangent lifts and, by $VX_{T^{\ast }\mathcal{Q}}^{c\ast },VY_{T^{\ast }%
\mathcal{Q}}^{c\ast }$ the vertical representatives of the latter. Following
Lie algebra isomorphism hold 
\begin{equation}
V\left[ X_{T^{\ast }\mathcal{Q}},Y_{T^{\ast }\mathcal{Q}}\right] ^{c\ast }=%
\left[ VX_{T^{\ast }\mathcal{Q}}^{c\ast },VY_{T^{\ast }\mathcal{Q}}^{c\ast }%
\right] _{pro}  \label{viso}
\end{equation}%
where the bracket $\left[ ,\right] _{pro}$ is defined in Eq.(\ref{probracet}%
).
\end{proposition}

\begin{remark}
Eq.(\ref{viso}) extends the isomorphism between vector fields on $T^{\ast }%
\mathcal{Q}$ and their complete cotangent lifts to an isomorphism%
\begin{equation*}
\mathfrak{X}\left( T^{\ast }\mathcal{Q}\right) \longleftrightarrow \mathfrak{%
X}^{c\ast }\left( T^{\ast }\mathcal{Q}\right) \subset \mathfrak{X}\left(
T^{\ast }T^{\ast }\mathcal{Q}\right) \longleftrightarrow V\mathfrak{X}%
^{c\ast }\left( T^{\ast }\mathcal{Q}\right) \subset VT^{\ast }T^{\ast }%
\mathcal{Q}
\end{equation*}%
between complete cotangent lifts and their vertical representatives. Here, $V%
\mathfrak{X}^{c\ast }\left( T^{\ast }\mathcal{Q}\right) $ is the space of
vertical representatives of cotangent lifts and%
\begin{equation*}
VT^{\ast }T^{\ast }\mathcal{Q=}\text{ker}\left\{ T\pi _{T^{\ast }\mathcal{Q}%
}:TT^{\ast }T^{\ast }\mathcal{Q\longrightarrow }TT^{\ast }\mathcal{Q}\right\}
\end{equation*}%
is the space of all vertical vectors on $T^{\ast }T^{\ast }\mathcal{Q}$.
\end{remark}

Following diagram summarizes the preceeding geometric constructions in which
we are intended to obtain the momentum-Vlasov equations

\begin{eqnarray*}
&&%
\begin{array}{ccc}
T^{\ast }T^{\ast }T^{\ast }\mathcal{Q} & 
\begin{array}{c}
\begin{array}{c}
\underleftarrow{\text{ \ \ \ \ \ \ \ }dH_{T^{\ast }T^{\ast }\mathcal{Q}}%
\text{ \ \ \ \ \ }}\text{ \ \ \ } \\ 
\overleftarrow{\text{ \ \ \ \ \ \ \ \ \ \ }T^{\ast }\pi _{T^{\ast }\mathcal{Q%
}}\text{ \ \ \ \ }}\text{\ \ \ }%
\end{array}
\\ 
\end{array}
& T^{\ast }T^{\ast }\mathcal{Q}=\mathfrak{g}^{\flat }\oplus \mathfrak{g}%
^{\ast }\text{\ \ } \\ 
\text{ \ }\Omega _{T^{\ast }T^{\ast }\mathcal{Q}}^{\sharp }{\huge \downarrow 
}\text{ } & \text{ \ \ \ }%
\begin{array}{c}
\\ 
\func{ver}{\Huge \swarrow \swarrow }X_{h}^{c\ast } \\ 
\end{array}
& 
\begin{array}{c}
\\ 
\begin{array}{c}
{\huge \uparrow }pr_{2}\text{\ } \\ 
\mathstrut%
\end{array}%
\end{array}
\\ 
\begin{array}{c}
\\ 
\begin{array}{c}
TT^{\ast }T^{\ast }\mathcal{Q} \\ 
\text{ \ \ }%
\end{array}%
\end{array}
& 
\begin{array}{c}
\\ 
\begin{array}{c}
\underrightarrow{\text{ \ \ \ }\func{ver}(\mathbf{\dot{\Pi}}_{id})=0\text{\
\ \ \ \ }}\text{\ } \\ 
\text{ \ }\overrightarrow{T\pi _{T^{\ast }\mathcal{Q}}\text{ }\times
_{T^{\ast }\mathcal{Q}}\tau _{T^{\ast }T^{\ast }\mathcal{Q}}}\text{\ \ \ }
\\ 
\end{array}%
\end{array}
& 
\begin{array}{c}
\text{ }TT^{\ast }\mathcal{Q}\times _{T^{\ast }\mathcal{Q}}T^{\ast }T^{\ast }%
\mathcal{Q} \\ 
\end{array}%
\end{array}
\\
&&\text{ . \ \ \ \ \ \ \ \ \ } \\
&&\text{ \ \ \ \ \ \ \ \ \ \ \ \ \ \ \ \ }%
\begin{array}{c}
{\Huge \searrow }\text{ }V=I-H\text{\ \ \ \ \ \ \ \ \ \ \ \ \ \ \ \ \ \ \ \ }%
H\text{\ \ }{\Huge \swarrow } \\ 
\end{array}%
\text{ \ \ \ } \\
&&\text{ \ \ \ \ \ \ \ \ \ \ \ \ \ \ \ \ \ \ \ \ \ \ \ \ \ \ \ }VT^{\ast
}T^{\ast }\mathcal{Q\oplus }HT^{\ast }T^{\ast }\mathcal{Q}\text{ .}
\end{eqnarray*}

\bigskip From a physical point of view, if components $\left( \Pi _{q}(%
\mathbf{z}),\Pi _{p}(\mathbf{z})\right) $ of $\Pi _{id}\left( \mathbf{z}%
\right) $ are solutions, parametrized by Eulerian coordinates $\mathbf{z}$
of particle motion, of the canonical Hamilton's equations on $T^{\ast
}T^{\ast }\mathcal{Q}$ and $X_{h}^{c\ast }$ is tangent to these curves on
fibers, vertical vector fields satisfy the tangency condition without
reference or restrictions to the particle motion on base manifold $T^{\ast }%
\mathcal{Q}$. The vertical representative of $X_{h}^{c\ast }(\mathbf{z},%
\mathbf{\Pi }_{id})$ is given by%
\begin{eqnarray}
VX_{h}^{c\ast }(\mathbf{z},\mathbf{\Pi }_{id}) &=&\left( e\left( \mathbf{\Pi 
}_{p}(\mathbf{z})\cdot \nabla _{q}\right) \left( \nabla _{q}\phi _{f}(%
\mathbf{q})\right) -X_{h}\left( \mathbf{\Pi }_{q}(\mathbf{z})\right) \right)
\cdot \nabla _{\Pi _{q}}  \notag \\
&&-({\frac{1}{m}}\mathbf{\Pi }_{q}(\mathbf{z})+X_{h}\left( \mathbf{\Pi }_{p}(%
\mathbf{z})\right) )\cdot \nabla _{\Pi _{p}}{,}  \label{verclift}
\end{eqnarray}%
where we denote the action of $X_{h}$ on components of $\mathbf{\Pi }_{id}$
by 
\begin{equation}
X_{h}\left( \mathbf{\Pi }_{p}(\mathbf{z})\right) ={\frac{1}{m}(}\mathbf{p}%
\cdot \nabla _{q})\mathbf{\Pi }_{p}(\mathbf{z})-e(\nabla _{q}\phi _{f}\left( 
\mathbf{q}\right) \cdot \nabla _{p})\mathbf{\Pi }_{p}(\mathbf{z})\text{.}
\end{equation}%
The components of the vector field in Eq.(\ref{verclift}) are precisely the
momentum-Vlasov equations \cite{gpd1} 
\begin{align*}
\mathbf{\dot{\Pi}}_{q}& =-X_{h}(\mathbf{\Pi }_{q})+e\left( \mathbf{\Pi }%
_{p}\cdot \nabla _{q}\right) \left( \nabla _{q}\phi _{f}\right) \\
\mathbf{\dot{\Pi}}_{p}& =-X_{h}(\mathbf{\Pi }_{p})-{\frac{1}{m}}\mathbf{\Pi }%
_{q}
\end{align*}%
given in Eqs.(\ref{mv1}) and (\ref{mv}). Being an element of the dual of Lie
algebra, $\mathbf{\Pi }_{id}$ and its time derivative are components of
one-forms in momentum-Vlasov equations whereas, in Eq.(\ref{verclift}) they
appear as components of vector fields. To make the connection between Eqs.(%
\ref{mv1}), (\ref{mv}) and (\ref{verclift}) precise we need the concept of
vertical lift of one-forms.

Consider the cotangent lift $T^{\ast }\pi _{T^{\ast }\mathcal{Q}}:T^{\ast
}T^{\ast }\mathcal{Q}\rightarrow T^{\ast }T^{\ast }T^{\ast }\mathcal{Q}$ of
the projection $\pi _{T^{\ast }\mathcal{Q}}:T^{\ast }T^{\ast }\mathcal{Q}%
\rightarrow T^{\ast }\mathcal{Q}$ and recall the musical isomorphism $\Omega
_{T^{\ast }T^{\ast }\mathcal{Q}}^{\sharp }:T^{\ast }T^{\ast }T^{\ast }%
\mathcal{Q}\rightarrow TT^{\ast }T^{\ast }\mathcal{Q}$ associated with the
symplectic two-form $\Omega _{T^{\ast }T^{\ast }\mathcal{Q}}$ on the
cotangent bundle $T^{\ast }T^{\ast }\mathcal{Q}.$ For $\pi \in T^{\ast
}T^{\ast }\mathcal{Q}$ define the Euler vector field 
\begin{equation}
\mathcal{X}_{E}:T^{\ast }T^{\ast }\mathcal{Q}\rightarrow TT^{\ast }T^{\ast }%
\mathcal{Q}:\pi \rightarrow \Omega _{T^{\ast }T^{\ast }\mathcal{Q}}^{\sharp
}\circ T^{\ast }\pi _{T^{\ast }\mathcal{Q}}\left( \pi \right) ,  \label{ver}
\end{equation}%
which is a vertical vector field, that is, $Im\left( \mathcal{X}_{E}\right)
\subset ker\left( T\pi _{T^{\ast }\mathcal{Q}}\right) .$ Define the vertical
lift of a one-from $\pi $ on $T^{\ast }\mathcal{Q}$ to be the vertical
vector field 
\begin{equation}
ver\left( \pi \right) =\mathcal{X}_{E}\circ \pi \circ \pi _{T^{\ast }%
\mathcal{Q}}:T^{\ast }T^{\ast }\mathcal{Q}\rightarrow TT^{\ast }T^{\ast }%
\mathcal{Q}
\end{equation}%
on $T^{\ast }T^{\ast }\mathcal{Q}$. In coordinates, the vertical lift of $%
\pi =\mathbf{\pi }_{q}\left( \mathbf{z}\right) \cdot d\mathbf{q}+\mathbf{\pi 
}_{p}\left( \mathbf{z}\right) \cdot d\mathbf{p}$ is $ver\left( \pi \right) =%
\mathbf{\pi }_{q}\left( \mathbf{z}\right) \cdot \nabla _{\Pi _{q}}+\mathbf{%
\pi }_{p}\left( \mathbf{z}\right) \cdot \nabla _{\Pi _{p}}$. Now, if we take 
$\pi =\dot{\Pi}_{id}=\mathbf{\dot{\Pi}}_{q}\cdot d\mathbf{q}+\mathbf{\dot{\Pi%
}}_{p}\cdot d\mathbf{p}$ then the exact relation between Eqs.(\ref{mv1}), (%
\ref{mv}) and (\ref{verclift}) can be expressed by

\begin{proposition}
Let $\dot{\Pi}_{id}=\mathbf{\dot{\Pi}}_{q}\cdot d\mathbf{q}+\mathbf{\dot{\Pi}%
}_{p}\cdot d\mathbf{p}$ be a one-form on $T^{\ast }\mathcal{Q}$ representing
time evolution of components of $\Pi _{id}$. Then, the momentum-Vlasov
equations can be written as%
\begin{equation}
ver\left( \dot{\Pi}_{id}\right) =VX_{h}^{c\ast }\left( \mathbf{z},\mathbf{%
\Pi }_{id}\right)  \label{vereq}
\end{equation}%
where the right hand side refers to Eqs.(\ref{mv1}) and (\ref{mv}).
\end{proposition}

Thus, we show that the momentum-Vlasov equations are generated by the
vertical representative of complete cotangent lift of particle motion on $%
T^{\ast }\mathcal{Q}$. The result in Proposition \ref{proviso} shows that
this process of lifting particle motion preserves algebraic structures.

\subsection{Lie-Poisson Hamiltonian operator}

We shall relate the geometric construction of the previous subsection to
Lie-Poisson structure. The Lie algebra $\mathfrak{g}$ acts on $\mathfrak{g}%
^{\ast }$ by coadjoint action given by the Lie derivative $ad_{X_{k}}^{\ast
}=\mathcal{L}_{X_{k}}$. The vertical lift $ver(-ad_{X_{k}}^{\ast })$ of
generator of coadjoint action is a vector field tangent to the fiber
coordinates $\left( \mathbf{\Pi }_{q},\mathbf{\Pi }_{p}\right) $ of $%
\mathfrak{g}^{\ast }\subset T^{\ast }T^{\ast }\mathcal{Q}$ and coincides
with the right hand side of Eq.(\ref{vereq}), namely, $ver(-ad_{X_{k}}^{\ast
}\left( \Pi _{id}\right) )=VX_{h}^{c\ast }\left( \mathbf{z},\mathbf{\Pi }%
_{id}\right) $. This observation leads us to connect a Lie algebra element,
that is, generator of particle motion, to corresponding generator of
coadjoint action. This connection is provided by an operator associated to
the Lie-Poisson structure on the dual space $\mathfrak{g}^{\ast }$. Recall
that $\mathfrak{g}^{\ast }$ is a Poisson manifold with the Lie-Poisson
bracket in Eq.(\ref{LP}) which may be written as 
\begin{equation*}
\left\{ K,H\right\} _{LP}=\int \frac{\delta K}{\delta \Pi _{id}}\cdot
J_{LP}(\Pi _{id})\frac{\delta H}{\delta \Pi _{id}}d\mu
\end{equation*}%
for the Hamiltonian operator%
\begin{equation}
J_{LP}(\Pi _{id})=-\left( 
\begin{array}{cc}
\Pi _{i}\dfrac{\partial }{\partial q^{j}}+\dfrac{\partial }{\partial q^{i}}%
\cdot \Pi _{j} & \text{ \ \ \ \ }\Pi ^{i}\dfrac{\partial }{\partial q^{j}}+%
\dfrac{\partial }{\partial p_{i}}\cdot \Pi _{j} \\ 
\Pi _{i}\dfrac{\partial }{\partial p_{j}}+\dfrac{\partial }{\partial q^{i}}%
\cdot \Pi ^{j} & \text{ \ \ \ \ }\Pi ^{i}\dfrac{\partial }{\partial p_{j}}+%
\dfrac{\partial }{\partial p_{i}}\cdot \Pi ^{j}%
\end{array}%
\right)
\end{equation}%
and for $\frac{\partial }{\partial q^{j}}\cdot \Pi _{j}=\frac{\partial \Pi
_{j}}{\partial q^{j}}+\Pi _{j}\frac{\partial }{\partial q^{j}}$ etc \cite%
{gpd1}. As the derivative $\delta K/\delta \Pi _{id}$ is in the Lie algebra $%
\mathfrak{g}$, the operator $J_{LP}(\Pi _{id})$ may be considered to be a map

\begin{eqnarray*}
J_{LP}(\Pi _{id}) &:&\mathfrak{g}\longrightarrow (ad_{\mathfrak{g}}^{\ast }:%
\mathfrak{g}^{\ast }\rightarrow \mathfrak{g}^{\ast })=End(\mathfrak{g}^{\ast
}\mathfrak{)} \\
&:&X_{h\text{ }}\longrightarrow J_{LP}(\Pi
_{id})(X_{h})=ver(-ad_{X_{h}}^{\ast }\left( \Pi _{id}\right) )
\end{eqnarray*}%
taking a generator $X_{h}$ in the Lie algebra $\mathfrak{g}$ to the
corresponding generator $ver(-ad_{X_{h}}^{\ast })$ of coadjoint action. Note
that $J_{LP}(\Pi _{id})$ is a map from the tangent space $TT^{\ast }\mathcal{%
Q}$ to $TT^{\ast }T^{\ast }\mathcal{Q}$ both of which are Tulczyjew.
Considering the representation 
\begin{equation}
\dot{\Pi}_{id}=J_{LP}(\Pi _{id})\frac{\delta H}{\delta \Pi _{id}}
\end{equation}%
of momentum-Vlasov equations on $\mathfrak{g}^{\ast }$ with $\delta H/\delta
\Pi _{id}=X_{h}$, we conclude that, the whole geometric process we described
to obtain the momentum-Vlasov equations from the generators of particle
motion is encoded in $J_{LP}(\Pi _{id})$ as \ 
\begin{equation*}
\left( 
\begin{array}{c}
\text{Lie-Poisson} \\ 
\text{Hamiltonian} \\ 
\text{operator at }X_{h\text{ }}%
\end{array}%
\right) =\left( 
\begin{array}{c}
\text{vertical} \\ 
\text{equivalence}%
\end{array}%
\right) \circ \left( 
\begin{array}{c}
\text{complete } \\ 
\text{cotangent} \\ 
\text{lift of }X_{h\text{ }}%
\end{array}%
\right)
\end{equation*}%
and this represents a geometric decomposition of Lie-Poisson Hamiltonian
operators. It also suggests a way to construct Lie-Poisson operator directly
from an arbitrary Lie algebra element \cite{og11}. This is the way we shall
relate the algebra of Hamiltonian vector fields to the algebra of strict
contact vector fields in the next section.

\begin{remark}
The Hamiltonian operator $J_{LP}(\Pi _{id})$ on $\mathfrak{g}^{\ast }$
transforms into the Hamiltonian operator%
\begin{equation}
J_{LP}(f)=\nabla _{p}f\cdot \nabla _{q}-\nabla _{q}f\cdot \nabla _{p}
\label{jlpf}
\end{equation}%
for the Vlasov equations in density variable on the space $Den(T^{\ast }%
\mathcal{Q})$ of densities under the correspondence in Eq.(\ref{density}%
).\bigskip 
\end{remark}

\pagebreak

\bigskip

\section{Moments of Momentum-Vlasov Dynamics}

The space $\mathfrak{T}\mathcal{Q}$ of all symmetric contravariant tensor
fields on a manifold $\mathcal{Q}$ carries a Lie algebra structure called
Schouten concomitant (symmetric Schouten braket) \cite{HoTr09},\cite{Sc40},%
\cite{kms93}. The dual $\mathfrak{T}^{\ast }\mathcal{Q}$ of the algebra $%
\mathfrak{T}\mathcal{Q}$ consists of the symmetric covariant tensor fields
and carries Kuperschmidt-Manin Lie-Poisson structure, \cite{gib81}. In \cite%
{GiHoTr08} and \cite{Tr08}, it was argued that, the kinetic moments of
plasma density function $f$ may be considered as elements of $\mathfrak{T}%
^{\ast }\mathcal{Q}$ and, it is established that, the operation taking the
plasma density $f$ to the moments of the dynamics is a Poisson mapping which
is the dual of a Lie algebra homomorphism from $\mathfrak{T}\mathcal{Q}$ to $%
\mathcal{F}\left( T^{\ast }\mathcal{Q}\right) $.

In this section, we first review the Schouten algebra of symmetric
contravariant tensors. Then, we will define a Lie algebra homomorphism from
the algebra $\mathfrak{T}\mathcal{Q}$ of symmetric contravariant tensor
fields to the algebra $\mathfrak{X}_{ham}\left( T^{\ast }\mathcal{Q}\right) =%
\mathfrak{g}$ of Hamiltonian vector fields as a generalization of the
complete cotangent lift introduced above. We will obtain the moments of
momentum-Vlasov dynamics. Finally, we will consider some subalgebras of $%
\mathfrak{T}\mathcal{Q}$, and derive plasma-to-fluid map in terms of
momentum variables.

\subsection{Schouten concomitant}

The direct product $\mathfrak{T}\mathcal{Q=\oplus }_{n=0}^{\infty }\mathfrak{%
T}^{n}\mathcal{Q}$ of spaces $\mathfrak{T}^{n}\mathcal{Q}$ of symmetric
contravariant tensor fields on a manifold $\mathcal{Q}$ of all orders
constitutes a vector space. In a local coordinate system $\left(
q^{i}\right) $, an element of $\mathfrak{T}\mathcal{Q}$ is in form 
\begin{equation}
\mathbb{X=}\dbigoplus\limits_{n=0}^{\infty }\mathbb{X}^{n}=\dbigoplus%
\limits_{n=0}^{\infty }\mathbb{X}^{i_{1}i_{2}...i_{n}}\left( \mathbf{q}%
\right) \partial _{q^{i_{1}}}\otimes ...\otimes \partial _{q^{i_{n}}},
\label{Xbb}
\end{equation}%
where $\mathbb{X}^{n}\in \mathfrak{T}^{n}\mathcal{Q}$ is a symmetric
contravariant tensor field of order $n$ and $X^{i_{1}i_{2}...i_{n}}\left( 
\mathbf{q}\right) $ are the real valued coefficient functions. The dual $%
\mathfrak{T}^{\ast }\mathcal{Q}$ of $\mathfrak{T}\mathcal{Q}$ is the direct
sum $\mathcal{\oplus }_{n=0}^{\infty }\mathfrak{T}_{n}\mathcal{Q}$ of
symmetric covariant tensor fields $\mathfrak{T}_{n}\mathcal{Q}$ of all
orders. In coordinates $\left( q^{i}\right) $, an element of $\mathfrak{T}%
^{\ast }\mathcal{Q}$ is given by%
\begin{equation*}
\mathbb{A=}\dbigoplus\limits_{n=0}^{\infty }\mathbb{A}_{n}=\dbigoplus%
\limits_{n=0}^{\infty }\mathbb{A}_{i_{1}i_{2}...i_{n}}\left( \mathbf{q}%
\right) dq^{i_{1}}\otimes ...\otimes dq^{i_{n}},
\end{equation*}%
where $\mathbb{A}_{n}\in \mathfrak{T}_{n}\mathcal{Q}$ is a symmetric
covariant tensor field of order $n.$ The pairing between $\mathfrak{T}^{\ast
}\mathcal{Q}$ and $\mathfrak{T}\mathcal{Q}$ is 
\begin{equation}
\left\langle \mathbb{A},\mathbb{X}\right\rangle =\dsum\limits_{n=0}^{\infty
}\left\langle \mathbb{A}_{n},\mathbb{X}_{n}\right\rangle
=\dsum\limits_{n=0}^{\infty }\int \mathbb{A}_{i_{1}i_{2}...i_{n}}\left( 
\mathbf{q}\right) \mathbb{X}^{i_{1}i_{2}...i_{n}}\left( \mathbf{q}\right) d%
\mathbf{q.}  \label{pair}
\end{equation}

If $\mathbb{X}^{n},$ $\mathbb{Y}^{m}$ and $\mathbb{Z}^{n+m-1}$ are
contravariant tensor fields of orders $n$, $m$ and $n+m-1$, respectively,
the Schouten concomitant 
\begin{equation}
\left[ \mathbb{X},\mathbb{Y}\right] _{SC}=\dbigoplus\limits_{n=0}^{\infty
}\dbigoplus\limits_{m=0}^{\infty }\left[ \mathbb{X}^{n},\mathbb{Y}^{m}\right]
_{SC}=\dbigoplus\limits_{n=0}^{\infty }\dbigoplus\limits_{m=0}^{\infty }%
\mathbb{Z}^{n+m-1}  \label{Sc}
\end{equation}%
defines a Lie algebra structure on the space $\mathfrak{T}\mathcal{Q}$ \cite%
{Sc40}, \cite{kms93}, \cite{Tr08}. The coefficient functions of $\mathbb{Z}%
^{n+m-1}$ in terms of those of $\mathbb{X}^{n}$ and $\mathbb{Y}^{m}$ are 
\begin{equation*}
\mathbb{Z}^{i_{1}...i_{n+m-1}}=n\mathbb{X}^{i_{m+1}...i_{m+n-1}l}\dfrac{%
\partial \mathbb{Y}^{i_{1}...i_{m}}}{\partial q^{l}}-m\mathbb{Y}%
^{i_{n+1}...i_{n+m-1}l}\dfrac{\partial \mathbb{X}^{i_{1}i_{2}...i_{n}}}{%
\partial q^{l}}.
\end{equation*}

\subsection{Generalized complete cotangent lift}

Let $\mathbb{X}^{n}$ be a (not necessarily symmetric) contravariant tensor
field of order $n$. Due to the canonical inclusion $\mathfrak{T}^{n}\mathcal{%
Q}\hookrightarrow \mathfrak{T}^{n}(T^{\ast }\mathcal{Q)}$, we may assume $%
\mathbb{X}^{n}$ as a tensor on the cotangent bundle $T^{\ast }\mathcal{Q}$.
We define a mapping%
\begin{equation}
\mathfrak{T}^{n}\mathcal{Q}\rightarrow \mathcal{F}\left( T^{\ast }\mathcal{Q}%
\right) :\mathbb{X}^{n}\rightarrow H_{\mathbb{X}^{n}}=\theta _{T^{\ast }%
\mathcal{Q}}^{n}\left( \mathbb{X}^{n}\right)  \label{cf}
\end{equation}%
from $\mathfrak{T}^{n}\mathcal{Q}$ to the space $\mathcal{F}\left( T^{\ast }%
\mathcal{Q}\right) $ of smooth functions on $T^{\ast }\mathcal{Q}$, by
contracting a contravariant tensor $\mathbb{X}_{n}\in \mathfrak{T}^{n}%
\mathcal{Q}$ with $n$-th tensor power $\theta _{T^{\ast }\mathcal{Q}%
}^{n}=\theta _{T^{\ast }\mathcal{Q}}\otimes ...\otimes \theta _{T^{\ast }%
\mathcal{Q}}$ of the canonical one-form $\theta _{T^{\ast }\mathcal{Q}}$.
Then, define the generalized complete cotangent lift 
\begin{equation}
^{c\ast }:\mathfrak{T}^{n}\mathcal{Q}\rightarrow \mathfrak{X}_{ham}\left(
T^{\ast }\mathcal{Q}\right) :\mathbb{X}^{n}\rightarrow \left( \mathbb{X}%
^{n}\right) ^{c\ast }=X_{H_{\mathbb{X}^{n}}}  \label{Xnc}
\end{equation}%
as an operation taking a contravariant tensor field $\mathbb{X}^{n}$ on $%
\mathcal{Q}$ to a Hamiltonian vector field $\left( \mathbb{X}^{n}\right)
^{c\ast }$ corresponding to the Hamiltonian function $H_{\mathbb{X}^{n}}=i_{%
\mathbb{X}^{n}}\theta _{T^{\ast }\mathcal{Q}}^{n},$ \cite{No93}. In
Darboux's coordinates $\left( q,p\right) $, the Hamiltonian function $H_{%
\mathbb{X}^{n}}$ is a polynomial in the fiber variables of $T^{\ast }%
\mathcal{Q}$%
\begin{equation}
H_{\mathbb{X}^{n}}\left( \mathbf{q},\mathbf{p}\right)
=p_{i_{1}}p_{i_{2}}...p_{i_{n}}\mathbb{X}^{i_{1}i_{2}...i_{n}}\left( \mathbf{%
q}\right)  \label{HXn}
\end{equation}%
and the complete cotengent lift is 
\begin{equation*}
\left( \mathbb{X}^{n}\right) ^{c\ast }=np_{i_{1}}p_{i_{2}}...p_{i_{n-1}}%
\mathbb{X}^{i_{1}...i_{n-1}l}\partial _{q^{l}}-p_{i_{1}}p_{i_{2}}...p_{i_{n}}%
\frac{\partial \mathbb{X}^{i_{1}i_{2}...i_{n}}}{\partial q^{l}}\partial
_{p_{l}}.
\end{equation*}

We further enhance the operations given in Eqs.(\ref{cf}) and (\ref{Xnc}) to
the product space $\mathfrak{T}\mathcal{Q}$ as follows. For $\mathbb{X=}%
\oplus \mathbb{X}^{n}\in \mathfrak{T}\mathcal{Q}$ define the function $H_{%
\mathbb{X}}$ on $T^{\ast }\mathcal{Q}$ as the sum%
\begin{equation}
\mathfrak{T}\mathcal{Q}\rightarrow \mathcal{F}\left( T^{\ast }\mathcal{Q}%
\right) :\mathbb{X}\rightarrow H_{\mathbb{X}}=\sum_{n=0}^{\infty }H_{\mathbb{%
X}_{n}},  \label{cf2}
\end{equation}%
\cite{DuMi95}. This infinite sum may be considered as the Taylor expansion
of the function $H_{\mathbb{X}}$ in terms of $p-$polynomials. A straight
forward calculation proves the following proposition.

\begin{proposition}
The map $\mathbb{X}\rightarrow H_{\mathbb{X}}$ is a Lie algebra
anti-homomorphism, that is 
\begin{equation*}
H_{\left[ \mathbb{X},\mathbb{Y}\right] _{SC}}=-\left\{ H_{\mathbb{X}},H_{%
\mathbb{Y}}\right\} _{T^{\ast }\mathcal{Q}},
\end{equation*}%
where the bracket on the left hand side is the Schouten concomitant of
covariant tensors and the bracket on the right hand side is the canonical
Poisson bracket of functions on $T^{\ast }\mathcal{Q}$. In particular, one
has%
\begin{equation*}
H_{\left[ \mathbb{X}^{n},\mathbb{Y}^{m}\right] _{SC}}=-\left\{ H_{\mathbb{X}%
^{n}},H_{\mathbb{Y}^{m}}\right\} _{T^{\ast }\mathcal{Q}},
\end{equation*}%
where $\left[ \mathbb{X}^{n},\mathbb{Y}^{m}\right] _{SC}\in \mathfrak{T}%
^{n+m-1}\mathcal{Q}$.
\end{proposition}

For the generalized complete cotangent lift 
\begin{equation}
^{c\ast }:\mathfrak{T}\mathcal{Q}\rightarrow \mathfrak{X}_{ham}\left(
T^{\ast }\mathcal{Q}\right) :\mathbb{X}=\dbigoplus\limits_{n=0}^{\infty }%
\mathbb{X}^{n}\rightarrow \mathbb{X}^{c\ast }=\dsum\limits_{n=0}^{\infty
}\left( \mathbb{X}^{n}\right) ^{c\ast },  \label{gccl}
\end{equation}%
we use the identity $\left[ X_{F},X_{G}\right] =-X_{\left\{ F,G\right\}
_{T^{\ast }\mathcal{Q}}}$ and the above proposition to have 
\begin{equation*}
\left[ \mathbb{X}^{c\ast },\mathbb{Y}^{c\ast }\right] =\left[ X_{H_{\mathbb{X%
}}},X_{H_{\mathbb{Y}}}\right] =-X_{\left\{ H_{\mathbb{X}},H_{\mathbb{Y}%
}\right\} _{T^{\ast }\mathcal{Q}}}=X_{H_{\left[ \mathbb{X},\mathbb{Y}\right]
_{SC}}}=\left[ \mathbb{X},\mathbb{Y}\right] _{SC}^{c\ast },
\end{equation*}%
which enables us to state the following result.

\begin{proposition}
The generalized complete cotangent lift in Eq.(\ref{gccl}) is a Lie algebra
isomorphism into $\mathbb{X}\rightarrow \mathbb{X}^{c\ast }:\mathfrak{T}%
\mathcal{Q}\rightarrow \mathfrak{g}$ 
\begin{equation}
\left[ \mathbb{X},\mathbb{Y}\right] _{SC}^{c\ast }=\left[ \mathbb{X}^{c\ast
},\mathbb{Y}^{c\ast }\right] _{JL},  \label{iso}
\end{equation}%
where $\left[ \mathbb{\ },\text{ }\right] _{SC}$ is the Schouten concomitant
of tensor fields in Eq.(\ref{Sc}) and $\left[ \mathbb{\ },\text{ }\right]
_{JL}$ is the Jacobi-Lie bracket of vector fields.
\end{proposition}

\subsection{Moments of momentum variables}

The dual map $\Phi :\mathfrak{g}^{\ast }\rightarrow \mathfrak{T}^{\ast }%
\mathcal{Q}$ of the homomorphism $\mathbb{X}\rightarrow \mathbb{X}^{c\ast }$
is a momentum and Poisson mapping. To compute it, take $\Pi _{id}=\mathbf{%
\Pi }_{q}\cdot d\mathbf{q}+\mathbf{\Pi }_{p}\cdot d\mathbf{p}\in \mathfrak{g}%
^{\ast }$, then the dual operation is%
\begin{equation}
\Phi \left( \Pi _{id}\right) =\dbigoplus\limits_{n=0}^{\infty }\int \left(
\theta _{T^{\ast }\mathcal{Q}}^{n-1}\otimes \vartheta \right) d^{3}\mathbf{p,%
}  \label{km}
\end{equation}%
where $\theta _{T^{\ast }\mathcal{Q}}^{n-1}$ is the $\left( n-1\right) $-th
tensor power of the canonical one form $\theta _{T^{\ast }\mathcal{Q}}$ and $%
\vartheta $ is a one-form on $T^{\ast }\mathcal{Q}$ given explicitly by 
\begin{equation*}
\vartheta \left( \mathbf{q},\mathbf{p}\right) =\left( n\mathbf{\Pi }%
_{q}+\left( \nabla _{q}\mathbf{\Pi }_{p}\right) \mathbf{p}\right) \cdot d%
\mathbf{q}
\end{equation*}%
The image of $\Pi _{id}$ under the dual map $\Phi $ consists of moments of
the momentum-Vlasov dynamics. Namely, the $n-$th moment of $\Pi _{id}$ is
given by 
\begin{equation*}
\mathbb{A}_{n}=\int \left( \theta _{T^{\ast }\mathcal{Q}}^{n-1}\otimes
\vartheta \right) d^{3}\mathbf{p.}
\end{equation*}%
In particular, for one dimensional plasma, using the momentum map 
\begin{equation*}
\mathfrak{g}^{\ast }\rightarrow \mathcal{F}\left( T^{\ast }\mathcal{Q}%
\right) :\Pi _{id}\rightarrow f\left( q,p\right)
\end{equation*}
in Eq.(\ref{density}) we have the kinetic moments 
\begin{equation*}
\mathbb{A}_{n}=\int p^{n}f\left( q,p\right) dp
\end{equation*}%
of the Vlasov density \cite{mwrss83}. The following proposition argues that
the moments are Poisson maps \cite{gib81}.

\begin{proposition}
The kinetic moments in Eq.(\ref{km}) is a Poisson map from the Lie-Poisson
bracket on $\mathfrak{g}^{\ast }$ to the Kuperschmidt-Manin bracket on $%
\mathfrak{T}^{\ast }\mathcal{Q}$.
\end{proposition}

To prove this, take a linear functional $\mathfrak{F}_{\mathbb{X}}$ on $%
\mathfrak{T}^{\ast }\mathcal{Q}$ of the form 
\begin{equation*}
\mathfrak{F}_{\mathbb{X}}\left( \mathbb{A}\right) =\left\langle \mathbb{A},%
\mathbb{X}\right\rangle =\dsum\limits_{n=0}^{\infty }\left\langle \mathbb{A}%
_{n},\mathbb{X}_{n}\right\rangle =\dsum\limits_{n=0}^{\infty }\int \mathbb{A}%
_{i_{1}i_{2}...i_{n}}\left( \mathbf{q}\right) \mathbb{X}%
^{i_{1}i_{2}...i_{n}}\left( \mathbf{q}\right) d^{3}\mathbf{q.}
\end{equation*}%
Its variation is $\delta \mathfrak{F}_{\mathbb{X}}/\delta \mathbb{A=X}$. The
pull-back $\Phi ^{\ast }\mathfrak{F}_{\mathbb{X}}$ of $\mathfrak{F}_{\mathbb{%
X}}$ to $\mathfrak{g}^{\ast }$ by the momentum map $\Phi $ in Eq.(\ref{km})
gives 
\begin{equation*}
\left( \Phi ^{\ast }\mathfrak{F}_{\mathbb{X}}\right) \left( \Pi _{id}\right)
=\dsum\limits_{n=0}^{\infty }\iint np_{i_{1}}p_{i_{2}}...p_{i_{n-1}}\left(
\Pi _{i_{n}}+p_{i_{n}}\frac{\partial \Pi ^{l}}{\partial q^{l}}\right) 
\mathbb{X}^{i_{1}i_{2}...i_{n}}\left( \mathbf{q}\right) d^{3}\mathbf{q}d^{3}%
\mathbf{p.}
\end{equation*}%
and the variation of this with respect to its argument $\Pi _{id}$ is 
\begin{equation*}
\frac{\delta \left( \Phi ^{\ast }\mathfrak{F}_{\mathbb{X}}\right) }{\delta
\Pi _{id}}=\left( \frac{\delta \left( \Phi ^{\ast }\mathfrak{F}_{\mathbb{X}%
}\right) }{\delta \Pi _{q}},\frac{\delta \left( \Phi ^{\ast }\mathfrak{F}_{%
\mathbb{X}}\right) }{\delta \Pi _{p}}\right) =X_{H_{\mathbb{X}}}=\mathbb{X}%
^{c\ast },
\end{equation*}%
where $X_{H_{\mathbb{X}}}$ is the Hamiltonian vector field corresponding to
the Hamiltonian function $H_{\mathbb{X}}$ in Eq.(\ref{cf2}). The Lie-Poisson
bracket on $\mathfrak{g}^{\ast }$ is 
\begin{equation}
\left\{ \Phi ^{\ast }\mathfrak{F}_{\mathbb{X}},\Phi ^{\ast }\mathfrak{F}_{%
\mathbb{Y}}\right\} _{\mathfrak{g}^{\ast }}\left( \Pi _{id}\right) =\iint 
\mathbf{\Pi }_{id}\cdot \mathbf{[}\frac{\delta \Phi ^{\ast }\mathfrak{F}_{%
\mathbb{X}}}{\delta \mathbf{\Pi }_{id}},\frac{\delta \Phi ^{\ast }\mathfrak{F%
}_{\mathbb{Y}}}{\delta \mathbf{\Pi }_{id}}\mathbf{]}d^{3}\mathbf{p}d^{3}%
\mathbf{q,}  \label{mV}
\end{equation}%
where the bracket inside the integral is minus the Jacobi-Lie bracket of
vector fields satisfying%
\begin{equation*}
\left[ \frac{\delta \left( \Phi ^{\ast }\mathfrak{F}_{\mathbb{X}}\right) }{%
\delta \Pi _{id}},\frac{\delta \left( \Phi ^{\ast }\mathfrak{F}_{\mathbb{Y}%
}\right) }{\delta \Pi _{id}}\right] =X_{\left\{ H_{\mathbb{X}},H_{\mathbb{Y}%
}\right\} }.
\end{equation*}%
On $\mathfrak{T}^{\ast }\mathcal{Q}$, the Kuperscmidt-Manin bracket is given
by%
\begin{equation}
\left\{ \mathfrak{F}_{\mathbb{X}},\mathfrak{F}_{\mathbb{Y}}\right\}
_{KM}=\int \left\langle \mathbb{A},\mathbf{[}\frac{\delta \mathfrak{F}_{%
\mathbb{X}}}{\delta \mathbb{A}},\frac{\delta \mathfrak{F}_{\mathbb{Y}}}{%
\delta \mathbb{A}}\mathbf{]}_{SC}\right\rangle d^{3}\mathbf{q}  \label{Km}
\end{equation}%
where the bracket inside the integral is the Schouten concomitant and the
pairing inside the integral is the one in Eq.(\ref{pair}) \cite{gib81}. The
fact that $\Phi ^{\ast }$ is a Poisson map

\begin{equation*}
\Phi ^{\ast }\left\{ \mathfrak{F}_{\mathbb{X}},\mathfrak{F}_{\mathbb{Y}%
}\right\} _{KM}=\left\{ \Phi ^{\ast }\mathfrak{F}_{\mathbb{X}},\Phi ^{\ast }%
\mathfrak{F}_{\mathbb{Y}}\right\} _{\mathfrak{g}^{\ast }}
\end{equation*}%
follows from direct substitutions.

\subsection{Plasma-to-fluid map in momentum variables}

The Lie algebra structure on $\mathfrak{T}\mathcal{Q}$ defined by the
Schouten concomitant has only three subalgebras; the space of smooth
functions $\mathfrak{T}^{0}\mathcal{Q=F}\left( \mathcal{Q}\right) ,$ the
space of vector fields $\mathfrak{T}^{1}\mathcal{Q}=\mathfrak{X}\left( 
\mathcal{Q}\right) $ and $\mathfrak{X}\left( \mathcal{Q}\right) \times 
\mathcal{F}\left( \mathcal{Q}\right) $. For the subalgebra $\mathcal{F}%
\left( \mathcal{Q}\right) ,$ Schouten concomitant reduces to the trivial
Poisson bracket of functions on $\mathcal{Q}$. The Lie algebra homomorphism
in Eq.(\ref{iso}) takes the particular form $\phi \rightarrow X_{\phi }$,
where $X_{\phi }=-\nabla _{q}\phi \cdot \nabla _{p}$ is the infinitesimal
generator of the action $\left( \mathbf{q},\mathbf{p}\right) \rightarrow
\left( \mathbf{q},\mathbf{p}-\nabla _{q}\phi \right) $ of additive group of
functions $\mathcal{F}(\mathcal{Q})$ on $T^{\ast }\mathcal{Q}$ by momentum
translations. For the subalgebra $\mathfrak{X}\left( \mathcal{Q}\right) $,
the concomitant reduces to the Jacobi-Lie bracket of vector fields and the
homomorphism in Eq.(\ref{iso}) reduces to the identity $\left[ X,Y\right]
_{JL}^{c\ast }=\left[ X^{c\ast },Y^{c\ast }\right] _{JL},$ where $X^{c\ast }=%
\mathbf{X}\cdot \nabla _{q}-\nabla _{q}\left( \mathbf{p}\cdot \nabla
_{q}\phi \right) \cdot \nabla _{p}$ is the complete cotangent lift of the
vector field $X=\mathbf{X}\cdot \nabla _{q}$. The lift $X^{c\ast }$ is an
infinitesimal generator of the right action of diffeomorphism group on $%
T^{\ast }\mathcal{Q}$.

On the third subalgebra $\mathfrak{X}\left( \mathcal{Q}\right) \times 
\mathcal{F}\left( \mathcal{Q}\right) $, the Schouten concomitant, for $%
\mathbb{X}=\left( X,\phi \right) $ and $\mathbb{Y}=\left( Y,\zeta \right)
\in \mathfrak{X}\left( \mathcal{Q}\right) \times \mathcal{F}\left( \mathcal{Q%
}\right) $ gives 
\begin{equation}
\left[ \mathbb{X},\mathbb{Y}\right] _{SC}=\left( \left[ X,Y\right] ,X\left(
\zeta \right) -Y\left( \phi \right) \right) ,
\end{equation}%
which turns $\mathfrak{X}\left( \mathcal{Q}\right) \times \mathcal{F}\left( 
\mathcal{Q}\right) $ into a semi-direct product algebra $\mathfrak{s}=%
\mathfrak{X}\left( \mathcal{Q}\right) \circledS \mathcal{F}\left( \mathcal{Q}%
\right) $ with the first factor acting on the second by Lie derivative.
Namely, $X\left( \zeta \right) $ is the directional derivative of $\zeta $
in the direction of $X$ and $\left[ X,Y\right] $ is the Jacobi-Lie bracket
of vector fields $X$ and $Y$. This semi-direct product algebra is the Lie
algebra of the group $S=Diff\left( \mathcal{Q}\right) \circledS \mathcal{F}%
\left( \mathcal{Q}\right) $ which is the configuration space of compressible
fluid. The dual space $\mathfrak{s}^{\ast }=\Lambda ^{1}\left( \mathcal{Q}%
\right) \times \mathcal{F}\left( \mathcal{Q}\right) $ of the Lie algebra $%
\mathfrak{s}$ is the product of one-forms and functions (identified with
three-forms) on $\mathcal{Q}$. The Lie-Poisson structure on $\mathfrak{s}%
^{\ast }$ is defined by 
\begin{eqnarray}
\left\{ F,G\right\} _{\mathfrak{s}^{\ast }}\left( \rho ,M\right) &=&\int 
\mathbf{M\cdot \lbrack }\frac{\delta F}{\delta \mathbf{M}},\frac{\delta G}{%
\delta \mathbf{M}}\mathbf{]}d\mathbf{q}  \notag \\
&&+\int \rho \left( \frac{\delta G}{\delta \mathbf{M}}\cdot \left( \nabla
_{q}\frac{\delta F}{\delta \rho }\right) -\frac{\delta F}{\delta \mathbf{M}}%
\cdot \left( \nabla _{q}\frac{\delta G}{\delta \rho }\right) \right) d%
\mathbf{q}  \label{Cf}
\end{eqnarray}%
and is known as the compressible fluid bracket. Here $\left( M=\mathbf{M}%
\cdot d\mathbf{q,}\rho \right) \in \mathfrak{s}^{\ast }$ and it is assumed
that $\delta F/\delta \mathbf{M},\delta G/\delta \mathbf{M}\in \mathfrak{X}%
\left( \mathcal{Q}\right) $.

The complete cotangent lift from the semi-direct product to the Lie algebra
of Hamiltonian vector fields 
\begin{equation*}
\mathfrak{s\rightarrow g}:\mathbb{X}=\left( \phi ,X\right) \rightarrow 
\mathbb{X}^{c\ast }=X^{c\ast }+X_{\phi }
\end{equation*}%
is the sum of $X^{c\ast }$ and $X_{\phi }.$ $\mathbb{X}^{c\ast }$ is the
Hamiltonian vector field 
\begin{equation*}
\mathbb{X}^{c\ast }=X^{i}\frac{\partial }{\partial q^{i}}-\left( p_{i}\frac{%
\partial X^{i}}{\partial q^{j}}+\frac{\partial \phi }{\partial q^{j}}\right) 
\frac{\partial }{\partial p_{j}}
\end{equation*}%
on $T^{\ast }\mathcal{Q}$ with the Hamiltonian function $H_{\mathbb{X}%
}\left( \mathbf{p},\mathbf{q}\right) =\mathbf{p}\cdot \mathbf{X}+\phi \left( 
\mathbf{q}\right) .$ In this case, the dual map 
\begin{equation*}
\Phi :\mathfrak{g}^{\ast }\rightarrow \mathfrak{s}^{\ast }:\Pi
_{id}\rightarrow \left( \rho ,M\right)
\end{equation*}%
of generalized complete cotangent lift $\mathbb{X}\rightarrow \mathbb{X}%
^{c\ast }$ is the first two moments of the momentum-Vlasov variables 
\begin{equation}
\rho \left( q\right) =\int \left( \nabla _{q}\cdot \mathbf{\Pi }_{p}\right)
d^{3}\mathbf{p}\text{, \ \ }\mathbf{M}=\int (\mathbf{\Pi }_{q}+\mathbf{p}%
\left( \nabla _{q}\cdot \mathbf{\Pi }_{p}\right) )d^{3}\mathbf{p}
\label{mvtocf}
\end{equation}%
and is the plasma-to-fluid map in the momentum variables $\Pi _{id}$. By
direct calculation, one can check that

\begin{proposition}
The mapping in Eq.(\ref{mvtocf}) is a momentum and a Poisson map from the
momentum-Vlasov dynamics on $\mathfrak{g}^{\ast }$ to the compresible fluid
dynamics on $\mathfrak{s}^{\ast }$.
\end{proposition}

The substitution $\mathfrak{g}^{\ast }\rightarrow \mathcal{F}\left( T^{\ast }%
\mathcal{Q}\right) :\Pi _{id}\rightarrow f\left( \mathbf{q},\mathbf{p}%
\right) $ in Eq.(\ref{density}) gives the usual plasma-to-fluid map (for $%
\mathbf{M=}\rho \mathbf{v}$ ) 
\begin{equation*}
f\left( \mathbf{q,p}\right) \mapsto (\rho (\mathbf{q})=\int f\left( \mathbf{%
q,p}\right) d^{3}\mathbf{p}\text{, \ }\mathbf{v(\mathbf{q})=}\int \mathbf{p}%
f\left( \mathbf{q,p}\right) d^{3}\mathbf{p})
\end{equation*}%
in density variable as described in \cite{mwrss83}.

\bigskip

\bigskip

\bigskip

\pagebreak

\bigskip

\section{Quantomorphisms for $1D$ Plasma}

Recall that Hamiltonian function of a Hamiltonian vector field is only
determined up to an additive constant. Based on this and with reference to
the work of Van Hove in \cite{vh51}, it was already (foot-)noted in \cite%
{mw82} that the correct configuration space for the Maxwell-Vlasov equations
is the group of transformations of $%
\mathbb{R}
^{6}\times 
\mathbb{R}
$ preserving the one-form $\mathbf{p}\cdot d\mathbf{q}+ds$. This is the
group of strict contact transformations. In \cite{sw94} it was shown how
certain geometric constructions underlying Vlasov-type equations require the
use of strict contact transformations, also known as quantomorphisms. In
particular, group of quantomorphisms was used in \cite{gv10} in order to
cast Euler's fluid equations on a geometric footing. In this section, we
shall indicate, for the simplest case of one-dimensional plasma, relations
between Vlasov dynamics and coadjoint motion on dual of Lie algebra of group
of quantomorphisms in three-dimensions. Our discussion will be based on
comparisons of the results obtained in section 3 for dynamics of plasma
particles and, in \cite{og11} for dynamics of particles moving with contact
diffeomorphisms, within the general framework outlined in, for example, \cite%
{gt11}. See also \cite{arn89}, \cite{ban97}, \cite{ms98}, \cite{LiMa} for
ingredients from contact geometry and \cite{RSA}, \cite{RaSc}, \cite{Vi97}, 
\cite{Vi} for group of quantomorphisms.

\subsection{Lie algebra of infinitesimal quantomorphisms}

Let $\mathcal{P}$ be a three dimensional manifold with a contact form $%
\sigma \in \Lambda ^{1}\left( \mathcal{P}\right) $ satisfying $d\sigma
\wedge \sigma \neq 0.$ The kernel of $\sigma $ determines a contact
structure on $\mathcal{P}$. A diffeomorphism on $\mathcal{P}$ is called a
contact diffeomorphism if it preserves the contact structure. We denote the
group of contact diffeomorphisms by 
\begin{equation*}
Diff_{con}\left( \mathcal{P}\right) =\left\{ \varphi \in Diff\left( \mathcal{%
P}\right) :\varphi ^{\ast }\sigma =\lambda \sigma ,\lambda \in \mathcal{F}%
\left( \mathcal{P}\right) \right\} .
\end{equation*}%
In Darboux's coordinates $\left( q,p,s\right) $ on $\mathcal{P}$, we take
the contact form to be $\sigma =pdq-ds.$ A vector field on the contact
manifold $\left( \mathcal{P},\sigma \right) $ is a contact vector field if
it generates one-parameter group of contact diffeomorphisms \cite{arn89},%
\cite{ms98}. We denote the space of contact vector fields by 
\begin{equation}
\mathfrak{X}_{con}\left( \mathcal{P}\right) =\left\{ X\in \mathfrak{X}\left( 
\mathcal{P}\right) :\mathcal{L}_{X}\sigma =\bar{\lambda}\sigma ,\bar{\lambda}%
\in \mathcal{F}\left( \mathcal{P}\right) \right\} .  \label{algcon}
\end{equation}

For each real valued function $H=H\left( q,p,s\right) $ on $\mathcal{P}$,
there corresponds a contact vector field 
\begin{equation}
X_{H}=\frac{\partial H}{\partial p}\frac{\partial }{\partial q}-\left( \frac{%
\partial H}{\partial q}+p\frac{\partial H}{\partial s}\right) \frac{\partial 
}{\partial p}+\left( p\frac{\partial H}{\partial p}-H\right) \frac{\partial 
}{\partial s}  \label{Xcon}
\end{equation}%
on $\mathcal{P}$ satisfying the identities%
\begin{equation}
i_{X_{H}}\sigma =H\text{ \ \ \ \ \ \ \ }i_{X_{H}}d\sigma =\left(
i_{R_{\sigma }}dH\right) \sigma -dH,  \label{contact}
\end{equation}%
where $R_{\sigma }=-\partial /\partial s$ is the Reeb vector field of $%
\sigma .$ $R_{\sigma }$ is the unique vector field satisfying $i_{R_{\sigma
}}\sigma =1$ and $i_{R_{\sigma }}d\sigma =0.$ The divergence $div_{d\mu
}X_{H}$ of $X_{H}$ with respect to the contact volume form $d\mu =d\sigma
\wedge \sigma $ can be computed to be $div_{d\mu }X_{H}=2R_{\sigma }H$.
Cartan's formula $\mathcal{L}_{X_{H}}=di_{X_{H}}+i_{X_{H}}d$ and Eq.(\ref%
{contact}) imply that the coefficient functions $\bar{\lambda}$ in Eq.(\ref%
{algcon}) must be of the form $\bar{\lambda}\left( q,p,s\right)
=i_{R_{\sigma }}dH$, that is, 
\begin{equation*}
\mathcal{L}_{X_{H}}\sigma =\left( i_{R_{\sigma }}dH\right) \sigma .
\end{equation*}%
Contact Poisson (or Lagrange) bracket 
\begin{equation}
\left\{ K,H\right\} _{c}=\frac{\partial H}{\partial p}\frac{\partial K}{%
\partial q}-\frac{\partial H}{\partial q}\frac{\partial K}{\partial p}+\frac{%
\partial K}{\partial s}\left( p\frac{\partial H}{\partial p}-H\right) +\frac{%
\partial H}{\partial s}\left( K-p\frac{\partial K}{\partial p}\right) ,
\label{Lagbra}
\end{equation}%
of two smooth functions $H$ and $K$ on $\mathcal{P}$ induces a Lie algebra
structure on $\mathcal{F}\left( \mathcal{P}\right) $. The identity $-\left[
X_{K},X_{H}\right] _{JL}=X_{\left\{ K,H\right\} _{c}}$ establishes the
isomorphism 
\begin{equation}
\left( \mathfrak{X}_{con}\left( \mathcal{P}\right) ,-\left[ \text{ },\text{ }%
\right] _{JL}\right) \longleftrightarrow \left( \mathcal{F}\left( \mathcal{P}%
\right) ,\left\{ \text{ },\text{ }\right\} _{c}\right)  \label{iso1}
\end{equation}%
between the Lie algebras of functions and contact vector fields. Here, $%
\left[ \text{ },\text{ }\right] _{JL}$ is the Jacobi-Lie bracket of vector
fields.

An element $\varphi \in Diff_{con}\left( \mathcal{P}\right) $ is called a
strict contact transformation or a quantomorphism if $\varphi ^{\ast }\sigma
=\sigma $. We denote the group of quantomorphisms by $Diff_{con}^{st}\left( 
\mathcal{P}\right) .$ The Lie algebra of $Diff_{con}^{st}\left( \mathcal{P}%
\right) $ is 
\begin{equation*}
\mathfrak{X}_{con}^{st}\left( \mathcal{P}\right) =\left\{ X_{H}\in \mathfrak{%
X}_{con}\left( \mathcal{P}\right) :\mathcal{L}_{X_{H}}\sigma =0\right\}
\end{equation*}%
which requires, for elements of $\mathfrak{X}_{con}^{st}\left( \mathcal{P}%
\right) ,$ the condition $\bar{\lambda}=i_{R_{\sigma }}dH=0$ on the function 
$H$. This gives $\partial H/\partial s=0.$ Hence, functions associated to
elements of the Lie algebra $\mathfrak{X}_{con}^{st}\left( \mathcal{P}%
\right) $ are independent of the last coordinate in $\left( q,p,s\right) .$
This condition reduces the Lagrange bracket in Eq.(\ref{Lagbra}) to the
nondegenerate Poisson bracket 
\begin{equation}
\left\{ K,H\right\} _{T^{\ast }Q}=\frac{\partial H}{\partial p}\frac{%
\partial K}{\partial q}-\frac{\partial H}{\partial q}\frac{\partial K}{%
\partial p}  \label{Poisson in2}
\end{equation}%
on a two dimensional cotangent bundle $T^{\ast }\mathcal{Q}$ with local
coordinates $\left( q,p\right) $.

Thus, we are led to consider the principal circle bundle 
\begin{equation}
S^{1}\rightsquigarrow (\mathcal{P},\sigma )\overset{pr}{\longrightarrow }%
\mathcal{(}T\mathcal{^{\ast }Q},\Omega _{T^{\ast }\mathcal{Q}}),
\label{qbundle}
\end{equation}%
which is the so called quantization bundle of the symplectic manifold $%
\left( T^{\ast }\mathcal{Q},\Omega _{T^{\ast }\mathcal{Q}}\right) $, with $%
\sigma =pr^{\ast }\theta _{T^{\ast }\mathcal{Q}}-ds$ where $\theta _{T^{\ast
}\mathcal{Q}}=pdq$ is the Liouville one-form on $T^{\ast }\mathcal{Q}$ and $%
pr^{\ast }\Omega _{T^{\ast }\mathcal{Q}}=d\sigma $ \cite{gt11}. Since $dim%
\mathcal{Q}=1$, $T^{\ast }\mathcal{Q}$ is the phase space of one-dimensional
plasma particles. The contact form $\sigma $ may be regarded as a connection
one-form on $\mathcal{P}\rightarrow T^{\ast }Q$. With respect to this, the
horizontal lift of a vector field $X$ on $T^{\ast }Q$ to a vector field on $%
\mathcal{P}$ is locally given by%
\begin{equation*}
X\rightarrow X^{hor}=X+\sigma \left( X\right) \frac{\partial }{\partial s}.
\end{equation*}%
For a Hamiltonian vector field $X_{h}$, the vector field 
\begin{equation}
X_{h}^{st}=X_{h}^{hor}+\left( h\circ pr\right) R_{\sigma }  \label{Xst}
\end{equation}%
is an infinitesimal quantomorphism, that is, an element of $\mathfrak{X}%
_{con}^{st}\left( \mathcal{P}\right) $. Here, we regard $h$ to be
restriction of a function $H$ on $\mathcal{P}$ associated to a strict
contact vector field. In Darboux's coordinate, we have 
\begin{equation*}
X_{h}^{st}=\frac{\partial h}{\partial p}\frac{\partial }{\partial q}-\frac{%
\partial h}{\partial q}\frac{\partial }{\partial p}+\left( p\frac{\partial h%
}{\partial p}-h\right) \frac{\partial }{\partial s}
\end{equation*}%
which can be obtained from the contact vector field in Eq.(\ref{Xcon}) by
imposing the condition $h\left( q,p\right) =H\left( q,p,s=0\right) .$ The
equality 
\begin{equation*}
\left[ X_{h}^{st},X_{k}^{st}\right] _{JL}=\left[ X_{h},X_{k}\right]
_{JL}^{st}=-X_{\left\{ h,k\right\} }^{st}{}
\end{equation*}%
reduces the isomorphism in (\ref{iso1}) to 
\begin{equation}
\left( \mathfrak{X}_{con}^{st}\left( \mathcal{P}\right) ,-\left[ \text{ },%
\text{ }\right] _{JL}\right) \longleftrightarrow \left( \mathcal{F}\left(
T^{\ast }\mathcal{Q}\right) ,\left\{ \text{ },\text{ }\right\} _{T^{\ast }%
\mathcal{Q}}\right) \longleftrightarrow \left( \mathfrak{X}_{ham}\left(
T^{\ast }\mathcal{Q}\right) \times 
\mathbb{R}
,-\left[ \text{ },\text{ }\right] _{JL}\right)  \label{iso2}
\end{equation}%
establishing, at the first step, the isomorphism between the Lie algebra of
infinitesimal quantomorphisms on $\mathcal{P}$ \ and the Poisson bracket
algebra of functions on $T^{\ast }\mathcal{Q}$ and, at the second step, the
isomorhism of Poisson bracket algebra to the Lie algebra of central
extension of Hamiltonian vector fields \cite{ilm06}, \cite{gv10}, \cite%
{gtv11}.

\begin{remark}
The group of quantomorphisms coincide with the automorphism group of the
quantization bundle (\ref{qbundle}). An automorphism of a principal $S-$%
bundle $\mathcal{P\longrightarrow M}$ is a diffeomorphism of $\mathcal{P}$
equivariant with respect to the action of the structure group $S$. If $\Phi :%
\mathcal{P\longrightarrow P}$ is such a diffeomorphism, then it induces a
diffeomorphism $\varphi :\mathcal{M\longrightarrow M}$ of the base manifold
via $pr\circ \Phi =\varphi \circ pr$. If we take $\mathcal{P}$ to be the
trivial bundle $S\times \mathcal{M}$ then the group $Aut(S\times \mathcal{M}%
) $ of automorphisms is isomorphic to $Diff(\mathcal{M})\circledS
Gau(S\times \mathcal{M})$ and the automorphism algebra is $\mathfrak{aut}%
(S\times \mathcal{M})\simeq \mathfrak{X}\left( \mathcal{M}\right) \circledS 
\mathcal{F}\left( \mathcal{M},\mathfrak{s}\right) $ where $\mathcal{F}\left( 
\mathcal{M},\mathfrak{s}\right) $ is the space of $\mathfrak{s-}$ valued
functions on $\mathcal{M}$ \cite{gr09}. Restricting to contact automorphisms
of the circle bundle above one arrives at the central extension $\mathfrak{X}%
_{ham}\left( T^{\ast }\mathcal{Q}\right) \circledS 
\mathbb{R}
$.
\end{remark}

\subsection{Kinetic equations of contact particles}

\begin{proposition}
The dual space $\mathfrak{X}_{con}^{\ast }\left( \mathcal{P}\right) $ of the
algebra $\mathfrak{X}_{con}\left( \mathcal{P}\right) $ of contact vector
fields is 
\begin{equation}
\mathfrak{X}_{con}^{\ast }\left( \mathcal{P}\right) =\left\{ \Gamma \otimes
d\mu \in \Lambda ^{1}\left( \mathcal{P}\right) \otimes Den\left( \mathcal{P}%
\right) :d\Gamma \wedge \sigma -2\Gamma \wedge d\sigma \neq 0\right\}
\label{condual}
\end{equation}%
where $\sigma $ is the contact form and $d\mu =d\sigma \wedge \sigma $ is
the contact volume form on $\mathcal{P}$.

\begin{proof}
This follows from the requirement that the pairing between $\mathfrak{X}%
_{con}\left( \mathcal{P}\right) $ and $\mathfrak{X}_{con}^{\ast }\left( 
\mathcal{P}\right) $ be nondegenerate. We compute%
\begin{eqnarray}
\int_{\mathcal{P}}\left\langle \Gamma ,X_{H}\right\rangle _{\mathcal{P}}d\mu
&=&\int_{\mathcal{P}}\Gamma \wedge i_{X_{H}}d\mu  \notag \\
&=&\int_{\mathcal{P}}\Gamma \wedge \left( i_{X_{H}}d\sigma \right) \wedge
\sigma +\int_{\mathcal{P}}\left( i_{X_{H}}\sigma \right) \Gamma \wedge
d\sigma  \notag \\
&=&\int_{\mathcal{P}}\Gamma \wedge \left( \left( i_{R_{\sigma }}dH\right)
\sigma -dH\right) \wedge \sigma +\int_{\mathcal{P}}H\Gamma \wedge d\sigma 
\notag \\
&=&\int_{\mathcal{P}}H\left( 2\Gamma \wedge d\sigma -d\Gamma \wedge \sigma
\right) ,
\end{eqnarray}%
where we used identities in Eq.(\ref{contact}) and integration by parts and,
omit divergence terms.
\end{proof}
\end{proposition}

The coadjoint action of $\mathfrak{X}_{con}\left( \mathcal{P}\right) $ on $%
\mathfrak{X}_{con}^{\ast }\left( \mathcal{P}\right) $ is given by 
\begin{equation}
ad_{X_{H}}^{\ast }\left( \Gamma \otimes d\mu \right) =\left( \mathcal{L}%
_{X_{H}}\Gamma +\left( div_{d\mu }X_{K}\right) \Gamma \right) \otimes d\mu
\end{equation}%
and the Lie-Poisson bracket on $\mathfrak{X}_{con}^{\ast }\left( \mathcal{P}%
\right) $ is 
\begin{equation}
\left\{ \mathcal{K},\mathcal{H}\right\} \left( \Gamma \right) =-\int_{%
\mathcal{P}}\left\langle \Gamma ,\left[ \frac{\delta \mathcal{K}}{\delta
\Gamma },\frac{\delta \mathcal{H}}{\delta \Gamma }\right] _{JL}\right\rangle
_{\mathcal{P}}d\mu =-\int_{\mathcal{P}}\left\langle \Gamma ,\left[
X_{K},X_{H}\right] _{JL}\right\rangle _{\mathcal{P}}d\mu ,  \label{conLP}
\end{equation}%
where $\mathcal{K},\mathcal{H}\ $are functionals on $\mathfrak{X}%
_{con}^{\ast }\left( \mathcal{P}\right) $ and we assume the reflexivity
condition $\delta \mathcal{K}/\delta \Gamma =X_{K},\delta \mathcal{H}/\delta
\Gamma =X_{H}\in \mathfrak{X}_{con}\left( \mathcal{P}\right) $. The
Hamiltonian operator $J_{LP}\left( \Gamma \right) $ associated to the
Lie-Poisson bracket in Eq.(\ref{conLP}) is defined as \cite{olv86} 
\begin{equation}
\left\{ \mathcal{K},\mathcal{H}\right\} \left( \Gamma \right) =\int_{%
\mathcal{P}}\left\langle X_{K},J_{LP}\left( \Gamma \right)
X_{H}\right\rangle _{\mathcal{P}}d\mu .
\end{equation}%
and the Lie-Poisson equations on $\mathfrak{X}_{con}^{\ast }\left( \mathcal{P%
}\right) $ take the form 
\begin{equation}
\dot{\Gamma}=J_{LP}\left( \Gamma \right) X_{H}=-ad_{X_{H}}^{\ast }\Gamma =-%
\mathcal{L}_{X_{H}}\Gamma -\left( div_{d\mu }X_{H}\right) \Gamma .
\label{conalp}
\end{equation}

\begin{proposition}
The Hamiltonian differential operator associated to the Lie-Poisson bracket
in Eq.(\ref{conLP}) is 
\begin{equation}
J_{LP}\left( \Gamma \right) =-%
\begin{pmatrix}
\Gamma _{q}\dfrac{\partial }{\partial q}+\dfrac{\partial }{\partial q}\cdot
\Gamma _{q} & \Gamma _{p}\dfrac{\partial }{\partial q}+\dfrac{\partial }{%
\partial p}\cdot \Gamma _{q} & \Gamma _{s}\dfrac{\partial }{\partial q}+%
\dfrac{\partial }{\partial s}\cdot \Gamma _{q} \\ 
\Gamma _{q}\dfrac{\partial }{\partial p}+\dfrac{\partial }{\partial q}\cdot
\Gamma _{p} & \Gamma _{p}\dfrac{\partial }{\partial p}+\dfrac{\partial }{%
\partial p}\cdot \Gamma _{p} & \Gamma _{s}\dfrac{\partial }{\partial p}+%
\dfrac{\partial }{\partial s}\cdot \Gamma _{p} \\ 
\Gamma _{q}\dfrac{\partial }{\partial s}+\dfrac{\partial }{\partial q}\cdot
\Gamma _{s} & \Gamma _{p}\dfrac{\partial }{\partial s}+\dfrac{\partial }{%
\partial p}\cdot \Gamma _{s} & \Gamma _{s}\dfrac{\partial }{\partial s}+%
\dfrac{\partial }{\partial s}\cdot \Gamma _{s}%
\end{pmatrix}%
,  \label{conLPdens}
\end{equation}%
where $\partial /\partial q\cdot \Gamma _{p}=\Gamma _{p}\partial /\partial
q+\partial \Gamma _{p}/\partial q$, etc.
\end{proposition}

In local coordinates $\left( q,p,s,\Gamma _{q},\Gamma _{p},\Gamma
_{s}\right) $ on $T^{\ast }\mathcal{P}$, the kinetic equations (\ref{conalp}%
) of contact particles take the form 
\begin{eqnarray}
\dot{\Gamma}_{q} &=&-X_{H}\left( \Gamma _{q}\right) +\left( A+\Gamma
_{s}\right) \frac{\partial H}{\partial q}+2\frac{\partial H}{\partial s}%
\Gamma _{q}  \notag \\
\dot{\Gamma}_{p} &=&-X_{H}\left( \Gamma _{p}\right) +A\frac{\partial H}{%
\partial p}+3\frac{\partial H}{\partial s}\Gamma _{p}  \notag \\
\dot{\Gamma}_{s} &=&-X_{H}\left( \Gamma _{s}\right) +A\frac{\partial H}{%
\partial s}+3\frac{\partial H}{\partial s}\Gamma _{s},  \label{kpkd}
\end{eqnarray}%
where $A$ stands for the linear differential operator 
\begin{equation}
A=\Gamma _{p}\frac{\partial }{\partial q}-\Gamma _{q}\frac{\partial }{%
\partial p}+p\Gamma _{p}\frac{\partial }{\partial s}-p\Gamma _{s}\frac{%
\partial }{\partial p}.
\end{equation}

With respect to $L^{2}$-pairing, the dual space of the Lie algebra $\left( 
\mathcal{F}\left( \mathcal{P}\right) ,\left\{ \text{ },\text{ }\right\}
_{c}\right) $ is the space of densities $Den\left( \mathcal{P}\right) $.
Using the Lie algebra isomorphism 
\begin{equation}
\mathcal{F}\left( \mathcal{P}\right) \rightarrow \mathfrak{X}_{con}\left( 
\mathcal{P}\right) :F\rightarrow X_{F},
\end{equation}%
the definition of an element of $Den\left( \mathcal{P}\right) $ can be
obtained from the dual map%
\begin{equation}
\mathfrak{X}_{con}^{\ast }\left( \mathcal{P}\right) \rightarrow Den\left( 
\mathcal{P}\right) :\Gamma \otimes d\mu \rightarrow 2\Gamma \wedge d\sigma
-d\Gamma \wedge \sigma  \label{condenmom}
\end{equation}%
and defines a real valued function $F$ on $\mathcal{P}$ by%
\begin{equation}
Fd\mu =2\Gamma \wedge d\sigma -d\Gamma \wedge \sigma .
\end{equation}%
This is the density of particles moving individually with the right action
of contact diffeomorphisms. In coordinates, we let $\Gamma =\Gamma
_{q}dq+\Gamma _{p}dp+\Gamma _{s}ds$ and recall that $d\mu =dq\wedge dp\wedge
ds$. Then, the density function of contact particles with momentum
coordinates $\Gamma \otimes d\mu \in \mathfrak{X}_{con}^{\ast }\left( 
\mathcal{P}\right) $ becomes 
\begin{equation}
F\left( q,p,s\right) =\frac{\partial \Gamma _{p}}{\partial q}-\frac{\partial
\Gamma _{q}}{\partial p}+p\frac{\partial \Gamma _{s}}{\partial p}+p\frac{%
\partial \Gamma _{p}}{\partial s}-2\Gamma _{s}.  \label{el}
\end{equation}%
With this definition, the Lie-Poisson bracket in Eq.(\ref{conLP}) reduces to
the one%
\begin{equation}
\left\{ \mathcal{K},\mathcal{H}\right\} \left( F\right) =\int_{\mathcal{P}%
}F\left\{ \frac{\delta \mathcal{K}}{\delta F},\frac{\delta \mathcal{H}}{%
\delta F}\right\} _{c}d\mu =\int_{\mathcal{P}}KJ_{LP}\left( F\right) Hd\mu ,
\label{conLP2}
\end{equation}%
on $Den\left( \mathcal{P}\right) $, where we assume $\delta \mathcal{H}%
\mathfrak{/}\delta F=H,$ $\delta \mathcal{K}/\delta F=K\in \mathcal{F}\left( 
\mathcal{P}\right) .$ Here, $J_{LP}\left( F\right) $ is the Hamiltonian
operator for the Lie-Poisson bracket.

\begin{proposition}
The Lie-Poisson equation on $Den\left( \mathcal{P}\right) \simeq \mathfrak{X}%
_{con}^{\ast }\left( \mathcal{P}\right) $, as kinetic equation of contact
particles, is 
\begin{equation}
\dot{F}=-ad_{H}^{\ast }F=J_{LP}\left( F\right) H=\left\{ H,F\right\}
_{c}-2div_{d\mu }\left( X_{H}\right) F  \label{conLPdens2}
\end{equation}%
where the Hamiltonian operator $J_{LP}\left( F\right) $ is 
\begin{equation}
J_{LP}\left( F\right) =X_{F}+4F\frac{\partial }{\partial s}+\frac{\partial F%
}{\partial s}.
\end{equation}

\begin{proof}
The coadjoint action is computed from 
\begin{eqnarray}
\left\langle ad_{H}^{\ast }F,K\right\rangle &=&\left\langle
F,ad_{H}K\right\rangle =\left\langle F,\left\{ H,K\right\} _{c}\right\rangle
\notag \\
&=&\int_{\mathcal{P}}F\left\{ H,K\right\} _{c}d\mu =-\int_{\mathcal{P}%
}F\left( X_{H}\left( K\right) +\frac{\partial H}{\partial s}K\right) d\mu \\
&=&\int_{\mathcal{P}}\left( X_{H}\left( F\right) +div_{d\mu }\left(
X_{H}\right) F-\frac{\partial H}{\partial s}F\right) Kd\mu  \notag \\
&=&\int_{\mathcal{P}}\left( \left\{ F,H\right\} _{c}+div_{d\mu }\left(
X_{H}\right) F-2\frac{\partial H}{\partial s}F\right) Kd\mu  \notag \\
&=&\int_{\mathcal{P}}\left( \left\{ F,H\right\} _{c}+2div_{d\mu }\left(
X_{H}\right) F\right) Kd\mu ,
\end{eqnarray}%
where we used integration by parts at the third line and the identities 
\begin{equation}
\left\{ H,K\right\} _{c}=X_{K}\left( H\right) +\frac{\partial K}{\partial s}%
H=-X_{H}\left( K\right) -\frac{\partial H}{\partial s}K
\end{equation}%
at the second and fourth lines. We single out $H$ in the expression 
\begin{equation}
ad_{H}^{\ast }F=\left\{ F,H\right\} _{c}+2div_{d\mu }\left( X_{H}\right) F,
\label{dencoad}
\end{equation}%
to find the Hamiltonian operator $J_{LP}\left( F\right) $. The verification
of Hamilton's equation Eq.(\ref{conLPdens2}) is a straightforward
calculation which follows directly from the Lie-Poisson in Eq.(\ref{kpkd})
together with the definition (\ref{el}) of $F$.
\end{proof}
\end{proposition}

\subsection{Quantomorphisms and momentum-Vlasov equations}

We shall obtain kinetic equations of a continuum consisting of particles in $%
\mathcal{P}\subset \mathbb{R}^{3}$ moving under the right action of
quantomorphisms. We shall do this by restricting the Lie-Poisson equations
for contact particles to strict contact transformations. We shall establish,
with a proper choice of Hamiltonian functional, the equivalence of kinetic
equations of quantomorphic particles in momentum and density variables to
the momentum-Vlasov and the Vlasov equations for one-dimensional plasma.

Coadjoint action of $\mathfrak{X}_{con}^{st}\left( \mathcal{P}\right) $ on $%
\left( \mathfrak{X}_{con}^{st}\left( \mathcal{P}\right) \right) ^{\ast }$ is%
\begin{equation*}
ad_{X_{h}^{st}}\left( \Gamma \otimes d\mu \right) =\mathcal{L}%
_{X_{h}^{st}}\Gamma \otimes d\mu
\end{equation*}%
because $div_{d\mu }X_{h}^{st}=0,$ $\forall X_{h}^{st}\in \mathfrak{X}%
_{con}^{st}\left( \mathcal{P}\right) .$ Accordingly, the Lie-Poisson
equations for momentum variables $\Gamma \in \left( \mathfrak{X}%
_{con}^{st}\left( \mathcal{P}\right) \right) ^{\ast }$ become 
\begin{eqnarray}
\dot{\Gamma}_{q} &=&-X_{h}^{st}\left( \Gamma _{q}\right) +\left( \tilde{A}%
+\Gamma _{s}\right) \frac{\partial h}{\partial q}  \notag \\
\dot{\Gamma}_{p} &=&-X_{h}^{st}\left( \Gamma _{p}\right) +\tilde{A}\frac{%
\partial h}{\partial p}  \notag \\
\dot{\Gamma}_{s} &=&-X_{h}^{st}\left( \Gamma _{s}\right) ,  \label{keqp}
\end{eqnarray}%
where $\tilde{A}$ stands for the linear differential operator 
\begin{equation*}
\tilde{A}=\Gamma _{p}\frac{\partial }{\partial q}-\Gamma _{q}\frac{\partial 
}{\partial p}-p\Gamma _{s}\frac{\partial }{\partial p}.
\end{equation*}
Using techniques of previous sections the following can readily be verified

\begin{proposition}
The complete cotangent lift $\left( X_{h}^{st}\right) ^{c\ast }$ of an
infinitesimal quantomorphism is 
\begin{equation*}
\left( X_{h}^{st}\right) ^{c\ast }=X_{h}^{st}+\left( \tilde{A}+\Gamma
_{s}\right) \frac{\partial h}{\partial q}\frac{\partial }{\partial \Gamma
_{q}}+\tilde{A}\frac{\partial h}{\partial p}\frac{\partial }{\partial \Gamma
_{p}}
\end{equation*}%
and the kinetic equation (\ref{keqp}) of quantomorphic particles can be
written as 
\begin{equation*}
ver\left( \dot{\Gamma}\right) =V\left( X_{h}^{st}\right) ^{c\ast }.
\end{equation*}
\end{proposition}

\begin{remark}
Eq.(\ref{keqp}) is a system of first order pde for three unkown functions of
essentially two variables $\left( q,p\right) $ because $s$ dependence of the
one-form $\Gamma $ cannot be determined from these equations. That means,
suppressing $s$ dependence, the flow defined by Eq.(\ref{keqp}) is actually
two dimensional.
\end{remark}

The density function of quantomorphic particles may be obtained as for
density of contact particles. In this case, since components of $X_{h}^{st}$
is independent of fiber variable $s$ we get 
\begin{equation}
f\left( q,p\right) =\int F\left( q,p,s\right) ds  \label{redf}
\end{equation}%
where $F\left( q,p,s\right) $ is given by Eq.(\ref{el}).

To relate the quantomorphisms to plasma motion, we will use the Lie algebra
isomorphism into $\mathfrak{X}_{ham}\left( T^{\ast }\mathcal{Q}\right)
\rightarrow \mathfrak{X}_{con}^{st}\left( \mathcal{P}\right) $ given in Eq.(%
\ref{Xst}). The dual of this is the momentum map%
\begin{equation*}
\mathbb{J}_{q}:\left( \mathfrak{X}_{con}^{st}\left( \mathcal{P}\right)
\right) ^{\ast }\rightarrow \mathfrak{X}_{ham}^{\ast }\left( T^{\ast }%
\mathcal{Q}\right) :\Gamma _{q}dq+\Gamma _{p}dp+\Gamma _{s}ds\rightarrow \Pi
_{q}dq+\Pi _{p}dp
\end{equation*}%
defined as $\left\langle \mathbb{J}_{q}\left( \Gamma \right)
,X_{h}\right\rangle =\left\langle \Gamma ,X_{h}^{st}\right\rangle $ and is
given by%
\begin{equation}
\Pi _{q}\left( q,p\right) =\int \Gamma _{q}\left( q,p,s\right) ds\text{, \ \ 
}\Pi _{p}\left( q,p\right) =\int \Gamma _{p}\left( q,p,s\right) ds,\text{ \
\ }\Gamma _{s}=0  \label{momo}
\end{equation}%
for which the Lie-Poisson equations (\ref{keqp}) reduces to 
\begin{eqnarray}
\dot{\Pi}_{q} &=&-X_{h}\left( \Pi _{q}\right) +\Pi _{p}\frac{\partial ^{2}h}{%
\partial q^{2}}-\Pi _{q}\frac{\partial ^{2}h}{\partial q\partial p}  \notag
\\
\dot{\Pi}_{p} &=&-X_{h}\left( \Pi _{p}\right) +\Pi _{p}\frac{\partial ^{2}h}{%
\partial q\partial p}-\Pi _{q}\frac{\partial ^{2}h}{\partial p^{2}}.
\end{eqnarray}%
In particular, choosing $h\left( q,p\right) =p^{2}/2m+e\phi \left( q\right) $
we obtain the momentum-Vlasov equations for one-dimensional plasma. The
density variable defined by Eq.(\ref{redf}) reduces the Lie-Poisson equation
(\ref{conalp}) to one-dimensional Vlasov equation.

\subsection{Hierarachy of Eulerian equations}

One can now expand on the relation between the Poisson-Vlasov equations and
the Euler equations of compressible fluid given by the plasma-to-fluid map.
Combining this with the results of previous sections, we have the following
diagram relating various kinetic and fluid theories

\begin{equation*}
\begin{array}{ccccc}
\underline{Configuration\ Space} &  & \underline{Lie\ Algebra} &  & 
\underline{Dual\ Space\ } \\ 
\text{ \ } & \text{ \ } &  &  &  \\ 
Diff\left( \mathcal{Q}\right) \circledS \mathcal{F}\left( \mathcal{Q}\right)
&  & \mathfrak{X}\left( \mathcal{Q}\right) \circledS \mathcal{F}\left( 
\mathcal{Q}\right) &  & \Lambda ^{1}\left( \mathcal{Q}\right) \times 
\mathcal{F}\left( \mathcal{Q}\right) \\ 
\text{ \ } &  &  &  &  \\ 
\downarrow \text{cotangent lift} &  & \downarrow \text{cotangent lift} &  & 
\uparrow \text{plasma to fluid} \\ 
\text{ \ } &  &  &  &  \\ 
Diff_{can}\left( T^{\ast }\mathcal{Q}\right) &  & \mathfrak{X}_{ham}\left(
T^{\ast }\mathcal{Q}\right) =\mathfrak{g} &  & \Lambda ^{1}\left( T^{\ast }%
\mathcal{Q}\right) /d\mathcal{F}\left( T^{\ast }\mathcal{Q}\right) \\ 
& \text{ \ } &  &  &  \\ 
\downarrow \text{horizontal lift} &  & \downarrow \text{homomorphism} &  & 
\uparrow \text{quanto to plasma} \\ 
& \text{ \ } &  &  &  \\ 
Diff_{con}^{st}\left( T^{\ast }\mathcal{Q\times 
\mathbb{R}
}\right) &  & \mathfrak{X}_{con}^{st}\left( T^{\ast }\mathcal{Q\times 
\mathbb{R}
}\right) &  & \Lambda _{con}^{st}\left( T^{\ast }\mathcal{Q\times 
\mathbb{R}
}\right) \\ 
& \text{ \ } &  &  &  \\ 
\downarrow \text{inclusion} &  & \downarrow \text{inclusion} &  & \uparrow 
\text{contacto to quanto} \\ 
& \text{ \ } &  &  &  \\ 
Diff_{con}\left( T^{\ast }\mathcal{Q\times 
\mathbb{R}
}\right) &  & \mathfrak{X}_{con}\left( T^{\ast }\mathcal{Q\times 
\mathbb{R}
}\right) &  & \Lambda _{con}\left( T^{\ast }\mathcal{Q\times 
\mathbb{R}
}\right)%
\end{array}%
\end{equation*}

In addition, the Poisson map generated by the action of semi-direct product $%
Diff(\mathcal{Q})\circledS \mathcal{F}(\mathcal{Q})$ reduces the
Maxwell-Vlasov equations to the Euler-Maxwell equations. In the limit that
the speed of light tends to infinity, these equations become the
Poisson-Vlasov and compressible fluid equations, respectively. Elimination
of the electric field in Euler-Maxwell equations results in the
magnetohydrodynamics equations \cite{mwrss83}.

\bigskip

\bigskip

\pagebreak

\bigskip

\section{Gauge Symmetries and Poisson Equation}

The general theory of reduction implies that constraints on Eulerian
dynamics can be described as momentum map associated to some gauge
symmetries of the underlying geometric structure. For the Maxwell-Vlasov
system the non-evolutionary Maxwell equations come out as constraints
resulting from the gauge symmetries of the electromagnetic field \cite{mw82}%
. In this section, following the reference \cite{gpd1}, we describe the
Poisson equation as a momentum map associated with the gauge symmetry $%
\mathcal{F}(\mathcal{Q})$ of Hamiltonian dynamics on phase space $T^{\ast }%
\mathcal{Q}$ of particle motion. Such a description is possible only if we
consider the semi-direct product space $\mathcal{F}(\mathcal{Q})\circledS
Diff_{can}(T^{\ast }\mathcal{Q})$, where $\mathcal{F}(\mathcal{Q})$ acts on $%
Diff_{can}(T^{\ast }\mathcal{Q})$ by composition on right. In obtaining
Poisson equation, we rely necessarily on the fact that the dual of Lie
algebra isomorphism into is a momentum map \cite{gs80}, \cite{mr94} because
the Lie algebra bracket on $d\mathcal{F}(\mathcal{Q})$, or equivalently, the
Lie-Poisson bracket on the dual $\Lambda ^{2}(\mathcal{Q})$ is trivial.

\subsection{Actions of $\mathcal{F}(\mathcal{Q})$}

The canonical symplectic structure on $T^{\ast }\mathcal{Q}$ is invariant
under the translation of fiber variable by an exact one-form over $\mathcal{Q%
}.$ This is the gauge transformation of canonical Hamiltonian formalism.
More precisely, let $\mathcal{F}(\mathcal{Q})$ be the additive group of
functions on $\mathcal{Q}.$ Define the Lie algebra of $\mathcal{F}(\mathcal{Q%
})$ to be the space $d\mathcal{F}(\mathcal{Q})$ of exact one-forms on $%
\mathcal{Q}$. $\mathcal{F}(\mathcal{Q})$ acts on $T^{\ast }\mathcal{Q}$ by
momentum translation $\mathbf{p}\rightarrow \mathbf{p}-\nabla _{q}\phi
\left( \mathbf{q}\right) $ for $\phi \in \mathcal{F}(\mathcal{Q}).$ The
generator is given by the vertical lift 
\begin{equation}
X_{\phi }\left( \mathbf{q,p}\right) =-\nabla _{q}\phi \left( \mathbf{q}%
\right) \cdot \nabla _{p}=ver\left( -d\phi \right) \left( \mathbf{q,p}\right)
\end{equation}%
of the one-form $d\phi $. This is a Hamiltonian vector on $T_{q}^{\ast }%
\mathcal{Q}$ with the Hamiltonian function $\phi $ regarded as an element of 
$\mathcal{F(}T^{\ast }\mathcal{Q)}$.

$\mathcal{F}\left( \mathcal{Q}\right) $ acts on $TT^{\ast }\mathcal{Q}$ by
the tangent lift of the fiber translation on $T^{\ast }\mathcal{Q}$. In
coordinates, this action is given by

\begin{equation}
(\mathbf{z},\mathbf{\dot{z})=}\left( \mathbf{q,p,\dot{q},\dot{p}}\right)
\mapsto \left( \mathbf{q,p}-\nabla _{q}\phi \left( \mathbf{q}\right) ,%
\mathbf{\dot{q}},\mathbf{\dot{p}-}(\mathbf{\dot{q}}\cdot \nabla _{q})\nabla
_{q}\phi \left( \mathbf{q}\right) \right)
\end{equation}%
and the generator is the complete tangent lift 
\begin{equation}
X_{\phi }^{c}\left( \mathbf{z},\mathbf{\dot{z}}\right) =-\nabla _{q}\phi
\left( \mathbf{q}\right) \cdot \nabla _{p}-\left( \mathbf{\dot{q}}\cdot
\nabla _{q}\right) \left( \nabla _{q}\phi \left( \mathbf{q}\right) \right)
\cdot \nabla _{\dot{p}}
\end{equation}%
of $X_{\phi }$. This is also a Hamiltonian vector field on $T_{z}T_{q}^{\ast
}\mathcal{Q}$ with respect to the Tulczyjew symplectic two-form $\Omega
_{TT^{\ast }\mathcal{Q}}$ and for the Hamiltonian function%
\begin{equation}
H_{TT^{\ast }\mathcal{Q}}(\mathbf{z},\mathbf{\dot{z}})=\mathbf{\dot{q}}\cdot
\nabla _{q}\phi \left( \mathbf{q}\right) .
\end{equation}%
Moreover, the one forms $\vartheta _{1}$ and $\vartheta _{2}$ in Eqs.(\ref%
{tet1}) and (\ref{tet2}) of the special symplectic structures on $TT^{\ast }%
\mathcal{Q}$ yield \ \ \ \ 
\begin{equation}
i_{X_{\phi }^{c}}\vartheta _{1}=H_{TT^{\ast }\mathcal{Q}},\ \ \ \ \ \
i_{X_{\phi }^{c}}\vartheta _{2}=0
\end{equation}%
upon contraction with the generator $X_{\phi }^{c}$ of the lifted action.

\begin{proposition}
The additive group $\mathcal{F(Q)}$ of functions on $\mathcal{Q}$ acts
symplectically on $T^{\ast }\mathcal{Q}$ and tensorial objects over $T^{\ast
}\mathcal{Q}$.
\end{proposition}

The induced action of $\mathcal{F}\left( \mathcal{Q}\right) $ on tensorial
objects over $T^{\ast }\mathcal{Q}$ includes, in particular, the Tulczyjew
symplectic space $TT^{\ast }\mathcal{Q}$. For another example, the action on 
$T^{\ast }T^{\ast }\mathcal{Q}$ is given by%
\begin{equation}
(\mathbf{q,p,\pi }_{q}\mathbf{,\pi }_{p})\rightarrow \left( \mathbf{q,p}%
-\nabla _{q}\phi \left( \mathbf{q}\right) ,\mathbf{\pi }_{q}+\left( \mathbf{%
\pi }_{p}\cdot \nabla _{q}\right) \nabla _{q}\phi \left( \mathbf{q}\right) ,%
\mathbf{\pi }_{p}\right) ,
\end{equation}%
whose infinitesimal generator is the complete cotangent lift 
\begin{equation}
X_{\phi }^{c\ast }(\mathbf{z},\mathbf{\pi }_{z})=-\nabla _{q}\phi \left( 
\mathbf{q}\right) \cdot \nabla _{p}+\left( \mathbf{\pi }_{p}\cdot \nabla
_{q}\right) \nabla _{q}\phi \left( \mathbf{q}\right) \cdot \nabla _{\pi _{q}}
\end{equation}%
of $X_{\phi }$. The lift $X_{\phi }^{c\ast }$ is a Hamiltonian vector field
on $T^{\ast }T^{\ast }\mathcal{Q}$ with the Hamiltonian function $H_{T^{\ast
}T^{\ast }\mathcal{Q}}(\mathbf{z},\mathbf{\pi })=-\mathbf{\pi }_{p}\cdot
\nabla _{q}\phi \left( \mathbf{q}\right) $ with respect to the canonical
symplectic two-form $\Omega _{T^{\ast }T^{\ast }\mathcal{Q}}$.

The action on zero-forms, that is, on space $\mathcal{F(}T^{\ast }\mathcal{Q)%
} $ of functions is obtained by composition and, the action on top forms, or
equivalently, the space $Den(T^{\ast }\mathcal{Q})$ of densities on $T^{\ast
}\mathcal{Q}$ will be given by composition of the volume density function
with the fiber translation once we choose the Liouville volume $d\mu =\Omega
_{T^{\ast }\mathcal{Q}}^{3}$ as a basis for the space of six-forms.

$X_{\phi }^{c}$ and $X_{\phi }^{c\ast }$ are generators of the action of $%
\mathcal{F}\left( \mathcal{Q}\right) $ on $\mathfrak{g=X}_{ham}(T^{\ast }%
\mathcal{Q)}$ and $\mathfrak{g}^{\ast }\subset \Lambda ^{1}(T^{\ast }%
\mathcal{Q)}$, respectively. As we identified $\mathfrak{g}^{\ast }$ as the
subspace $(\mathfrak{g}^{\ast })^{\sharp }$ of $TT^{\ast }\mathcal{Q}$, it
will be convenient to consider the corresponding generator on this subspace.
Since $\mathbf{\pi }^{\sharp }=(-\mathbf{\pi }_{p},\mathbf{\pi }_{q})$, we
find%
\begin{equation}
X_{\phi }^{c}\left( \mathbf{z},\mathbf{\dot{z}}\right) |_{(\mathfrak{g}%
^{\ast })^{\sharp }}=-\nabla _{q}\phi \left( \mathbf{q}\right) \cdot \nabla
_{p}+\left( \mathbf{\pi }_{p}\cdot \nabla _{q}\right) \left( \nabla _{q}\phi
\left( \mathbf{q}\right) \right) \cdot \nabla _{\dot{p}}
\end{equation}%
and this is Hamiltonian with $-\mathbf{\pi }_{p}\cdot \nabla _{q}\phi \left( 
\mathbf{q}\right) $ with respect to the Tulczyjew symplectic structure.

\subsection{Poisson Equations}

Having the Hamiltonian actions of gauge group $\mathcal{F}\left( \mathcal{Q}%
\right) $ on various spaces over $T^{\ast }\mathcal{Q}$, we can now compute
the momentum maps into the dual space $Den(\mathcal{Q})$ of densities
(three-forms) on $\mathcal{Q}$. To this end, we recall that the true
configuration space of the Poisson-Vlasov dynamics is the semi-direct
product space $\mathcal{F}\left( \mathcal{Q}\right) \circledS
Diff_{can}(T^{\ast }\mathcal{Q})$ with the action of $\mathcal{F}\left( 
\mathcal{Q}\right) $ on second factor given by fiber translation. On the
other hand, the Lie algebra of vector fields generating the Hamiltonian
action of $\mathcal{F}\left( \mathcal{Q}\right) $ is commutative. That
means, we have a trivial Lie-Poisson structure for the first factor. So, we
first consider a convenient framework for the adjoint action of $\mathcal{F}%
\left( \mathcal{Q}\right) $ on its algebra and the corresponding momentum
map \cite{gpd1}.

Let $\Lambda ^{1}(\mathcal{Q})$ be the space of one-forms on $\mathcal{Q}$.
We regard the algebra $d\mathcal{F}\left( \mathcal{Q}\right) $ as a subspace
of $\Lambda ^{1}(\mathcal{Q})$. We obtain the action on $\Lambda ^{1}(%
\mathcal{Q})$ by identifying it with $T^{\ast }\mathcal{Q}$ and take the
action on $d\mathcal{F}\left( \mathcal{Q}\right) $ to be the one induced
from $\Lambda ^{1}(\mathcal{Q})$. Thus, we have 
\begin{equation}
\Lambda ^{0}(\mathcal{Q})\times \Lambda ^{1}(\mathcal{Q})\rightarrow \Lambda
^{1}(\mathcal{Q}):\left( \phi \left( \mathbf{q}\right) ,\mathbf{p\cdot }d%
\mathbf{q}\right) \mapsto \mathbf{p\cdot }d\mathbf{q}-d\phi \left( \mathbf{q}%
\right)
\end{equation}%
where we denote $\Lambda ^{0}(\mathcal{Q})\equiv \mathcal{F}(\mathcal{Q}).$
From an algebraic point of view, the exterior derivative $d:\Lambda ^{0}(%
\mathcal{Q})\rightarrow \Lambda ^{1}(\mathcal{Q})$ can be interpreted as a
map describing a Lie algebra isomorphism (up to addition of constants) of
the additive algebra of functions $\mathcal{F}(\mathcal{Q})$ into the
additive algebra of one-forms $\Lambda ^{1}(\mathcal{Q})$ \cite{gs}. We
define the dual spaces of $d\mathcal{F}(\mathcal{Q})\subset \Lambda ^{1}(%
\mathcal{Q})$ and $\Lambda ^{0}(\mathcal{Q})$ to be the space of two forms $%
\Lambda ^{2}(\mathcal{Q})$ and the space $Den(\mathcal{Q})$ of densities
(three-forms), respectively. The additive algebras $d\mathcal{F}(\mathcal{Q}%
) $ and $\Lambda ^{0}(\mathcal{Q})$ can be identified with their duals by
the Hodge duality operator $\ast $ associated to a Riemannian metric on $%
\mathcal{Q}$. Then, the $L^{2}-$pairing between them becomes%
\begin{equation}
\left\langle \ast d\phi ,d\phi \right\rangle =-\int \phi d\ast d\phi \text{,
\ \ \ \ \ }\left\langle \ast \phi ,\phi \right\rangle =\int \phi ^{2}\ast 1
\end{equation}%
the first of which is non-degenerate for functions satisfying $d\ast d\phi
\neq 0$. We can now compute the momentum map%
\begin{equation}
\mathbb{J}_{\mathcal{F}(\mathcal{Q})}:\Lambda ^{2}(\mathcal{Q})\rightarrow
Den(\mathcal{Q})
\end{equation}%
for the action of gauge group $\mathcal{F}(\mathcal{Q})$ on its Lie algebra
from 
\begin{equation}
\left\langle \mathbb{J}_{\mathcal{F}(\mathcal{Q})}\left( \ast d\phi \left( 
\mathbf{q}\right) \right) ,\phi \left( \mathbf{q}\right) \right\rangle \text{%
\ }=-\int_{\mathcal{Q}}\phi \left( \mathbf{q}\right) d\ast d\phi \left( 
\mathbf{q}\right)  \label{momform}
\end{equation}%
which, for the Euclidean metric on $\mathcal{Q}$, gives $d\ast d\phi \left( 
\mathbf{q}\right) =\nabla _{q}^{2}\phi \left( \mathbf{q}\right) d^{3}\mathbf{%
q}$.

For the momentum map $\mathfrak{g}^{\ast }\rightarrow d\mathcal{F}(\mathcal{Q%
})^{\ast }$ we first recall a property of the vertical lift of one-forms.
The vertical lift of an exact one-form $d\phi \left( \mathbf{q}\right)
=\nabla _{q}\phi \cdot d\mathbf{q}$ is given by $ver\left( d\phi \right)
=\nabla _{q}\phi \cdot \nabla _{p}$. For any function $\phi :\mathcal{Q}%
\rightarrow \mathbb{R}$, $ver\left( d\phi \right) $ is a Hamiltonian vector
field with respect to the canonical two-form $\Omega _{T^{\ast }\mathcal{Q}}$
for the Hamiltonian function $-\phi $. Conversely, any Hamiltonian vector
field which is also vertical can be identified with its Hamiltonian function
on $\mathcal{Q}$. Therefore, we have the identification 
\begin{equation}
\mathcal{F(Q)\leftrightarrow }ver(d\mathcal{F(Q))}=\mathfrak{X}%
_{ham}(T^{\ast }\mathcal{Q})\cap VT^{\ast }\mathcal{Q}.
\end{equation}%
Obviously, the algebra of vertical Hamiltonian vector fields is commutative.
So, we have the commutative subalgebra 
\begin{equation}
\left[ ver(d\mathcal{F}(\mathcal{Q})),ver(d\mathcal{F}(\mathcal{Q}))\right]
=0
\end{equation}%
in the algebra $\mathfrak{X}_{ham}(T^{\ast }\mathcal{Q)}$ of all Hamiltonian
vector fields. If we regard $d\mathcal{F(Q)}$ as a Poisson algebra with zero
Poisson bracket, then the map $ver:d\mathcal{F}(\mathcal{Q})\longrightarrow 
\mathfrak{X}_{ham}(T^{\ast }\mathcal{Q})$ may be interpreted as a Lie
algebra isomorphism-into. Hence, the dual map

\begin{equation}
ver^{\ast }:\mathfrak{g}^{\ast }=\mathfrak{X}_{ham}^{\ast }(T^{\ast }%
\mathcal{Q})\rightarrow d\mathcal{F}(\mathcal{Q})^{\ast }=\Lambda ^{2}(%
\mathcal{Q})
\end{equation}%
is a Poisson map into a trivial Lie-Poisson structure. More conveniently, we
take $ver\circ d:\mathcal{F}(\mathcal{Q})\rightarrow \mathfrak{g}$ and the
dual $\ $map $(ver\circ d)^{\ast }:\mathfrak{g}^{\ast }\rightarrow Den(%
\mathcal{Q})$ is a momentum map given by 
\begin{eqnarray}
\left\langle (ver\circ d)^{\ast }\left( \Pi _{id}\right) ,\phi \right\rangle
&=&\left\langle \Pi _{id},ver\left( d\phi \right) \right\rangle
=\left\langle \Pi _{id}(\mathbf{z}),\nabla _{q}\phi \left( \mathbf{q}\right)
\cdot \nabla _{p}\right\rangle  \notag \\
&=&\int_{T^{\ast }\mathcal{Q}}\mathbf{\Pi }_{p}\left( \mathbf{z}\right)
\cdot \nabla _{q}\phi \left( \mathbf{q}\right) d\mu \left( \mathbf{z}\right)
\notag \\
&=&\int_{T^{\ast }\mathcal{Q}}-\phi \left( \mathbf{q}\right) \nabla
_{q}\cdot \mathbf{\Pi }_{p}\left( \mathbf{z}\right) d\mu \left( \mathbf{z}%
\right) .  \label{vermom}
\end{eqnarray}%
Combining this with the momentum map in Eq.(\ref{momform}) with $\ast $
being defined by the Euclidean metric, we have%
\begin{eqnarray}
\mathbb{J}_{\mathcal{P}} &:&\ast d\mathcal{F}(\mathcal{Q})\times \mathfrak{g}%
^{\ast }\rightarrow Den(\mathcal{Q})  \notag \\
\mathbb{J}_{\mathcal{P}}\left( \ast d\phi \left( \mathbf{q}\right) ,\Pi
_{id}(\mathbf{z})\right) &=&(\nabla _{q}^{2}\phi \left( \mathbf{q}\right)
-\int \nabla _{q}\cdot \mathbf{\Pi }_{p}\left( \mathbf{z}\right) d^{3}%
\mathbf{p)}d^{3}\mathbf{q}
\end{eqnarray}%
whose zero value gives the Poisson equation 
\begin{equation}
\nabla _{q}^{2}\phi \left( \mathbf{q}\right) =\int \nabla _{q}\cdot \mathbf{%
\Pi }_{p}\left( \mathbf{z}\right) d^{3}\mathbf{p}\text{.}  \label{pipoi}
\end{equation}

\begin{proposition}
The zero value of momentum map $\mathbb{J}_{\mathcal{P}}:\ast d\mathcal{F}(%
\mathcal{Q})\times \mathfrak{g}^{\ast }\longrightarrow Den(\mathcal{Q})$ for
the action of gauge group $\mathcal{F}(\mathcal{Q})$ constrains the dynamics
of the momentum-Vlasov equations (\ref{mv1}), (\ref{mv}) on $\mathfrak{g}%
^{\ast }$. Similarly, the zero value of $\ast d\mathcal{F}(\mathcal{Q}%
)\times Den(T^{\ast }\mathcal{Q})\rightarrow Den(\mathcal{Q})$ constrains
the dynamics of the Vlasov equation (\ref{denvlasov}) on $Den(T^{\ast }%
\mathcal{Q})$.
\end{proposition}

To obtain the usual Poisson equation as given by Eq.(\ref{poi}), we think of 
$\mathcal{F}(T^{\ast }\mathcal{Q})$ equipped with the Poisson bracket to be
an algebra isomorphic to the Lie algebra of Hamiltonian vector fields. Then, 
$\mathcal{F}(\mathcal{Q})$ is a commutative subalgebra of $(\mathcal{F}%
(T^{\ast }\mathcal{Q}),\{\;,\;\}_{T^{\ast }\mathcal{Q}})$ corresponding to
the generators of action of $\mathcal{F}(\mathcal{Q})$. Thus, we have the
Lie algebra isomorphism from the additive algebra of functions $\mathcal{F}(%
\mathcal{Q})$ into the Poisson bracket algebra on $\mathcal{F}(T^{\ast }%
\mathcal{Q})$. Using dualization, the momentum map $\mathbb{J}%
_{den}:Den(T^{\ast }\mathcal{Q})\rightarrow Den(\mathcal{Q})$ is \cite{mar82}
\begin{equation}
\left\langle \mathbb{J}_{den}\left( fd\mu \right) ,\phi \right\rangle
=-\int_{T^{\ast }\mathcal{Q}}f\left( \mathbf{z}\right) \phi \left( \mathbf{q}%
\right) d\mu \left( \mathbf{z}\right) .
\end{equation}%
Combining with $\mathbb{J}_{\mathcal{F}(\mathcal{Q})}$ in Eq.(\ref{momform}%
), we have 
\begin{eqnarray}
\mathbb{J}_{\mathcal{P}} &:&\ast d\mathcal{F}(\mathcal{Q})\times Den(T^{\ast
}\mathcal{Q})\rightarrow Den(\mathcal{Q})  \notag \\
\mathbb{J}_{\mathcal{P}}\left( \ast d\phi ,efd\mu \right) &=&-(\nabla
_{q}^{2}\phi \left( \mathbf{q}\right) +e\int_{T^{\ast }\mathcal{Q}}f\left( 
\mathbf{z}\right) d^{3}\mathbf{p})d^{3}\mathbf{q}
\end{eqnarray}%
whose zero value results in Eq.(\ref{poi}). To show that this is equivalent
to the Poisson equation (\ref{pipoi}) one can use the definition of the
density in Eq.(\ref{density}) and omit the divergence terms \cite{gpd1}.

\begin{remark}
The Poisson part of the Poisson-Vlasov system involves as a kinematical
constraint into the variation of the Hamiltonian functional \cite{gpd1}. The
dynamical Vlasov part arises as a non-canonical Hamiltonian system in
Eulerian variables. The constraint imposed by the Poisson equation is
essentially obtained from the action of \ $\mathcal{F}(\mathcal{Q})$ on the
cotangent bundle $T^{\ast }\mathcal{Q}$. In this case, the zero level sets
of momentum mappings are coisotropic \cite{agw89}. In the language of Dirac
formalism, the constraint is first class and hence does not affect the
Poisson bracket on the reduced space. Thus, in obtaining equivalent
dynamical formulations in alternative Eulerian variables we must use the
same constraint.
\end{remark}

\pagebreak

\bigskip

\section{Discussion and Conclusions}

We outline the results of the present work and summarize them
diagrammatically. We comment on implications for other kinetic theories of
the way we obtain the Poisson equation. Finally, we conclude with a
discussion on how the ingredient of this paper may be used to study the
orbital dynamics of plasma.

The configuration space is $G=Diff_{can}(T^{\ast }\mathcal{Q})$. The
individual motion of particles is generated by the Hamiltonian vector field $%
X_{h}\in \mathfrak{g}$. The space of sections of the tangent space $TT^{\ast
}\mathcal{Q}$ of particle phase space admits the direct sum decomposition $%
\mathfrak{g}\mathbf{\circledS }\left( \mathfrak{g}^{\ast }\right) ^{\sharp }$%
. $TT^{\ast }\mathcal{Q}$ is symplectic with the Tulczyjew's symplectic
two-form $\Omega _{TT^{\ast }\mathcal{Q}}$. Since any Hamiltonian vector
field defines a Lagrangian submanifold of $TT^{\ast }\mathcal{Q}$, $%
\mathfrak{g}$ is isomorphic to the space of all Lagrangian submanifolds of $%
(TT^{\ast }\mathcal{Q},\Omega _{TT^{\ast }\mathcal{Q}})$. Complete cotangent
lift $X_{h}^{c\ast }$ of $X_{h}$ is canonically Hamiltonian on $T^{\ast
}T^{\ast }\mathcal{Q}$ with the Hamiltonian function $i_{X_{h}}\Pi _{id}$.
Its vertical representative $VX_{h}^{c\ast }$ gives momentum-Vlasov
equation. Both the complete lifts and their vertical representatives are Lie
algebra isomorphisms into.

The configuration space can be described as the space of Lagrangian
submanifolds of sections of a trivial bundle%
\begin{equation*}
G=Diff_{can}(T^{\ast }\mathcal{Q})\simeq Lag\Gamma (pr_{0},\Omega _{-}%
\mathcal{)}
\end{equation*}%
and the corresponding representation of its Lie algebra is by the space of
Lagrangian submanifolds of Tulczyjew symplectic space

\begin{equation*}
\mathfrak{g}=(\mathfrak{X}_{ham}(T^{\ast }\mathcal{Q});-[,])\simeq
Lag(TT^{\ast }\mathcal{Q},\Omega _{TT^{\ast }\mathcal{Q}}).
\end{equation*}%
This Lie algebra of Hamiltonian vector fields is isomorphic to the algebra
of non-constant functions

\begin{equation*}
(\mathfrak{X}_{ham}(T^{\ast }\mathcal{Q});-[,])\simeq \left( \mathcal{F}%
\left( T^{\ast }\mathcal{Q}\right) /\text{constants},\left\{ \text{ },\text{ 
}\right\} _{T^{\ast }\mathcal{Q}}\right)
\end{equation*}%
with canonical Poisson bracket.

The generalized complete cotangent lift for symmetric contravariant tensor
fields gives%
\begin{equation}
\left[ \mathbb{X},\mathbb{Y}\right] _{SC}^{c\ast }=\left[ \mathbb{X}^{c\ast
},\mathbb{Y}^{c\ast }\right] _{JL},
\end{equation}%
which is a Lie algebra isomorphism into $\mathbb{X}\rightarrow \mathbb{X}%
^{c\ast }:\mathfrak{T}\mathcal{Q}\rightarrow \mathfrak{g}$ with $\left[ 
\mathbb{\ },\text{ }\right] _{SC}$ being the Schouten concomitant of tensor
fields. The dual map gives kinetic moments in momentum variables and is a
Poisson map from the Lie-Poisson bracket on $\mathfrak{g}^{\ast }$ to the
Kuperschmidt-Manin bracket on $\mathfrak{T}^{\ast }\mathcal{Q}$. For the
subalgebra $\mathfrak{X}\left( \mathcal{Q}\right) \times \mathcal{F}\left( 
\mathcal{Q}\right) $ in $\mathfrak{T}\mathcal{Q}$ this construction results
in plasma-to-fluid map in momentum variables.

$\mathfrak{g}$ admits a Lie algebra isomorphism into the Lie algebra of
infinitesimal quantomorphisms, or strict contact transformations%
\begin{equation*}
\mathfrak{X}_{con}^{st}\left( \mathcal{P}\right) =\left\{ X_{H}\in \mathfrak{%
X}_{con}\left( \mathcal{P}\right) :\mathcal{L}_{X_{H}}\sigma =0\right\} ,
\end{equation*}%
of the quantization bundle $S^{1}\rightsquigarrow (\mathcal{P},\sigma )%
\overset{pr}{\longrightarrow }\mathcal{(}T\mathcal{^{\ast }Q},\Omega
_{T^{\ast }\mathcal{Q}})$ over particle phase space. This relates the
kinetic theory of plasma particles to that of particles moving with contact
diffeomorphisms. The Lie algebra of infinitesimal quantomorphisms is
included in the Lie algebra of contact vector fields on $(\mathcal{P},\sigma
)$ which, in turn, is isomorphic to the algebra of functions on $\mathcal{P}$
with Lagrange bracket%
\begin{equation*}
\left( \mathfrak{X}_{con}\left( \mathcal{P}\right) ,-\left[ \text{ },\text{ }%
\right] _{JL}\right) \longleftrightarrow \left( \mathcal{F}\left( \mathcal{P}%
\right) ,\left\{ \text{ },\text{ }\right\} _{c}\right) .
\end{equation*}%
Dualizing these relations one obtains the hierarchy of kinetic theories for
contact particles, quantomorphic particles which can be considered to be
plasma particles and, compressible fluid. As is well-known, the last
relation with fluids follows from the fact that the semi-direct product of
diffeomorphisms and functions on $\mathcal{Q}$ can be lifted to canonical
diffeomorphisms on $T^{\ast }\mathcal{Q}$.

Formulation of Vlasov dynamics with the quantomorphism group includes also
particle phase space translations which is missing, at the infinitesimal
level, in present treatment. Infinitesimal quantomorphisms, as central
extension of the algebra of Hamiltonian vector fields, arises naturally in
the dual pair construction of \cite{gtv11} from the requirement that the
action of $\mathfrak{X}_{ham}\left( T^{\ast }\mathcal{Q}\right) $ on certain
space of embeddings be equivariant. The problem caused by constant functions
on the Lie algebra side (no particle dynamics) also arises for the dual
space of densities. In this case, the quotient in the space of densities 
\begin{equation*}
\left( \mathcal{F}\left( T^{\ast }\mathcal{Q}\right) /\mathbb{R}\right)
\otimes \Lambda ^{6}\left( T^{\ast }\mathcal{Q}\right) \simeq \mathfrak{X}%
_{ham}^{\ast }\left( T^{\ast }\mathcal{Q}\right)
\end{equation*}%
is identified with homotheties of particle phase space.

Hamiltonian dynamics of particles has gauge symmetries $\mathcal{F}(\mathcal{%
Q})$. The algebra of these symmetries can be realized as a subalgebra of $%
\mathfrak{g}$. It, thus, acts on $TT^{\ast }\mathcal{Q}$, $T^{\ast }T^{\ast }%
\mathcal{Q}$ and $\mathcal{F}(T^{\ast }\mathcal{Q})$ by Hamiltonian actions.
Combined with the coadjoint action on the dual space $\Lambda ^{2}(\mathcal{Q%
})$, these actions give Poisson equations as zero values of momentum maps
into $Den(\mathcal{Q})$. Accordingly, the relations between particle motion
and its symmetries with the Eulerian dynamical equations may be summarized
by the diagrams 
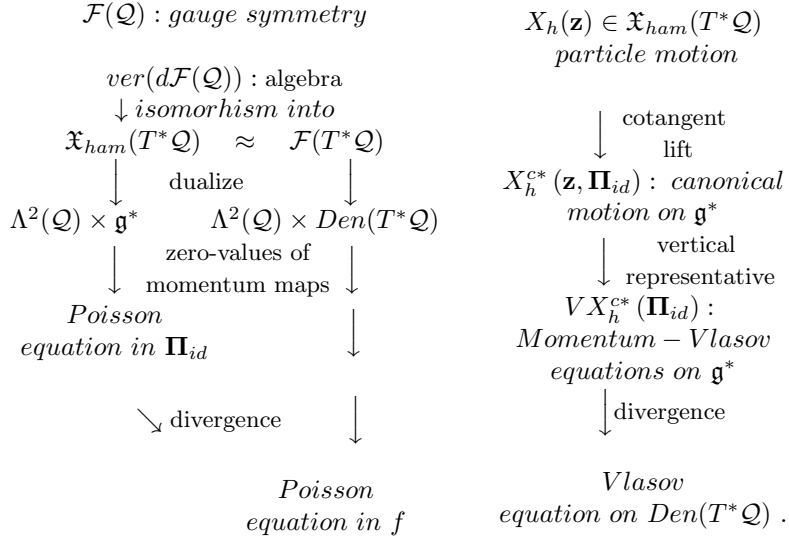
\begin{equation*}
\begin{array}{c}
\begin{array}{c}
\mathcal{F}(\mathcal{Q}):gauge\text{ }symmetry \\ 
\\ 
ver(d\mathcal{F}(\mathcal{Q})):\text{{\small algebra}} \\ 
\downarrow isomorhism\text{ }into \\ 
\mathfrak{X}_{ham}(T^{\ast }\mathcal{Q})\text{ \ \ }\approx \text{ \ \ }%
\mathcal{F}(T^{\ast }\mathcal{Q})%
\end{array}
\\ 
\text{ \ }\Big\downarrow\text{ \ \ \ \ {\small dualize \ \ \ \ \ \ \ \ \ \ }%
\ }\Big\downarrow \\ 
\Lambda ^{2}(\mathcal{Q})\times \mathfrak{g}^{\ast }\text{ \ \ \ \ \ \ \ }%
\Lambda ^{2}(\mathcal{Q})\times Den(T^{\ast }\mathcal{Q}) \\ 
\text{\ \ }\Big\downarrow\text{ }%
\begin{array}{c}
\text{{\small zero-values\ of}} \\ 
\text{{\small momentum maps}}%
\end{array}%
\Big\downarrow \\ 
\begin{array}{c}
Poisson \\ 
equation\text{ }in\text{ }\mathbf{\Pi }_{id}%
\end{array}%
\text{ \ \ \ \ \ \ \ \ \ \ \ \ \ }\Big\downarrow\text{\ \ \ \ \ \ \ \ \ } \\ 
\text{\ \ \ \ \ \ \ \ \ } \\ 
\text{ \ \ \ \ }\searrow \text{{\small divergence \ \ \ \ \ \ \ }}%
\Big\downarrow \\ 
\\ 
\text{ \ \ \ \ \ \ \ \ \ \ \ \ \ \ \ \ \ \ \ \ \ \ }%
\begin{array}{c}
Poisson \\ 
equation\text{ }in\text{ }f%
\end{array}%
\end{array}%
\text{ \ }%
\begin{array}{c}
\begin{array}{c}
X_{h}(\mathbf{z})\in \mathfrak{X}_{ham}(T^{\ast }\mathcal{Q}) \\ 
\;particle\text{ }motion%
\end{array}
\\ 
\\ 
\text{ \ \ \ \ \ }\Big\downarrow%
\begin{array}{c}
\text{{\small cotangent\ }} \\ 
\text{{\small lift}}%
\end{array}
\\ 
\begin{array}{c}
X_{h}^{c\ast }\left( \mathbf{z},\mathbf{\Pi }_{id}\right) :\;canonical \\ 
motion\;on\;\mathfrak{g}^{\ast }%
\end{array}
\\ 
\text{ \ \ \ \ \ \ \ \ \ \ }\Big\downarrow%
\begin{array}{c}
\text{{\small vertical\ }} \\ 
\text{{\small representative}}%
\end{array}
\\ 
\begin{array}{c}
VX_{h}^{c\ast }\left( \mathbf{\Pi }_{id}\right) :\; \\ 
Momentum-Vlasov \\ 
equations\text{ }on\text{ }\mathfrak{g}^{\ast }%
\end{array}
\\ 
\text{ \ \ \ }\Big\downarrow\text{{\small divergence}} \\ 
\\ 
\begin{array}{c}
Vlasov \\ 
equation\text{ }on\text{ }Den(T^{\ast }\mathcal{Q})\text{ .}%
\end{array}%
\end{array}%
\end{equation*}
The Poisson equations (\ref{poi}) and (\ref{pipoi}) constrain the regions in
the product spaces $\Lambda ^{2}(\mathcal{Q})\times Den(T^{\ast }\mathcal{Q}%
) $ and $\Lambda ^{2}(\mathcal{Q})\times \mathfrak{g}^{\ast }$ for
consideration of the plasma dynamics in the Eulerian variables $(\phi
_{f},f) $ and $(\phi _{\Pi },\Pi _{id})$, respectively. More generally, one
can consider canonical Hamiltonian motions of an ensemble of mutually
interacting identical particles. The potential energy acting on individual
particles will be a function of density of particles. Gauge symmetries of
canonical Hamiltonian formulation will then have an action on particle
density. Reduction of Eulerian dynamics of density by gauge symmetries of
individual particle motion will result in Poisson-like equation. The Vlasov
equation is the collisionless limit for one-particle density function of the
more general BBGKY hierarchy of equations governing the evolution of
many-particle density functions \cite{mmw84}. It will be interesting to see
implications, if any, of constraints arising from gauge symmetries of
particle motions\ to\ this hierarchy.

The geometric treatment, on higher order tangent and cotangent bundles over $%
T^{\ast }\mathcal{Q}$, of the momentum-Vlasov equations will act as a model
for an application of Tulczyjew construction for motions on coadjoint
orbits. The coadjoint orbit $\mathcal{O}_{\Pi _{id}}^{\ast }$ through $\Pi
_{id}\in \mathfrak{g}^{\ast }$ admits a symplectic structure induced from
the Lie-Poisson structure. Let $\mathcal{O}_{X_{k}}$ denote the adjoint
orbit in $\mathfrak{g}$ through $X_{k}\in \mathfrak{g}$. Being cotangent
spaces, $T^{\ast }\mathcal{O}_{\Pi _{id}}^{\ast }$ and $T^{\ast }\mathcal{O}%
_{X_{k}}$ are canonically symplectic as well. The situation is similar to
the case of particle motion treated in section 2.3 once we replace the dual
spaces $T^{\ast }\mathcal{Q}$ and $T\mathcal{Q}$ with $\mathcal{O}_{\Pi
_{id}}^{\ast }$ and $\mathcal{O}_{X_{k}}$, respectively. So, we expect the
tangent space $T\mathcal{O}_{\Pi _{id}}^{\ast }$ to the coadjoint orbit to
admit Tulczyjew symplectic structure. As $\mathfrak{g}$ and $\mathfrak{g}%
^{\ast }$ are vector subspaces of $TT^{\ast }\mathcal{Q}$ and $T^{\ast
}T^{\ast }\mathcal{Q}$, respectively, Tulczyjew symplectic structure of $T%
\mathcal{O}_{\Pi _{id}}^{\ast }$ will be the one related to $TT^{\ast
}T^{\ast }\mathcal{Q}$ described in section 3.3. To make these ideas
precise, we shall aim, in our next publication \cite{gpd3}, to construct the
Tulczyjew triple

\begin{equation}
\begin{array}{c}
T^{\ast }\mathcal{O}_{\Pi _{id}}^{\ast }\text{\ }%
\begin{array}{c}
\underleftarrow{\text{ \ \ \ \ \ \ }} \\ 
\mathstrut%
\end{array}%
\text{ \ }T\mathcal{O}_{\Pi _{id}}^{\ast }\text{\ }%
\begin{array}{c}
\underrightarrow{\text{ \ \ \ \ \ \ }}\text{\ } \\ 
\mathstrut%
\end{array}%
\text{ \ \ }T^{\ast }\mathcal{O}_{X_{k}} \\ 
\searrow \text{ \ \ }\swarrow \text{ \ \ \ }\searrow \text{\ \ \ }\swarrow
\\ 
\text{ \ \ \ \ \ }\mathcal{O}_{\Pi _{id}}^{\ast }\text{ \ \ \ \ \ \ \ \ \ \
\ \ }\mathcal{O}_{X_{k}}\text{ \ }%
\end{array}%
\end{equation}%
for orbits of canonical diffeomorphisms.

\section*{Acknowledgement}

We thank one of the anonymous referees for questions and comments that help
us to improve the manuscript and, in particular, raising the question about
quantomorphisms.

\bigskip

\bigskip

\pagebreak

\end{document}